\documentclass[prc,twocolumn]{revtex4-1}
\usepackage[english]{babel}
\usepackage{bm}
\usepackage{soul}
\usepackage{array}
\usepackage{lipsum}
\usepackage{relsize}
\usepackage{dsfont}
\usepackage{mathrsfs}
\usepackage{amssymb}
\usepackage{amsmath}
\usepackage{graphicx}
\usepackage[usenames]{color}
\usepackage{pslatex}
\usepackage{booktabs}

\newcommand{\Bigepsilon}{{\mathlarger{\epsilon}}}

\begin{document}
\title{The Nuclear Physics of Neutron Stars}
\thanks{This is a draft of an article that has been accepted for 
publication by Oxford University Press in the Oxford Research 
Encyclopedia of Physics edited by Brian Foster due for publication 
in 2022.}
\author{J. Piekarewicz}\email{jpiekarewicz@fsu.edu}

\affiliation{Department of Physics, Florida State University,
                Tallahassee, FL 32306, USA}
\date{\today}
\begin{abstract}
Neutron stars---compact objects with masses similar to that 
of our Sun but radii comparable to the size of a city---contain the 
densest form of matter in the universe that can be probed in terrestrial 
laboratories as well as in earth- and space-based observatories. 
The historical detection of gravitational waves from a binary neutron 
star merger has opened the brand new era of multimessenger 
astronomy and has propelled neutron stars to the center of a variety 
of disciplines, such as astrophysics, general relativity, nuclear physics, 
and particle physics. The main input required to study the structure 
of neutron stars is the pressure support generated by its constituents 
against gravitational  collapse. These include neutrons, protons, electrons, 
and perhaps even more exotic constituents. As such, nuclear physics 
plays a prominent role in elucidating the fascinating structure, dynamics, 
and composition of neutron stars.
\end{abstract}
\smallskip

\maketitle

\section{Introduction}
\label{sec:Introduction}

Among the most interesting questions animating nuclear astrophysics
today are: \emph{What are the new states of matter at exceedingly high 
density and temperature?} and \emph{how were the elements from iron 
to uranium made?} In one clean sweep, the historical detection of
gravitational waves from the binary neutron star merger GW170817 by 
the LIGO-Virgo collaboration\,\cite{Abbott:PRL2017} has provided 
critical insights into the nature of dense matter\,\cite{Bauswein:2017vtn, 
Fattoyev:2017jql,Annala:2017llu, Abbott:2018exr, Most:2018hfd, 
Tews:2018chv, Malik:2018zcf,Radice:2017lry, Radice:2018ozg, 
Tews:2019cap, Capano:2019eae,Tsang:2019mlz,Tsang:2020lmb,
Drischler:2020hwi,Landry:2020vaw,Xie:2020rwg,Essick:2021kjb,
Chatziioannou:2021tdi} and on the synthesis of the heavy elements 
in the cosmos\,\cite{Drout:2017ijr,Cowperthwaite:2017dyu,
Chornock:2017sdf,Nicholl:2017ahq}.

The term neutron star appears in writing for the first time in the 1933 
proceedings of the American Physical Society by Baade and  Zwicky 
who wrote:  \emph{With all reserve we advance the view that supernovae 
represent the transition from ordinary stars into ``neutron stars'', 
which in their final stages consist of extremely closed packed 
neutrons}\,\cite{Baade:1934}. It appears, however, that a couple of 
years earlier Landau speculated on the existence of dense stars 
that look like giant atomic nuclei\,\cite{Yakovlev:2012rd}. The first
quantitative calculation of the structure of neutron stars was
performed by Oppenheimer and Volkoff in 1939 by employing the 
full power of Einstein's theory of general relativity\,\cite{Opp39_PR55}. 
Using what it is now commonly referred to as the Tolman-Volkoff-Oppenheimer 
(TOV) equations\,\cite{Opp39_PR55,Tol39_PR55}, Oppenheimer and Volkoff 
predicted that a neutron star supported exclusively by quantum
mechanical pressure from its constituent neutrons will collapse into a 
black hole once its mass exceeds seven tenths of a solar mass, or
$M_{\rm max}\!=\!0.7\,{\rm M}_{\odot}$. Almost 30 years later, Jocelyn 
Bell  discovered ``pulsars'' which, after a period of great confusion in 
which they were mistaken as potential beacons from an extraterrestrial 
civilization, were finally identified as rapidly rotating neutron 
stars\,\cite{Hewish:1968}. 

Since then, the field has evolved by leaps and bounds, and to date, a few 
thousands neutron stars have been identified\,\cite{Lattimer:2004pg}. 
Among the most impactful discoveries since then is the 
identification of three neutron stars with masses in the vicinity of 
two solar masses\,\cite{Demorest:2010bx,Antoniadis:2013pzd,
Cromartie:2019kug,Fonseca:2021wxt}, although a recent publication 
has reported the observation of a neutron star with an even larger mass of 
about $M_{\rm max}\!=\!2.35\,{\rm M}_{\odot}$\,\cite{Romani:2022jhd}.
These discoveries suggest that the original prediction by Oppenheimer 
and Volkoff underestimates the current limit on the maximum neutron star 
mass by at least a factor of three. It is now known that such dramatic 
discrepancy is associated with the unrealistic assumption that neutrons 
behave as a non-interacting gas of fermions. That is, Oppenheimer and 
Volkoff incorrectly assumed that neutrons in a neutron star behave 
as electrons in a white-dwarf star. Thus, the mere existence of massive 
neutron stars highlights the vital role of nuclear interactions in explaining 
the structure, dynamics, and composition of neutron stars. In this manner, 
the detailed understanding of neutron stars has been transferred to the 
domain of nuclear physics.

The TOV equations are model independent as the only assumption 
behind them is the validity of Einstein's theory of general relativity. 
However, the solution of the TOV equation requires a detailed 
description of the relation between the energy density---which 
provides the source of gravity---and the pressure---which prevents 
the gravitational collapse of the star. This relation is enshrined in the 
equation of state (EOS). Unlike a classical ideal gas where the density 
is low and the temperature is high, the relevant domain for neutron 
stars involves high densities and low temperatures. That is, in neutron 
stars the inter-particle separation is small relative to the thermal de 
Broglie wavelength, thereby highlighting the role of quantum 
correlations. That the inter-particle separation in the core of neutron stars 
is small implies that the dynamics is dominated by the short-distance
repulsion of the nucleon-nucleon interaction. Such a repulsion provides
an additional source of pressure support above and beyond the Fermi
(or Pauli) pressure assumed by Oppenheimer and Volkoff. It is this
additional support that becomes essential in supporting massive 
neutron stars against gravitational collapse. Further, the quantum 
regime of low temperatures and high densities suggests that 
neutron stars may effectively be treated as zero-temperature 
systems. Thus the challenge is to infer the absolute ground state 
of the system over the enormous range of densities and pressures
present in a neutron star.

\vfill

\section{Tomography of a Neutron Star}
\label{sec:Tomography}

\begin{figure}[h]
  \includegraphics[width=0.75\columnwidth]{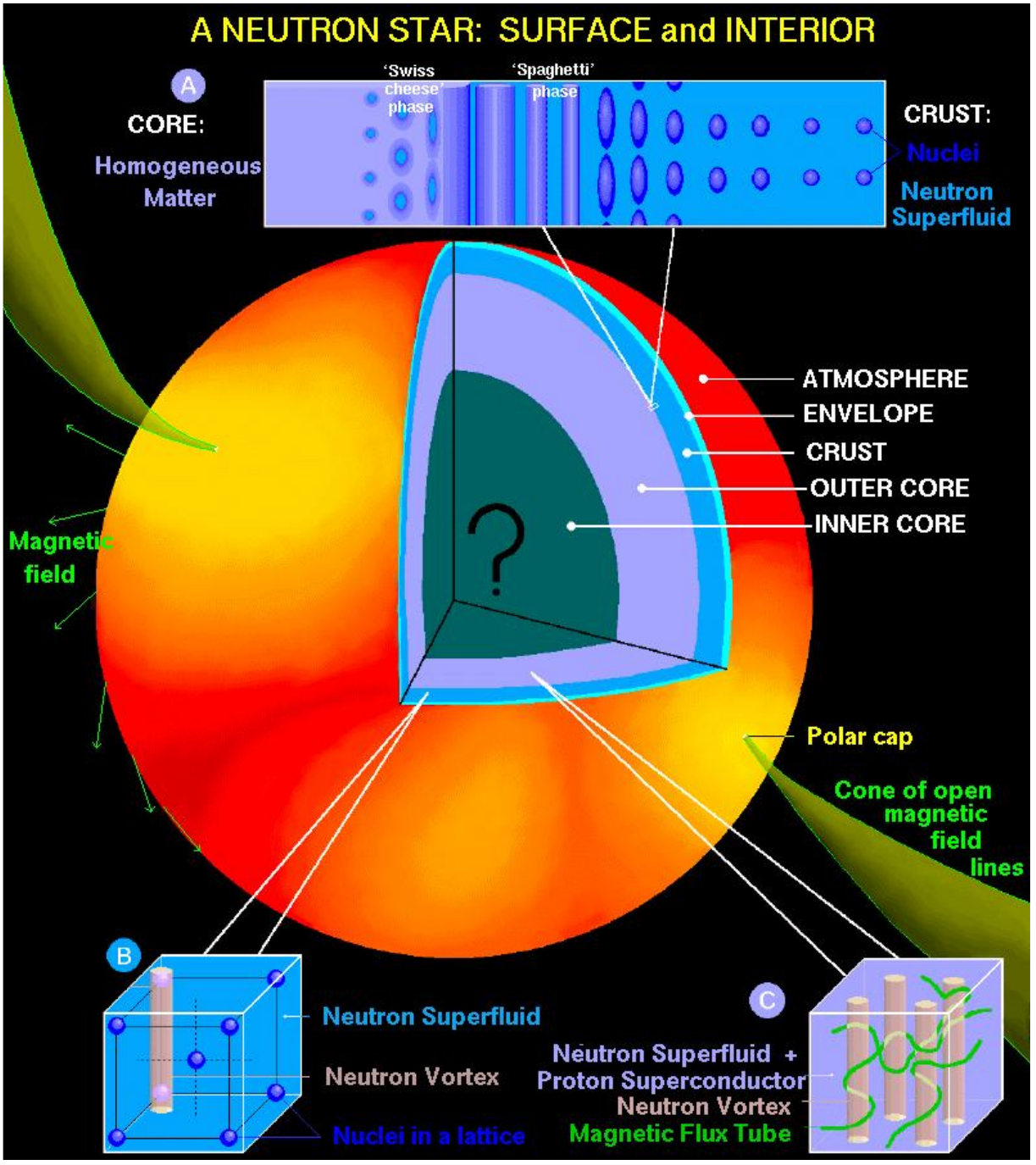}
  \caption{A theoretically-informed rendition of the structure and 
                phases of a neutron star. Courtesy of Dany 
                Page\,\cite{,Page:2006ud}.}
 \label{Fig1}
\end{figure}

The structure of neutron stars is both interesting and complex. Figure\,\ref{Fig1} 
encapsulates our current understanding of the structure and composition of 
neutron stars\,\cite{Page:2004fy,Page:2006ud,Page:2009fu}. The outermost 
surface of the neutron star contains a very thin atmosphere of only a few
centimeters thick that is composed of Hydrogen, but may also contain
heavier elements such as Helium and Carbon. The electromagnetic
radiation emitted from the surface and collected on earth- and
spaced-based telescopes is instrumental in our determination of some
critical parameters of the neutron star. For example, under the
assumption of blackbody emission from the stellar surface, one can
determine the effective surface temperature $T$ of the star. Moreover,
by invoking the Stefan-Boltzmann law, $L\!=\!4\pi\sigma R^{2}T^{4}$,
one can in principle infer the radius of the star, a quantity that
provides critical information on the equation of state. Unfortunately,
complications associated with distance measurements that are required
to determine the absolute stellar luminosity $L$, as well as
atmospheric distortions of the black-body spectrum, make the accurate
determination of stellar radii a challenging
task\,\cite{Ozel:2010fw,Steiner:2010fz,Suleimanov:2010th}. 

Fortunately, the situation has improved  through a better
understanding of systematic uncertainties, important theoretical
developments, and the implementation of robust statistical
methods\,\cite{Guillot:2013wu,Lattimer:2013hma,Heinke:2014xaa,
Guillot:2014lla,Ozel:2015fia,Watts:2016uzu,Steiner:2017vmg,Nattila:2017wtj}. 
More recently, the tidal deformability of neutron stars extracted from  
GW170817\,\cite{Abbott:PRL2017,Abbott:2018exr} together with the
monitoring of hot spots in neutron stars\,\cite{Riley:2019yda,
Miller:2019cac,Miller:2021qha,Riley:2021pdl,Raaijmakers:2021uju}, 
are providing new tools for the accurate determination of
stellar radii. 

Just below the atmosphere lies a thick envelope that acts as a ``blanket'' 
between the hot interior and the relatively cold surface\,\cite{Page:2004fy}. 
Matter in this region is not yet fully degenerate so details associated 
with both the complexity of the atmosphere and large magnetic fields 
are beyond the nuclear physics purview.
However, the layers below, starting with the solid crust and
ending with the high density core, are the main focus of this
article. The non-uniform crust represents a region of about 1\,km that
is composed of a lattice of exotic neutron-rich nuclei immersed in a
uniform, neutralizing electron gas and a quantum liquid of superfluid
neutrons\,\cite{Chamel:2008ca,Bertulani:2012}. Even deeper in the
crust, the universal phenomenon of Coulomb frustration favors the
formation of a complex set of topological phases dubbed ``nuclear
pasta"\,\cite{Ravenhall:1983uh,Hashimoto:1984}. The non-uniform crust
sits above a uniform liquid core that consists of neutrons, protons,
electrons, and muons. The core accounts for practically all the mass
and for about 90\% of the size of a neutron star. Finally, depending 
on the presently unknown largest density that may be attained in the 
stellar core, there is also a possibility, marked with a question mark 
in Fig.\ref{Fig1}, of a transition into a phase made of exotic particles, 
such as hyperons, meson condensates, and deconfined quark
matter\,\cite{Page:2006ud}.

\subsection{The Outer Crust}
\label{OuterCrust}

The typical densities encountered in the outer crust of the neutron
star span about seven orders of magnitude, from $10^{4}{\rm g/cm^{3}}$ 
to about $10^{11}{\rm g/cm^{3}}$\,\cite{Baym:1971pw}. In this
region the average inter-particle separation is significantly larger
than the range of the nucleon-nucleon interaction. As such, the
system---if arranged in a uniform configuration---would not be
able to benefit from the intermediate-range attraction of the 
nucleon-nucleon interaction. Hence, at these densities it becomes 
energetically favorable for the system to break the uniform spatial 
symmetry and to cluster into neutron-rich nuclei segregated into a 
Coulomb lattice that is immersed in a uniform electron gas. 
Electrons are an essential component of the neutron star as they
enforce the overall charge neutrality of the system. Thus, the 
constituents of the outer crust are neutrons, protons, and electrons 
in chemical (or beta) equilibrium. 

In the outer crust, and indeed throughout the entire star, the basic 
question that one needs to answer is how do the relevant constituents 
organize themselves to minimize the energy over the enormous range
of densities encountered in a neutron star. Fortunately, the dynamics 
of the outer crust is relatively simple, as its dependence on the strong 
nuclear interaction is fully encapsulated in the masses of a few atomic 
nuclei. The dynamics of the outer crust is entirely contained in three 
distinct contributions to the Gibbs free energy of the system: lattice, 
electronic, and nuclear. The energy associated with the lattice contribution 
is complicated due to the long-range nature of the Coulomb interaction. 
It consists of divergent contributions that must be carefully canceled as 
required by the overall charge neutrality of the system. However, accurate 
numerical calculations for the electron gas have been available for a long 
time\,\cite{Coldwell:1960,Sholl:1967}, which have since been adapted to the 
problem of interest\,\cite{Baym:1971pw,Ruester:2005fm,RocaMaza:2008ja}. 
In turn, the electronic contribution is encoded in the known energy per particle 
of a relativistic Fermi gas of electrons. All that is left is the nuclear contribution. 
The nuclear contribution is obtained by identifying the optimal nucleus whose
mass per nucleon minimizes the Gibbs free energy per particle at a given 
pressure. In essence, all one needs to compute the composition of the outer 
crust is an accurate nuclear mass table.

\begin{figure}[h]
  \includegraphics[width=0.85\columnwidth]{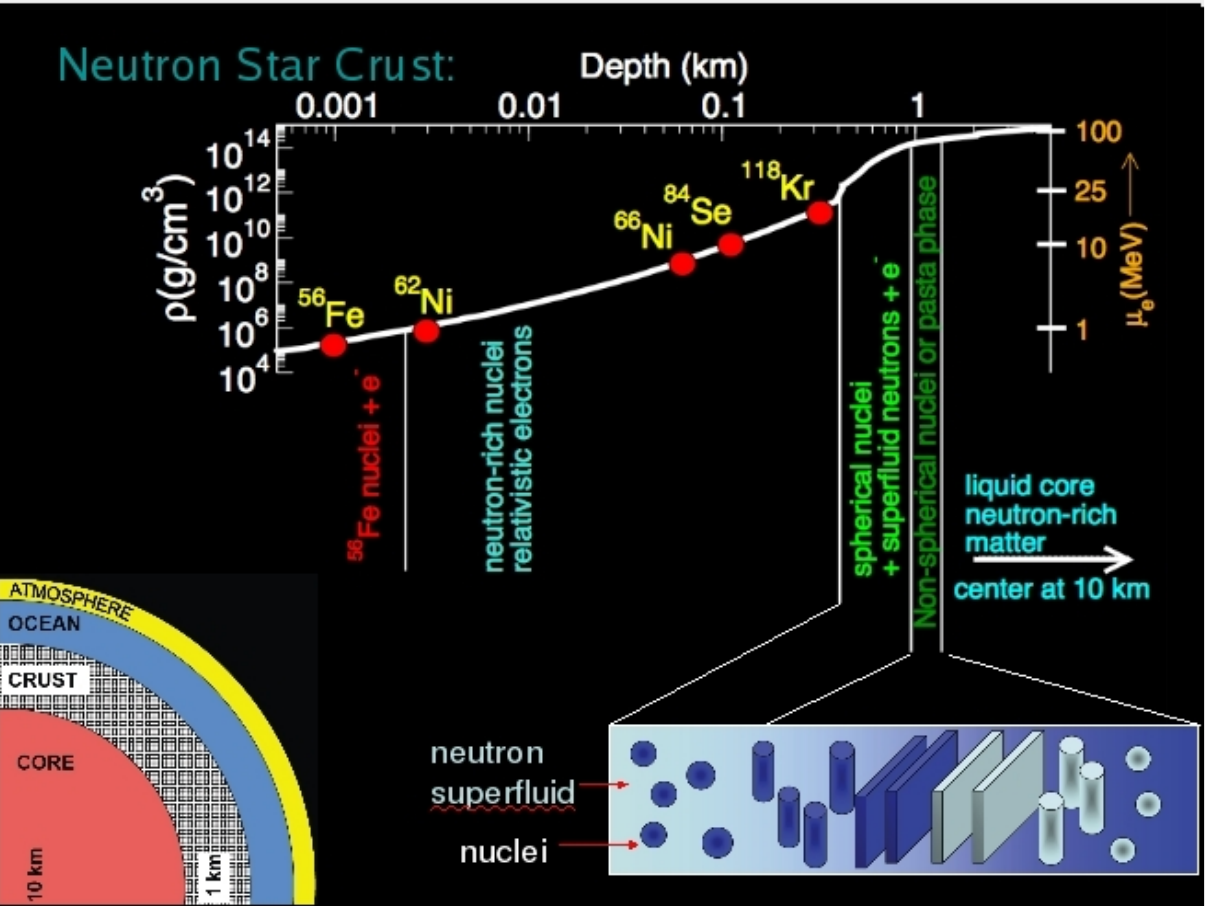}
  \caption{A theoretically-informed rendition of a neutron star with 
            	emphasis on the crustal composition. Courtesy of Sanjay 
	        Reddy.}
 \label{Fig2}
\end{figure}

At the top layers of the outer crust where the density is lowest, it
is energetically favorable for nucleons to cluster into ${}^{56}$Fe
nuclei, which arranged themselves in a face-centered cubic
lattice\,\cite{Baym:1971pw}; see Fig.\,\ref{Fig2}. Iron is the optimal
nucleus at low density because it has the lowest mass per nucleon in
the entire nuclear chart. However, as the density and pressure
increase, ${}^{56}$Fe ceases to be the preferred nucleus. The reason
for this behavior is that the electronic contribution to the Gibbs free 
energy grows rapidly with increasing density/pressure. Thus, it becomes 
energetically advantageous for the energetic electrons to
capture onto protons, making the matter neutron rich. At densities of
about $10^{6}{\rm g/cm^{3}}$, ${}^{62}$Ni becomes the most favorable
nucleus. Note that although ${}^{62}$Ni has a lower binding energy per 
nucleon than ${}^{56}$Fe, it is ${}^{56}$Fe which has a lower mass
per nucleon because its slightly smaller neutron fraction; the entire 
difference amounts to only 12 keV. For a neutron star in hydrostatic equilibrium, 
the pressure and density continue to increase as one moves towards 
the deeper layers of the outer crust. As the density increases, the 
system evolves into a Coulomb lattice of progressively more exotic, 
neutron-rich nuclei\,\cite{Ruester:2005fm,RocaMaza:2008ja}.  Finally, 
at a neutron-drip density of about $10^{11}{\rm g/cm^{3}}$, atomic nuclei 
are incapable of binding any more neutrons. As suggested 
in Fig.\,\ref{Fig2}, most of the existing mass models predict that the 
sequence of progressively more exotic nuclei ends with ${}^{118}$Kr---a 
nucleus with 36 protons and 82 neutrons. 

Ultimately, the determination of the optimal nucleus is set by a competition 
between the electron fraction and the nuclear symmetry energy, a concept 
that appears for the first time in the semi-empirical mass model of Bethe and 
Weizs\"acker\,\cite{Bethe:1936,Weizsacker:1935}. The symmetry energy 
quantifies the energy cost of turning symmetric nuclear matter, with equal 
number of protons and neutrons, into pure neutron matter. Low electron 
fractions diminish the electronic contribution to the Gibbs free energy
but at the expense of creating very neutron-rich nuclei. In turn, a large 
neutron excess  is disfavored by the nuclear symmetry energy. Knowledge 
of the symmetry energy---and especially its density dependence---is one 
of the most important contributions of nuclear physics to our understanding 
of neutron stars.

Given that the last krypton isotope with a well measured mass is ${}^{97}$Kr,
which is still 21 neutrons away from the predicted drip-line nucleus
${}^{118}$Kr, one must rely on mass models that are often hindered by
uncontrolled extrapolations. To mitigate this problem, two important
developments have taken place. First, the highly anticipated Facility
for Rare Isotope Beams (FRIB) is now up and running. FRIB will surpass
current limits of precision and sensitivity that will enable the
production of exotic nuclei at or near the drip lines. Second,
state-of-the-art mass models\,\cite{Duflo:1994,Duflo:1995,
Moller:1997bz,Moller:2012,Goriely:2009mw,Goriely:2013nxa,
Kortelainen:2010hv} have been refined by capitalizing on the
explosion of machine learning techniques\,\cite{Gernoth:1993,
Clark:2006ua,Athanassopoulos:2003qe,Utama:2015hva,
Utama:2017wqe,Utama:2017ytc,Neufcourt:2018syo,
Neufcourt:2019qvd,Neufcourt:2020nme,Niu:2018csp,
Lovell:2022pkw}. To overcome some of the intrinsic limitations 
of existing mass models, machine learning approaches have 
been implemented with the goal of optimizing mass residuals 
between theory and experiment. Beyond a significant improvement 
in the predictions of existing models, the Bayesian refinement provides 
predictions that are accompanied by statistical uncertainties.

Finally, we note that there are three important regions in the nuclear
chart that are critical for our understanding of the outer crust: (a)
the iron-nickel region, (b) the region around the $N\!=\!50$ isotones,
and (c) the region around the $N\!=\!82$ isotones. In the case of the
iron-nickel region, the masses of all relevant nuclei have been
measured with enormous precision. Although not with the same level of
precision as in the iron-nickel region, nuclear masses in the
neighborhood of the $N\!=\!50$ isotonic chain have for the most part
been measured\,\cite{Giraud:2022cgb}. For example, in a landmark
experiment at ISOLTRAP at CERN, the mass of ${}^{82}$Zn 
($Z\!=\!30$  and $N\!=\!52$) was measured for the very first
time\,\cite{Wolf:2013ge}. The addition of this one mass value alone
resulted in an interesting modification to the composition of the
outer crust\,\cite{Wolf:2013ge,Pearson:2011zz}. Lastly, very little
(if at all!) is known about the masses of the very neutron-rich nuclei
in the neighborhood of the $N\!=\!82$ isotonic chain. It is precisely
in this region that machine learning refinements to mass models
informed by experiments at FRIB and other rare-isotope facilities
will become of outmost importance.

\subsection{The Inner Crust}
\label{Inner Crust}

The inner stellar crust comprises the region from neutron-drip density 
up to the density at which uniformity in the system is restored. The 
exact crust-to-core transition density is unknown, but calculations 
predict that it lies between a third to a half of nuclear matter saturation 
density. Based on the nearly uniform interior density of heavy nuclei,
nuclear matter saturation density has been determined to
be around $n_{{}_{\!0}}\!=\!0.15\,{\rm fm^{-3}}$\,\cite{Horowitz:2020evx}, 
corresponding to a mass density of 
$\rho_{{}_{\!0}}\!\approx\!10^{14}{\rm g/cm^{3}}$. 

On the top layers of the inner crust, nucleons continue to cluster into 
a Coulomb crystal of neutron-rich nuclei embedded in a uniform electron 
gas. However, beyond neutron drip, the crystal is in chemical equilibrium 
with a superfluid neutron vapor. Most of the evidence in favor of neutron 
pairing in the crust is obtained from studying its early thermal relaxation 
period\,\cite{Page:2009fu}. An interesting consequence of superfluid pairing 
is the appearance of pulsar ``glitches". Rotation-powered pulsars tend to 
spin down slowly and steadily due to the emission of magnetic dipole 
radiation, making pulsars one of nature's most accurate clocks. For 
example, the 33\,ms Crab pulsar spins down by about $13\,\mu{\rm s}$ 
per year. Despite this remarkable regularity, young pulsars often glitch, 
a unique phenomenon characterized by a sudden and abrupt spin-up 
in their rotational frequency. Pulsar timing has revealed that glitches are
recurrent, with some of the more active glitchers being the Vela and
the Crab pulsars\,\cite{JBPGC}. Although the precise details of the
pulsar glitch mechanism are
unclear\,\cite{Andersson:2012iu,Chamel:2012ae}, it is believed that
the glitch mechanism is intimately related to the formation of
superfluid vortices in the inner crust of the rotating neutron
star; see Refs.\cite{Anderson:1975zze,Alpar:1984,Pines:1985kz}
and references contained therein. As the pulsar slows down, the 
initial distribution of vortices that are believed to be pinned to the 
crystal lattice falls out of equilibrium. This induces a differential rotation 
between the slower neutron star and the faster superfluid vortices. When 
the differential lag is too large, then suddenly and abruptly some fraction 
of the vortices unpin, migrate outwards, and transfer their angular 
momentum to the solid crust---which is detected as a glitch. After this
transient event, the star continues to slow down, stresses between the
crust and the superfluid start to build up again until eventually more
vortices unpin, transfer angular momentum to the solid crust, and
ultimately generate another glitch.

From the perspective of nuclear structure, the most fascinating
component of the inner crust is the nuclear pasta phase.  In the outer
crust and in the top layers of the inner crust, the short-distance
scales associated with the strong interaction is well separated from
the large distance scales characteristic of the Coulomb repulsion. In
this density regime, nucleons bind onto nuclei that are well
segregated in a crystal lattice. However, at the bottom layers of the
inner crust, these length scales become comparable. Competition
between short-range attraction and long-range repulsion leads to a 
universal phenomenon known as ``Coulomb frustration". Such a
competition becomes responsible for the emergence of complex 
structures of various topologies that are collectively known as nuclear
pasta\,\cite{Ravenhall:1983uh,Hashimoto:1984}.

\begin{figure}[h]
  \includegraphics[width=0.75\columnwidth]{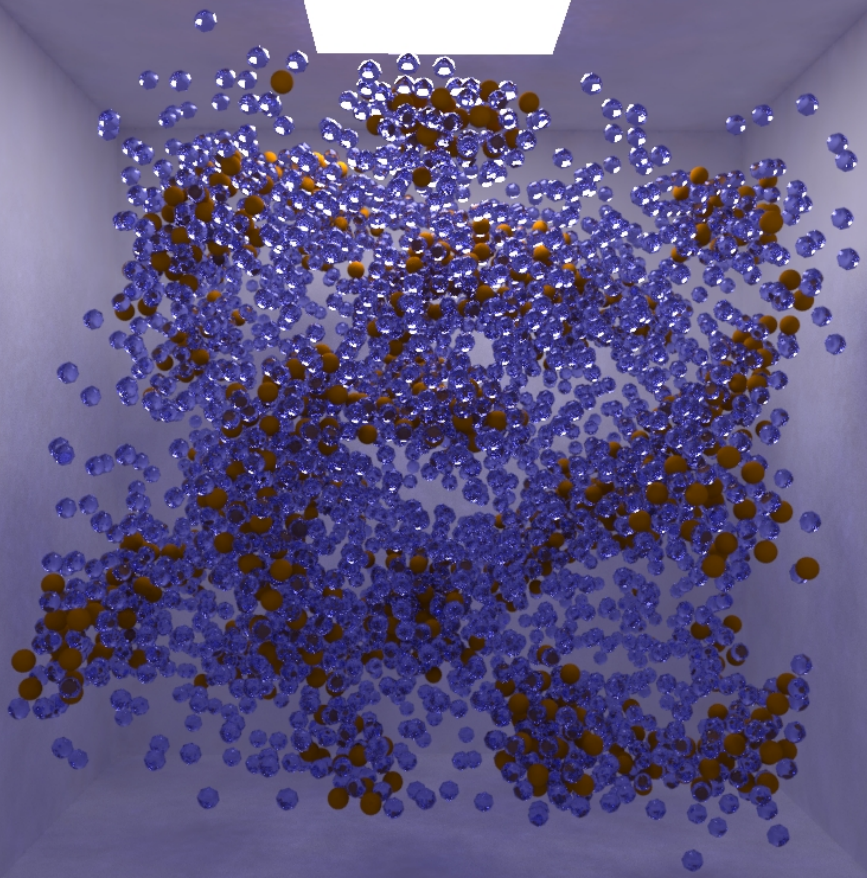}
 \caption{A snapshot of a Monte-Carlo simulation for a system containing 4000 nucleons 
 (3200 neutrons--800 protons) at a baryon density of $n\!=\!n_{{}_{\!0}}/6$ and an 
 effective temperature of $T\!=\!1$\,\,MeV\,\cite{Horowitz:2004yf}.} 
 \label{Fig3}
\end{figure}

Most theoretical approaches to the structure and dynamics of the inner
crust fall into two broad categories: (a) Mean-field models using
accurately calibrated interactions that incorporate quantum effects
but miss some important many-body correlations\,\cite{Maruyama:2005vb,
Avancini:2008zz,Newton:2009zz,Avancini:2012bj,Schuetrumpf:2015nza,
Fattoyev:2017zhb,Newton:2021vyd} and (b) semiclassical Monte Carlo (MC) 
and Molecular Dynamics (MD) formulations where many-body correlations 
are treated properly but quantum effects are incorporated in an approximate
way\,\cite{Watanabe:2003xu,Watanabe:2004tr,Horowitz:2004yf,Watanabe:2009vi,
Piekarewicz:2011qc,Caplan:2014gaa,Schneider:2018yby,Lopez:2020zne,
Shafieepour:2022meo}. Among the virtue of the semi-classical MC/MD 
approaches is that existing computational resources afford the possibility 
of simulating an enormous number of particles, thereby minimizing the 
finite-size effects. For example, Schneider and collaborators have studied 
the formation of multi domains in the nuclear pasta using a simulation box
containing about three million nucleons\,\cite{Schneider:2018yby}. 
Figure\,\ref{Fig3} displays a snapshot of an early Monte-Carlo simulation 
for a system of 4,000 nucleons at a density that clearly identifies the 
formation of non-spherical clusters embedded in a dilute neutron
vapor\,\cite{Horowitz:2004yf,Horowitz:2004pv}. Another great virtue of
the semiclassical simulations is that pasta formation is studied in an
unbiased way without assuming any particular set of shapes. Rather,
the system evolves dynamically into these complex configurations from 
an underlying two-body interaction consisting of a short-range nuclear
attraction and a long-range Coulomb repulsion. This fact is of great
relevance given that one of the main characteristics of
Coulomb-frustrated systems, such as the nuclear pasta, is the
preponderance of quasi-degenerate ground states. That is, small
changes in either the baryon density or the proton fraction can induce
dramatic changes in topology, but with only a very modest gain in
energy\,\cite{Kycia:2017ibr}.  However, such small changes may have a
profound impact on the crustal dynamics and transport
properties\,\cite{Horowitz:2004yf,Horowitz:2004pv,Horowitz:2005zb,
Horowitz:2008vf,Horowitz:2009ya,Grygorov:2010zz,Chamel:2012zn,
Horowitz:2014xca,Caplan:2016uvu,Nandi:2017aqq,Lin:2020nxy}. 

Finally, whereas the study of nuclear pasta is fascinating in its own
right, it may be the key to explain some interesting astrophysical
phenomena such as the lack of X-ray-emitting isolated pulsars with
long spin periods\,\cite{Pons:2013nea} and the slow cooling rate of
MXB 1659-29\,\cite{Brown:2017gxd}, a binary system consisting of
a neutron star accreting from a low-mass companion star. However, 
it is unclear 
how thick the crustal region is that is occupied by the nuclear pasta. 
This feature---as well as the transition density from the non-uniform 
crust to the uniform core---depend critically on the poorly known density
dependence of the symmetry energy\,\cite{Horowitz:2000xj}. If the
symmetry energy is small at the densities of relevance for pasta
formation, then the imbalance between neutrons and protons may be
large, thereby reducing the overall long-range repulsion and thus
hindering the emergence of the pasta phase; after all, a system
made entirely of neutrons does not cluster.

\subsection{The Outer Core}
\label{Outer Core}

At densities of about $10^{14}{\rm g/cm^{3}}$ the nuclear pasta 
``melts'' and uniformity in the system is restored.  Above the 
crust-core transition density the original perception of Baade and 
Zwicky\,\cite{Baade:1934} of a neutron star is finally realized: 
a uniform assembly of extremely closed packed neutrons. At these 
densities the neutron star consists of neutrons, protons, electrons, 
and muons in chemical equilibrium. As a result of chemical 
equilibrium, the proton fraction in the core hovers around 10\% 
with a corresponding fraction of charged leptons required to maintain 
charge neutrality. Muons are important once the Fermi energy of the 
electron equals the rest mass of the muon, which happens in the 
vicinity of nuclear matter saturation density.

Whereas the stellar crust displays fascinating and intriguing
dynamics, its structural impact on the star is rather modest. 
Indeed, with a crustal radius of about 1-2\,km, the core 
accounts for about 90\% of the size and most of the mass
of the neutron star. Thus, the stellar core provides a unique 
laboratory for the study of the ground state of uniform, 
neutron-rich matter over conditions that are unattainable in 
terrestrial laboratories. By the same reason, the equation of 
state of neutron-rich matter is poorly constrained by laboratory
observables. As articulated later in Sec.\ref{EOS}, the most
powerful constraints on the EOS between the core-crust transition
density and about twice saturation density emerge from a combination 
of theory and laboratory experiments. The stellar composition in this 
region is also controlled by the density dependence of the symmetry 
energy. A stiff symmetry energy, namely, one that increases rapidly 
with baryon density, disfavors a large neutron-proton asymmetry. This 
has interesting consequences as a stiff symmetry energy favors  
large proton fractions that are required for the onset of the direct Urca
process\,\cite{Horowitz:2002mb}. The direct Urca process provides a
very efficient mechanism for cooling the star by neutrino emission
from direct beta decay and electron capture\,\cite{Lattimer:1991ib,
Page:2004fy,Page:2006ud,Page:2009fu}. A recent laboratory experiment 
that extracted the neutron skin thickness of ${}^{208}$Pb seems to suggest 
that the symmetry energy in the vicinity of nuclear saturation density 
is stiff\,\cite{Adhikari:2021phr}. One of the many implications of this
experiment is the possibility that the onset of direct Urca process
happens at relatively low central densities and correspondingly 
for relatively low mass neutron stars\,\cite{Reed:2021nqk}. This is
particularly interesting and timely given that x-ray observations
suggest that some neutron stars---for example, the neutron star in the
low-mass x-ray binary MXB 1659-29---may need a fast neutrino-cooling
process\,\cite{Brown:2017gxd}. Although there is still much to learn
about the dynamics of neutron rich matter in terms of conventional
degrees of freedom, perhaps even more exciting is the possible 
emergence of exotic states of matter, besides the nuclear pasta and 
Coulomb crystals that populate the stellar crust. At densities 
above twice saturation density, a region denoted by a question mark
in Fig.\ref{Fig1}, it is plausible that a system of nucleons and charged 
leptons evolves into others form of matter, such as hyperonic matter, 
meson condensates, deconfined quark mater, and/or color 
superconductors. In some cases, the argument in favor for such a 
transition is based on simple qualitative arguments but in other cases 
it requires detailed calculations.

\subsection{The Inner Core}
\label{Inner Core}

Qualitatively, the presence of hyperons in the core of neutron star is 
easy to understand. Hyperons are baryons containing strange quarks and
are therefore heavier than the neutrons and protons that populate the
outer core. About 20\% more massive than a neutron, the
$\Lambda$-hyperon is a neutral baryon with one each of an up, a 
down, and a strange quark. At densities exceeding those found in 
the outer core, it becomes energetically favorable to produce 
$\Lambda$-hyperons rather than to continue to add neutrons
to the system. The emergence of hyperons, analogous to the
emergence of muons, is thus a simple consequence of the
Pauli exclusion principle. Naively, the onset of hyperons should
happen around six times nuclear-matter saturation density. However,
unlike leptons in high density matter, both nucleons and hyperons are
subject to strong interactions in the stellar medium, which are poorly
constrained at the short distances of relevance to the inner 
core. Thus, whereas it is plausible to expect the formation of hyperonic 
matter in the inner stellar core, the quantitative details of the transition 
are highly uncertain. Given that transitions to a new state of matter are 
usually accompanied by a softening of the equation of state, the 
emergence of hyperons yields maximum neutron-star masses that are 
inconsistent with observation\,\cite{Demorest:2010bx,Antoniadis:2013pzd,
Cromartie:2019kug,Fonseca:2021wxt}. This problem is commonly 
referred to in the literature as the ``hyperon puzzle"\cite{Vidana:2000ew,
Sammarruca:2009wn,Weissenborn:2011ut,Massot:2012pf,Lonardoni:2014bwa,
Oertel:2016xsn,Oertel:2016bki,Tolos:2016hhl,Fortin:2017cvt,Negreiros:2018cho,
Logoteta:2021iuy}.

Among the other exotic states of matter that have been proposed to
populate the inner stellar core are meson condensates, particularly 
pion\,\cite{Sawyer:1972,Migdal:1973jkf,Baym:1974vzp,Weise:1975tk,
Migdal:1978} and kaon condensates\,\cite{Kaplan:1986yq,Brown:1992ib,
Thorsson:1993bu,Glendenning:1998zx,Ramos:2000dq}. Pions, kaons,
and in general mesons are bound states containing one quark and
one antiquark. The explanation for the possible emergence of meson 
condensates in neutron stars is relatively simple. The propagation of 
a pion or a kaon in the stellar medium can be drastically different from 
its propagation in free space because of its coupling to ``particle-hole" 
excitations. Unlike most atomic transitions, nuclear excitations often lead 
to collective modes involving the coherent superposition of many 
particle-hole pairs. If the effective particle-hole interaction in the medium 
becomes strong and attractive, the assumed ground state may become 
unstable against the formation of pion- or kaon-like excitations. Such an 
instability signals the restructuring of the ground state into a new state of 
matter in which a fundamental symmetry of the underlying theory is no 
longer realized in the ground state. In this particular case, the broken 
symmetry is parity and the new ground state is a pion or a kaon condensate. 
Given the difficulty in identifying robust astronomical signatures of such exotic 
states of matter and inspired by early theoretical calculations of pion-like 
excitations in atomic nuclei\,\cite{Toki:1980zp,Alberico:1980vb,Alberico:1981sz}, 
experiments have searched---unsuccessfully---for precursors of pion condensation 
in laboratory experiments\,\cite{McClelland:1992yb,Chen:1993fs,Wang:1994zzd,
Pandharipande:1994jc}. As in the case of hyperons, the difficulty in identifying the 
emergence of such exotic states of matter is our poor knowledge of the strong 
interaction under the extreme conditions present in the stellar core.

Lastly, we address the possibility of matter composed of quarks and 
gluons populating the inner stellar core. Quarks and gluons are the
basic constituents of Quantum Chromo Dynamics (QCD) the fundamental 
theory of the the strong interaction. On purely geometrical grounds one
would expect a transition from hadronic matter, in which the quarks and 
gluons are confined within baryons and mesons, to deconfined quark
matter. Given that protons and neutrons have a finite size, one would 
anticipate a transition to some form of quark matter around a critical 
density ($n_{c}$) at which the nucleons start to touch. Assuming a 
close packing of spheres of radius $r_{0}$, one obtains a surprisingly 
low critical density of 
\begin{equation}
 n_{c} = \left(\frac{3}{4\pi r_{0}^3}\right)\left(\frac{\pi}{3\sqrt{2}}\right)
            	= \frac{1}{4\sqrt{2}r_{0}^{3}}\approx 2n_{0},
\end{equation}
where the expression in the second bracket ($\sim\!0.741$) was shown by 
Gauss in 1831 to be the highest packing fraction, and as an estimate of 
$r_{0}$ we used the charge radius of the proton $r_{0}\!\approx\!0.84\,{\rm fm}$.
At $2n_{0}$, QCD is notoriously difficult to solve in the non-perturbative 
regime, even using some of the most sophisticated theoretical approaches
available today. Perhaps a more useful length scale is that at 
which the average inter-quark separation becomes smaller than the QCD 
confining scale, so quarks could roam freely throughout the stellar 
core\,\cite{Lattimer:2004pg}. However, the naive picture of free-roaming
quarks must be revisited because of the attractive interaction between a
pair of quarks\,\cite{Wilczek:2000}. As shown by Bardeen, Cooper, and 
Schrieffer in their seminal work on superconductivity, an arbitrarily 
weak attractive force may cause a dramatic restructuring of the normal 
ground state. Under such a paradigm, QCD predicts that at ultra-high 
densities and low temperatures---where the up, down, and strange quarks 
are effectively massless---the ground state is a color superconductor with 
a unique ``color-flavor-locking" (CFL) pairing scheme\,\cite{Alford:1998mk,
Alford:1997zt,Rajagopal:2000ff,Wilczek:2000,Alford:2007xm}. In the CFL 
phase the exact color symmetry and the approximate flavor symmetry become 
intertwined\,\cite{Wilczek:2000}. Moreover, in this limit of massless quarks, 
three-flavor quark matter is automatically neutral without the need to add
neutralizing leptons; indeed, ``no electrons are required and none are 
admitted"\,\cite{Rajagopal:2000ff}. Unfortunately, it is now believed that 
the extreme densities required for the CFL phase to emerge can not be 
reached in the stellar cores. So assessing the impact of QCD at the densities 
of relevance to neutron stars remains an important challenge that is starting
to be addressed in the regime in the very high-density regime where QCD
is amenable to perturbation theory\,\cite{Annala:2019puf}.

\section{Taming Gravity: The Equation of State}
\label{EOS}

The structure of neutron stars emerges from a fierce competition between gravity 
and the internal pressure support that prevents the collapse of the star. In hydrostatic 
equilibrium, these two opposing forces are perfectly balanced. The law of universal 
gravitation formulated by Newton in 1687 posits that two massive particles attract 
each other with a force that is proportional to the product of their masses and 
inversely proportional to the square of their separation. From all  fundamental 
forces of nature---gravity, electromagnetism, weak, and strong---only gravity is 
always attractive. Such a picture finds its natural explanation in Einstein's
theory of general relativity that attributes the attraction to the curvature
of space-time that all objects---both massive and massless---experience as
they travel through space. Although Newton's law of universal gravitation is adequate
for the description of the motion of planets around the Sun, the gravitational effects 
around a neutron star are so strong that modifications to Newtonian gravity are 
required.

\subsection{The Tolman-Oppenheimer-Volkoff Equations}
\label{TOVEqs}

The structure of neutron stars is encoded in the Tolman-Volkoff-Oppenheimer (TOV) 
equations\,\cite{Opp39_PR55,Tol39_PR55}, which represent the generalization of 
Newtonian gravity to the domain of general relativity. For static, spherically symmetric 
stars in hydrostatic equilibrium the TOV equations may be written as a pair of coupled,
first order differential equations:
\begin{widetext}
\begin{subequations}
 \begin{align}
   \frac{dP(r)}{dr} & = -\frac{G}{c^{2}} \frac{\Big({\Bigepsilon}(r)+P(r)\Big)
      \left(M(r)+4\pi r^{3}\displaystyle{\frac{P(r)}{c^{2}}}\right)}
      {r^{2}\Big(1-2GM(r)/c^{2}r\Big)} \longrightarrow
      -\frac{G\,\rho(r)M(r)}{r^{2}} \;, \\ 
    \frac{dM(r)}{dr} & = 4\pi r^{2}\frac{{\Bigepsilon}(r)}{c^{2}} 
    \rightarrow 4\pi r^{2}\rho(r),
  \end{align}
 \label{TOV}
\end{subequations}
\end{widetext}
where $G$ is Newton's gravitational constant and $c$ is the speed of light in 
vacuum. The structural information is contained in $M(r)$, $\Bigepsilon(r)$, 
$P(r)$, and $\varrho(r)$, which represent the enclosed mass, energy density, 
pressure, and mass density profiles, respectively. The arrows in the above 
expressions represent the Newtonian limit of the TOV equations. These limits 
are obtained by assuming that the corrections from general relativity are negligible, 
as in the case of white-dwarf stars\,\cite{Phillips1998,Weber:1999,
Glendenning:2000,Piekarewicz2016}. 

Once two boundary conditions are provided in terms of a central pressure 
$P(0)\!=\!P_{\!c}$ and an enclosed mass at the origin $M(0)\!=\!0$, the TOV 
equations may be solved numerically using a suitable differential-equation 
solver, such as the Runge-Kutta algorithm\,\cite{NumericalRecipes}. The 
solution of the TOV equations over a wide range of central pressures can 
then be used to generate the entire mass-radius relation, often regarded as 
the ``holy grail"of neutron star structure. Once these equations are solved, 
the radius $R$ and mass $M$ of the neutron star are obtained as 
$P(R)\!=\!0$ and $M\!=\!M(R)$. That is, the radius is determined as the 
point at which the pressure goes to zero and the total mass as the value 
of the enclosed mass at such radius. We note in passing that with the 
exception of a few (mostly academic choices), the TOV equations must be 
solved numerically. This, however, presents some non-trivial challenges 
associated with the diversity of scales involved in the problem. 
Neutrons, with a mass of $m\!\approx\!2\!\times\!10^{-27}\,{\rm kg}$, provide 
most of the pressure support against the gravitational collapse of a 
neutron star with a typical mass comparable to that of our own Sun 
$M_{\odot}\!\approx\!2\!\times\!10^{30}\,{\rm kg}$. This represents a 
mismatch in mass scales of 57 orders of magnitude. Thus, in order 
to avoid computer overflows---and as important, to gain critical insights 
into the problem---one must properly rescale the TOV 
equations\,\cite{Piekarewicz2016}. It is interesting to point out that there 
are approximately $10^{57}$ neutrons in a neutron star, an astronomical
number that makes Avogadro's number of $\sim10^{23}$ pale in 
comparison!  

The validity of general relativity and the existence of a source of pressure 
support is all that is needed to derive the TOV equations. However, to solve 
the equations one must in addition provide an equation of state to connect 
the pressure to the energy density at zero temperature. The best known EOS 
is that of a classical ideal gas. However, unlike the low densities and high 
temperatures that define a classical ideal gas, neutron stars are 
highly-degenerate compact objects where quantum effects are critical. 
In the quantum regime applicable to neutron stars, the average inter-particle 
separation is considerably smaller than the thermal de Broglie wavelength, 
so the fermionic (and bosonic if relevant) nature of the constituents plays a 
critical role. Moreover, whereas treating the constituents as a non-interacting 
Fermi gas may be adequate in the description of white-dwarf stars, assuming 
that neutrons/protons in the stellar interior may be described as a non-interacting 
Fermi gas is woefully inadequate. Indeed, it was precisely such an assumption 
by Oppenheimer and Volkoff that led them to predict a maximum neutron star 
mass of only 
$M_{\rm max}\!=\!0.7\,{\rm M}_{\odot}$\,\cite{Opp39_PR55,Tol39_PR55},
a prediction that is known to be in stark disagreement with the observation 
of  $\gtrsim\!2\,{\rm M}_{\odot}$ neutron stars\,\cite{Demorest:2010bx,
Antoniadis:2013pzd,Cromartie:2019kug,Fonseca:2021wxt}. This underscores
the critical role that strong interactions play in the description of neutron stars. 
In what follows, we discuss the various sources of pressure support and how
recent theoretical, experimental, and observational advances are providing a
new windows into the EOS.

\subsection{The Equation of State Ladder}
\label{Ladder}

\begin{figure}[h]
  \includegraphics[width=0.99\columnwidth]{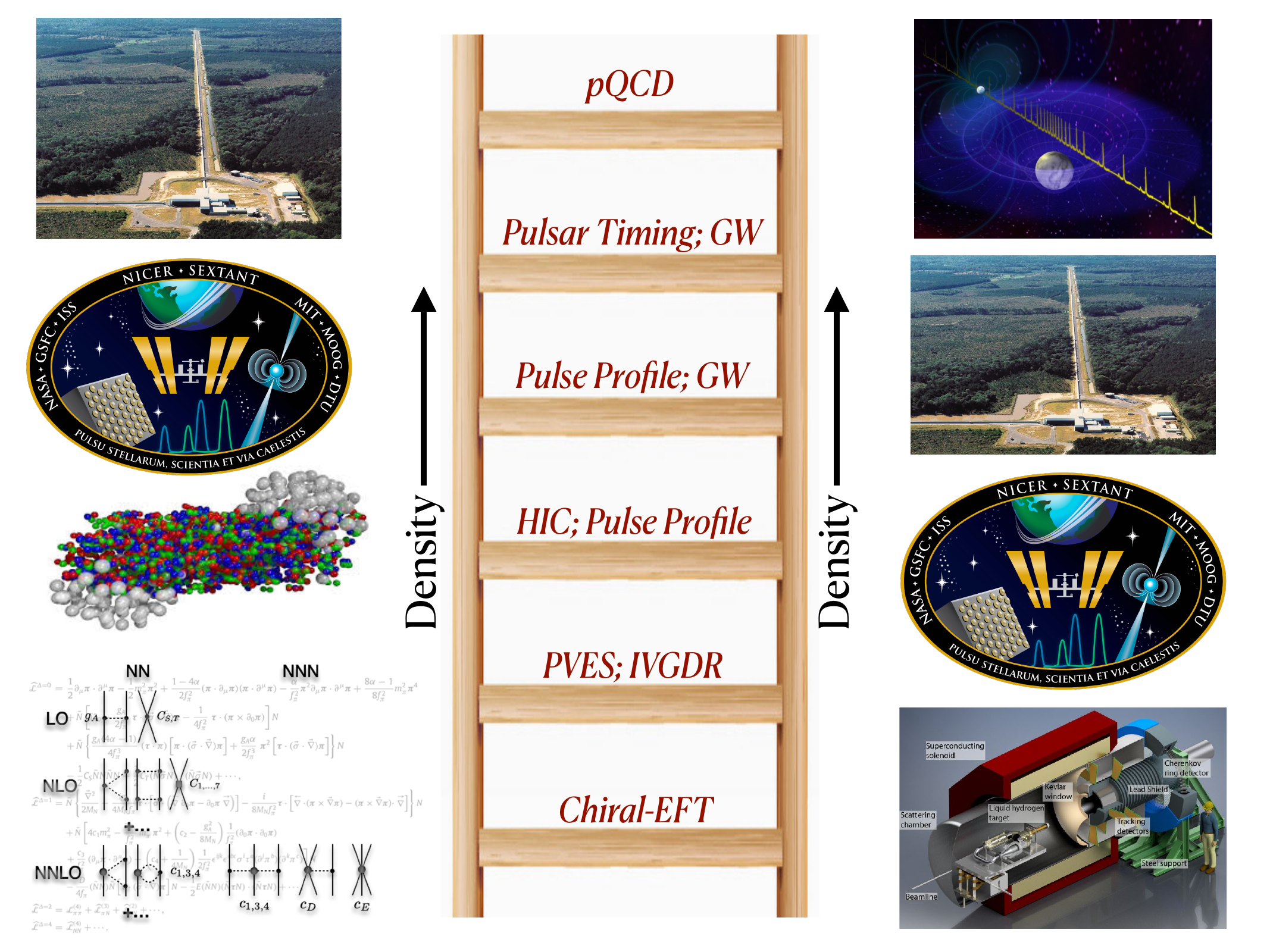}
 \caption{The nuclear equation of state ladder. Akin to the cosmic distance 
 	       ladder in cosmology, the equation of state ladder represents a 
	       succession of theoretical, experimental and observational methods  
	       to determine how the pressure changes with increasing density. 
	       Given that the range of densities spanned in a neutron star is enormous, 
	       no single method can determine the entire EOS. Instead, each rung 
	       of the ladder provides information that can be used to determine the 
	       EOS at a neighboring rung.} 
 \label{Fig4}
\end{figure}

Given that neutron stars would inevitably collapse into black holes in the 
absence of pressure support, a detailed understanding of the equation of 
state is critical. To do so, it is convenient to frame the discussion in terms 
of an ``EOS ladder" akin to the cosmic distance ladder used in cosmology.
In Fig.\ref{Fig4}, each rung in the ladder represents a theoretical, experimental, 
or observational technique that informs the EOS in a suitable density regime. 
The first rung in the ladder invokes chiral effective field theory, a purely 
theoretical approach rooted in the symmetries of QCD\,\cite{Weinberg:1990rz} 
and which has its strongest impact on  the EOS at and below saturation density. 
The next rung in the ladder includes laboratory experiments that constrain the 
EOS in the vicinity of saturation density. Beyond twice saturation density but
still below the maximum density found in neutron stars, the EOS is informed by 
both electromagnetic observations and gravitational wave detections. For the 
highest densities anticipated to exist in the stellar core, the most stringent 
constraints are obtained from the identification of the most massive neutron 
stars. Finally, the last rung in the ladder is motivated by a recent claim that 
perturbative QCD may impose some constraints on the EOS of neutron-star 
matter\,\cite{Annala:2019puf}. The result is somehow surprising given that 
QCD is estimated to become perturbative above 40 times nuclear matter 
saturation density. Still, perturbative QCD may be effective in excluding 
some equations of state displaying some extreme behavior. In the next few 
sections we elaborate on the role of each each rung in the ladder. 
First, however, we start by explaining the marginal role that nuclear interactions 
play in generating the pressure support in the neutron star crust.

Although the structure of the crust with all its exotic phases is fascinating, the
role of nuclear degrees of freedom and their interactions on supporting the 
crust is marginal, since most of the pressure support is provided by the degenerate 
electrons. Still, there are some interesting nuclear-structure effects that must
be taken into consideration. For example, lack of experimental information on 
the masses of the neutron-rich $N\!=\!82$ isotones generates some uncertainty 
in the composition of the outer crust. Given our limited knowledge of the density 
dependence of the symmetry energy, this affects the nuclear composition in the 
bottom layers of the outer crust which, in turn, affects the electron fraction and 
consequently the pressure support. However, the impact of an uncertain
nuclear composition on the EOS is very small. Uncertainties in the inner crust 
are more severe because the ground state after neutron drip is complex and 
because no guidance can be obtained from experiment. However, whereas the 
existence of these exotic phases influence the transport properties in the stellar 
crust, their impact on the EOS is believed to be rather small\,\cite{Chamel:2008ca}. 
Henceforth, we focus exclusively on the study of the EOS above the crust-core 
transition density where the system is uniform and nuclear interactions play a 
predominant role.

\subsection{Theoretical Constraints}
\label{Theory}

The saturation of symmetric nuclear matter, namely, the existence of a nearly
uniform density in the nuclear interior of medium to heavy nuclei is one of the 
hallmarks of the nuclear dynamics. By the same token, finite nuclei can only
provide a very narrow window into the equation of state. Thus, one is forced 
to resort to theoretical approaches to compute the EOS below saturation density. 
Using a Quantum Monte Carlo approach, the equation of state of dilute neutron 
matter has been computed with a simplified interaction that matches smoothly 
to the known analytic results at very low densities and provides important 
information at higher densities\,\cite{Gezerlis:2009iw}. The choice of interaction 
has since been refined by adopting a chiral interaction that is motivated by the 
underlying symmetries of QCD. One of the great virtues of chiral effective field 
theory (EFT) is that it provides a systematic and improvable expansion in terms 
of a suitable small parameter\,\cite{Weinberg:1990rz,vanKolck:1994yi,
Ordonez:1995rz}; for a recent review on the application of chiral EFT to the 
nucleon-nucleon interaction see Ref.\,\cite{RodriguezEntem:2020jgp}. During 
the last decade, enormous progress has been made in our understanding
of the EOS of pure neutron matter by systematically improving the chiral 
expansion\,\cite{Hebeler:2009iv,Tews:2012fj,Kruger:2013kua,Lonardoni:2019ypg,
Drischler:2021kxf,Sammarruca:2021mhv,Sammarruca:2022ser}. Although the 
chiral expansion is known to break down above saturation density, high-order 
chiral EFT calculations of pure neutron matter provide vital constraints on the 
EOS at and below saturation density\,\cite{Alford:2022bpp}.

\subsection{Laboratory Constraints}
\label{Laboratory}

Laboratory experiments on finite nuclei play an essential role in constraining 
the EOS of neutron rich matter in the vicinity of saturation density. Although 
no single technique can be used to determine the EOS over the enormous 
range of densities spanned in a neutron star, several methods provide insights 
over a common range of densities, which can then be used to test for consistency 
and for extending the EOS to the next density rung. In particular, laboratory
experiments provide valuable information on the density dependence of the
symmetry energy. Recall that the symmetry energy is related to the energy 
cost required for converting symmetric nuclear matter into pure neutron matter. 
In particular, the slope of the symmetry energy at saturation density is closely 
related to the corresponding pressure of pure neutron matter---a quantity that
has already been predicted using chiral EFT\,\cite{Hebeler:2009iv,Tews:2012fj,
Kruger:2013kua,Lonardoni:2019ypg,Drischler:2021kxf,Sammarruca:2021mhv,
Sammarruca:2022ser}. 

The slope of the symmetry energy, a quantity denoted as $L$, may also be 
extracted from laboratory experiments. Although not directly an observable, 
$L$ is strongly correlated to the thickness of the neutron skin of heavy 
nuclei\,\cite{Brown:2000,Furnstahl:2001un,RocaMaza:2011pm}. Because 
of the repulsive Coulomb interaction between protons, heavy nuclei tend to 
be neutron rich. For example, the nucleus of ${}^{208}$Pb contains 126 
neutrons but only 82 protons. Hence, the root-mean-square radius of the 
neutron distribution tends to be larger than the radius of the corresponding 
proton distribution. The difference between the neutron and proton radii is 
defined as the neutron skin. 

But what controls the thickness of the neutron skin?\,\cite{Piekarewicz:2019ahf}. 
Although the quantitative value of the neutron skin requires detailed calculations, 
a qualitative explanation of its size and its strong connection to the slope of 
the symmetry energy are largely model independent. Surface tension favors the 
formation of a spherical liquid drop containing all 208 nucleons. However, the 
symmetry energy increases monotonically in the density region of relevance 
to atomic nuclei. So the symmetry energy favors moving some of the excess 
neutrons to the surface where the symmetry energy is lower than in the interior. 
This, however, increases the surface tension. Ultimately then, the thickness of 
the neutron skin emerges from a competition between the surface tension and 
the difference between the value of the symmetry energy in the interior relative 
to its value at the surface. Such a difference is encoded in the slope of the
symmetry energy $L$. If this difference is large, namely, there is a significant 
energy gain by moving some excess neutrons to the surface, then the 
neutron skin will be thick. Conversely, if the energy cost is low, then the
excess neutrons will remain in the core, leading to a thin neutron skin.
This suggests a powerful correlation: the larger the value of $L$ the thicker 
the neutron skin\,\cite{Horowitz:2001ya}. Such a robust correlation has been 
verified using a large set of models that have been calibrated to reproduce 
well-known properties of finite nuclei\,\cite{RocaMaza:2011pm}. 

Motivated by the possible impact of a terrestrial experiment in constraining
the EOS, a proposal was submitted to the Thomas Jefferson 
National Accelerator Facility more than 20 years ago. The proposed goal of 
the experiment was to measure the neutron skin thickness of ${}^{208}$Pb 
using parity violating electron scattering; denoted by PVES in Fig.\ref{Fig4}. 
By measuring the neutron skin thickness of ${}^{208}$Pb, one would be
able to provide valuable information on $L$. Unlike experiments involving 
strongly-interacting probes, such as pions and protons, PVES is clean 
because it relies exclusively on electroweak probes. Indeed, the parity-violating 
asymmetry emerges from the interference of Feynman diagrams involving 
only the exchange of photons and $Z^{0}$ bosons. Moreover, unlike the 
photon, which is largely insensitive to the neutron distribution, the neutral 
weak vector boson $Z^{0}$ couples strongly to the neutrons, making it an 
ideal probe to map the neutron density. 

The first Lead Radius Experiment (``PREX'') established the existence of 
a neutron-rich skin in ${}^{208}$Pb\,\cite{Abrahamyan:2012gp,Horowitz:2012tj}, 
suggesting that the pressure of pure neutron matter is ``stiff", namely, that 
the pressure increases rapidly with increasing density in the vicinity of 
saturation density. Unfortunately, the impact of the experiment was hindered
by unanticipated large error bars. A second effort aimed at improving the 
precision of the original experiment was successfully completed and reported 
a neutron skin thickness that confirmed---with much better precision---that the 
neutron skin of ${}^{208}$Pb is thick\,\cite{Adhikari:2021phr} and, in turn, 
that the symmetry energy is stiff\,\cite{Reed:2021nqk}. Relying on the strong 
correlation between the neutron skin thickness of ${}^{208}$Pb and the slope 
of the symmetry energy\,\cite{RocaMaza:2011pm}, one was able to infer a value 
for $L$ that systematically overestimates existing limits based on both theoretical 
approaches and experimental measurements involving strongly-interacting probes. 
Hence, PREX provides the strongest evidence to date in favor of a stiff EOS around 
saturation density. Moreover, because neutron-star radii are mostly sensitive to the 
pressure around twice saturation density\,\cite{Lattimer:2006xb}, PREX constrains 
both, the thickness of the neutron skin and, although to a lesser extent, the radius 
of a neutron star\,\cite{Carriere:2002bx}. This is a remarkable connection involving 
objects that differ in size by more than 18 orders of magnitude\,\cite{Yang:2019fvs}.
Other laboratory experiments, such as those that excite the isovector giant 
dipole resonance or heavy-ion collisions (IVGDR and HIC in Fig.\ref{Fig4}) 
are also sensitive to the density dependence of the symmetry energy. A
discussion of the advantages and disadvantages of these and many other
experimental techniques may be found in Refs.\,\cite{Tsang:2012se,
Horowitz:2014bja,Thiel:2019tkm}.

\subsection{Observational Constraints I}
\label{Observational-I}

The first rung in the ladder relies on chiral EFT to determine the EOS at low densities. 
Piggybacking on this rung, laboratory experiments extend our knowledge of the EOS
perhaps as far as twice saturation density. In turn, measurements of stellar radii by the
Neutron Star Interior Composition Explorer (NICER) inform the EOS at 
even higher densities. NICER, launched in June 2017 aboard SpaceX's Falcon 9 
rocket and deployed to the International Space Station, is part of NASA's first program 
dedicated to the study of the exotic structure and composition of neutron stars. NICER
relies on the powerful technique of Pulse Profile Modeling to monitor electromagnetic
emission from the hot spots located on the surface of the neutron 
star\,\cite{Psaltis:2013fha,Watts:2016uzu}. Magnetic fields in pulsars are so strong 
and complex that charged particles that are ripped away from the star often crash 
back into the stellar surface creating hot spots, regions within the star that glow 
brighter than the rest of the star. As the neutron star spins, the hot spots come in 
and out of view, producing periodic variations in the brightness, or a distinct pulse 
profile, that are recorded by NICER's powerful instruments. Because of gravitational 
light-bending---one of the most dramatic consequence of the general theory of 
relativity---hot spots emissions are detected by NICER, even when the emission is 
coming from the ``back" of the star. The more compact the star, the more dramatic 
the effect. In this manner, NICER provides critical information on the EOS by 
constraining the stellar compactness:
\begin{equation}
 \frac{{v}^{2}}{c^{2}} = \frac{2GM}{c^{2}R} = \frac{R_{s}}{R},
\end{equation}
where ${v}$, $M$, $R$, and $R_{s}$ are the escape velocity, mass, radius, 
and Schwarzschild radius of the star, respectively. The Schwarzschild radius 
represents the radius at which the neutron star would become a black hole, or
equivalently, when the escape velocity from the star would equal the speed of 
light. For example, a one-solar mass object like our Sun would become a black
hole when its radius reaches about 3\,km.  

Before the deployment of NICER, no single neutron star had both their mass and 
radius simultaneously determined. This changed in 2019 with the determination
of the mass and radius of the millisecond pulsar PSR J0030+0451. Two different
groups, one based in Europe\,\cite{Riley:2019yda} and the other one in the United 
States\,\cite{Miller:2019cac}, have provided independent---and fully 
consistent---estimates. Both groups found that a neutron star with a mass of about 
1.4 solar masses has a radius of approximately 
13\,km\,\cite{Riley:2019yda,Miller:2019cac}.
The second target of the NICER mission was the millisecond pulsar PSR J0740+6620
that was selected because, unlike PSR J0030+0451, its mass was previously 
determined using pulsar timing\,\cite{Cromartie:2019kug,Fonseca:2021wxt}; 
see Section\,\ref{Observational-II}. For this massive pulsar with a mass of more than
twice the mass of the Sun, the extracted neutron star radius was about 
12.4\,km\,\cite{Riley:2021pdl,Miller:2021qha}. In tandem, these two pioneering observations 
suggest that the equation of state in the relevant density domain is relatively  stiff. Moreover, 
these two measurements appear to validate an earlier conjecture that suggests that neutron 
stars have approximately the same radius over a wide mass range \,\cite{Guillot:2013wu}.

\subsection{Gravitational-Wave Constraints}
\label{GravitationalWave}

The historical detection of gravitational waves emitted from the binary 
neutron star merger GW170817 has opened a brand new window into the 
Universe\,\cite{Abbott:PRL2017}. Gravitational waves offer a treasure trove 
of new information that complements electromagnetic observations of myriad 
of astrophysical phenomena. In particular, GW170817 started to provide
answers to two fundamental questions in nuclear science: (i) What are the 
new states of matter at exceedingly high density and temperature? and (ii) 
how were the elements from iron to uranium made? 

Just a few hours after the gravitational-wave detection, ground- and spaced-based 
telescopes identified the associated electromagnetic transient, or ``kilonova'', that 
is believed to have been powered by the radioactive decay of the heavy elements 
synthesized in the rapid neutron-capture process\,\cite{Drout:2017ijr,
Cowperthwaite:2017dyu,Chornock:2017sdf,Nicholl:2017ahq}. 

Concerning the equation of state, GW170817 provides a few critical observables,
such as the chirp mass, the mass ratio, and the tidal deformability. The tidal 
deformability describes the tendency of a neutron star to develop a mass quadrupole 
in response to the tidal field generated by its companion star\,\cite{Damour:1991yw,
Flanagan:2007ix}. In the linear regime of small tidal disturbances, the constant of 
proportionality connecting the tidal forces to the mass quadrupole is the dimensionless 
tidal deformability $\Lambda$ defined as
\begin{equation}
 \Lambda = \frac{2}{3}k_{2}\left(\frac{c^{2}R}{GM}\right)^{5}
                 =\frac{64}{3}k_{2}\left(\frac{R}{R_{s}}\right)^{5}\;,
 \label{Lambda}
\end{equation}
where $k_{2}$ is the second Love number\,\cite{Binnington:2009bb,Damour:2012yf}.
Although $k_{2}$ is sensitive to the underlying EOS, most of the sensitivity is contained
in the compactness parameter\,\cite{Hinderer:2007mb,Hinderer:2009ca,
Damour:2009vw,Postnikov:2010yn,Fattoyev:2012uu,Steiner:2014pda,Fattoyev:2017jql,
Piekarewicz:2018sgy}. Indeed, for a given mass, the tidal deformability scales approximately
as the fifth power of the stellar radius, a quantity that has been notoriously difficult to 
constrain before NICER and LIGO-Virgo\,\cite{Ozel:2010fw,Steiner:2010fz,Suleimanov:2010th,
Guillot:2013wu,Nattila:2017wtj}. In particular, an estimate by the LIGO-Virgo collaboration 
of the tidal deformability of a 1.4\,$M_{\rm sun}$ neutron star yields the relatively small 
value of $\Lambda_{1.4}\!\lesssim\!580$, suggesting that neutron stars are compact objects 
that are difficult to tidally deform\,\cite{Abbott:2018exr}. Such a small value for the tidal 
deformability translates into a stellar radius of approximately 
$R_{1.4}\!=\!11.4\,{\rm km}$\,\cite{Essick:2020flb}. Although consistent within error bars, 
such a low central value for $R_{1.4}$ differs significantly from the $R_{1.4}\!=\!13\,{\rm km}$ 
radius extracted from the first NICER observation. To our knowledge, GW170817 provides 
one of the very few indications that the EOS is soft.

Besided the tidal deformability, other observables of relevance to the EOS are encoded 
in the gravitational waveform. To leading order in the post-Newtonian expansion---an expansion 
in terms of the ratio of the orbital velocity to the speed of light---the gravitational-wave signal 
is determined by the ``chirp'' mass,  which involves a linear combination of the individual stellar 
masses\,\cite{Abbott:2016wyt}. To determine the individual masses one must invoke higher orders 
in the post-Newtonian expansion\,\cite{Cutler:1994ys}. Although the individual masses of GW170817 
could not be determined very precisely, the extracted upper limit on the most massive of the two 
stars\,\cite{Abbott:PRL2017} did not challenge the present maximum-mass limit of about 
$2.1\,M_{\odot}$. This situation could have changed dramatically with the detection of gravitational 
waves from the coalescence of a binary system GW190814 with the most extreme mass ratio ever 
observed: a 23 solar mass black hole and a 2.6 solar mass compact object\,\cite{Abbott:2020khf}. 
The LIGO-Virgo collaboration suggested that GW190814 is unlikely to have originated from a 
neutron star-black hole coalescence, yet left open the possibility that improved knowledge of 
the equation of state or future observations could alter this assessment. 

Within a theoretical framework uniquely suited to investigate both the properties of finite 
nuclei and neutron stars, it was concluded that the low deformability demanded by GW170817 
combined with heavy-ion data, make it highly unlikely that neutron stars can have masses as 
large as $2.6\,{\rm M}_{\odot}$\,\cite{Fattoyev:2020cws}. The absence of very massive neutron stars 
is also consistent with the analysis by Margalit and Metzger who argue against their formation 
based on the lack of evidence of a large amount of rotational energy in the ejecta during the 
spin-down phase of GW170817\,\cite{Margalit:2017dij}. Interestingly, the suggested upper 
limit of $M_{\rm max}\!\lesssim\!2.17\,M_{\odot}$\,\cite{Margalit:2017dij} is in full agreement 
with the recent observation by Cromartie and collaborators of the most massive neutron star 
to date\,\cite{Cromartie:2019kug,Fonseca:2021wxt}. Yet the mystery remains unsolved, as
the existence of black holes with a mass as small as $2.6\,{\rm M}_{\odot}$ is also
problematic. It has even been suggested that GW190814 may have been a second-generation 
merger from a hierarchical triple system: a binary neutron star coalescence leading to the
2.6\,$M_{\rm \odot}$ compact object that ultimately merges with the 23\,$M_{\rm \odot}$
black hole\,\cite{Lu:2020gfh}. Finally, the detection of massive neutrons stars from the binary 
coalescence complements pulsar timing techniques that provide the most precise information 
to date on massive neutron stars and forms the last observational rung in the EOS ladder. 

\subsection{Observational Constraints II}
\label{Observational-II}

According to Newton's law of universal gravitation, all that can be determined 
from the motion of two orbiting objects is their combined mass. This fact is
encapsulated in Kepler's third law of planetary motion:
\begin{equation}
 P^{2} = \frac{4\pi^{2}}{G(M_{1}+M_{2})}a^{3},
 \label{Kepler3}
\end{equation}
where $P$ is the orbital period, $a$ is the length of the semi-major axis of the 
ellipse, and $M_{1}$ and $M_{2}$ are the individual masses of the two orbiting 
bodies. To determine the individual masses one must go beyond Newtonian
mechanics and invoke general relativity. The most successful observational
technique used to determine the individual masses is pulsar timing. The success 
of pulsar timing in extracting orbital parameters stems 
from the remarkable precision of pulsars in keeping time. These celestial
timekeepers often rival the precision of the most sophisticated atomic clocks. 
Millisecond pulsars form a distinct group of  recycled neutron stars that rotate 
extremely fast. It is believed that the rapid rotation involves a spin-up during
mass accretion from a companion star. Further, this class of pulsars generate 
weak magnetic fields of ``only" about 100 million Gauss, leading to very small 
energy losses due to magnetic dipole radiation. The fast rotation and the 
weak magnetic fields make pulsars extremely stable clocks with minuscule 
spin-down rates of about 1 second every trillion years! Long-time observations 
of millisecond pulsars that account for each rotation provide a precise 
determination of the orbital parameters. 

Shapiro delay\,\cite{Shapiro:1964}, regarded as the fourth test of general 
relativity, is a powerful technique that has been used in combination with pulsar 
timing to infer the mass of the companion star, thereby breaking the mass 
degeneracy encountered in Newtonian gravity. The main concept behind the 
Shapiro delay is that even massless particles, such as photons, bend as they
pass near massive objects because of the curvature of space-time. This causes 
a time delay in the arrival of the electromagnetic radiation emitted by the neutron 
star as its passes near the companion star on its way to the observer. In 
accounting for every rotation of the neutron star over long periods of time, 
pulsar timing together with Shapiro delay provide a highly precise value for the 
mass of the companion star. Once the mass degeneracy has been broken, 
Kepler's law may then be used to extract the mass of the neutron star. This 
powerful technique was used back in 2010 to determine both the mass of the 
pulsar J1614-2230 and of its companion white-dwarf star\,\cite{Demorest:2010bx}. 
In particular, the mass of the neutron star was found to be close to two solar
masses, providing the first serious challenge to models that predict soft equations 
of state due to the emergence of exotic states of matter. Recently, Shapiro delay 
was used to measure the mass of the millisecond pulsar J0740+6620, that with 
a mass of about $M\!=\!2.1\,M_{\odot}$ is the most massive---well measured---neutron 
start to date\,\cite{Cromartie:2019kug,Fonseca:2021wxt}; although 
see\,\cite{Romani:2022jhd}. Note that the millisecond pulsar J0740+6620 is 
one of the two neutron stars that was targeted by the NICER mission and 
for which a radius measurement now exists\,\cite{Riley:2021pdl,Miller:2021qha}.

\subsection{Conclusions}
\label{Conclusions}

\begin{figure}[h]
  \includegraphics[width=0.85\columnwidth]{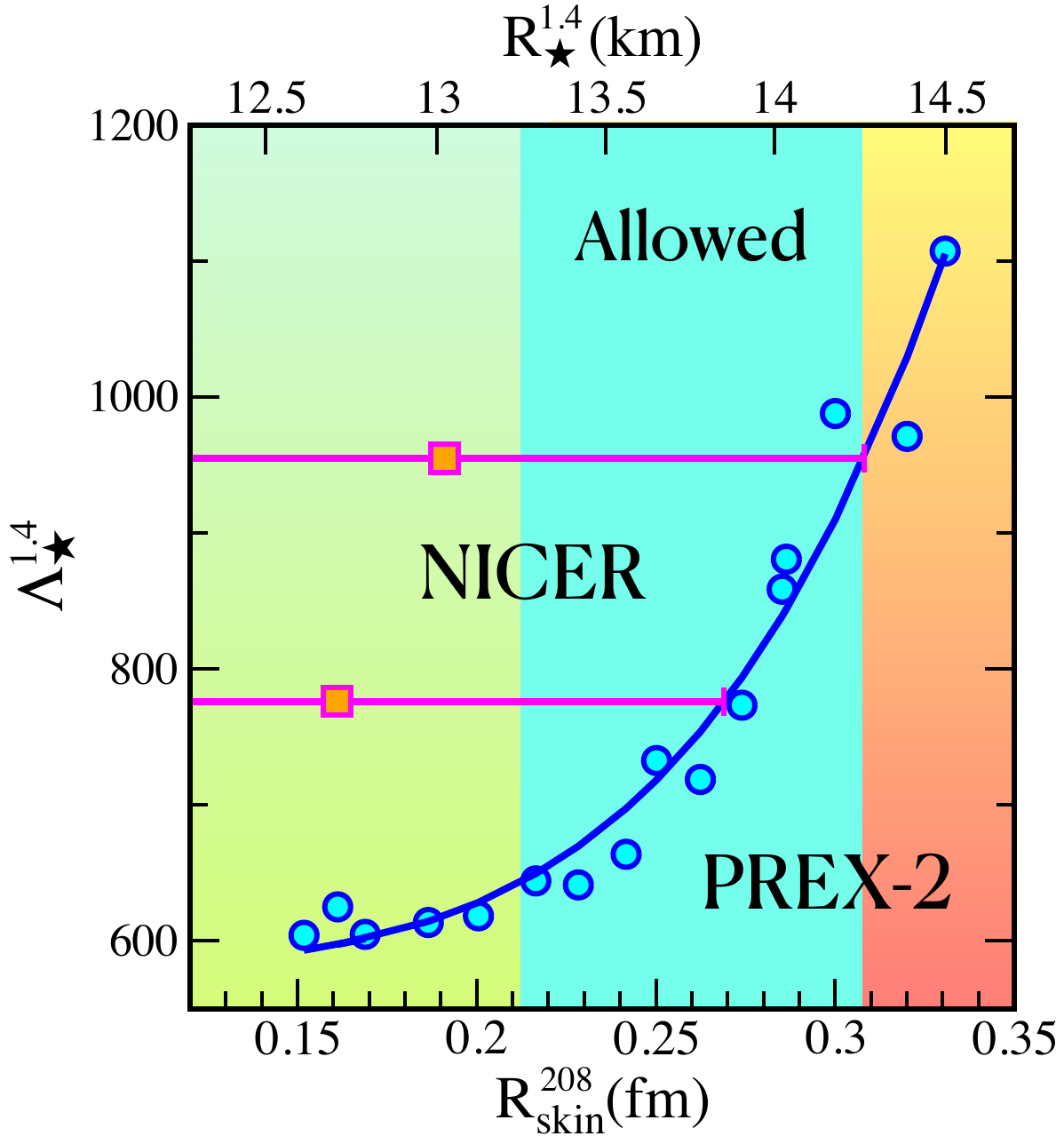}
  \caption{Experimental and Observational constraints on theoretical models of the
                equation of state, with its predictions denoted by the blue dots. Results 
                are shown for the neutron skin thickness of ${}^{208}$Pb, together with 
                the tidal deformability and radius of a $M\!=\!1.4\,M_{\odot}$ neutron star. 
                The combined PREX-II result together with NICER constraints on the stellar 
                radius is depicted by the small (blue) window of models 
                allowed\,\cite{Reed:2021nqk}.}
 \label{Fig5}
\end{figure}

So what can we conclude from the wealth of information that we have gathered during
the last five years? Figure\,\ref{Fig5} encapsulates some of the information obtained 
from laboratory experiments (PREX-2), electromagnetic observations (NICER), and gravitational-wave
detections (LIGO-Virgo). Predictions are displayed from a set of models that reproduce 
properties of finite nuclei and generate an equation of state stiff enough to support neutron 
stars with masses of at least 2$M_{\odot}$. At the 1$\sigma$ level, PREX-2 disfavors 
models that predict a neutron skin thickness in ${}^{208}$Pb of less than 
$R_{\rm skin}\!=\!0.21\,{\rm fm}$\,\cite{Adhikari:2021phr}, suggesting that the EOS in the 
vicinity of saturation density is stiff\,\cite{Reed:2021nqk}. In turn, the two magenta 
horizontal lines indicate the radius of a $M\!=\!1.4\,M_{\odot}$ neutron star as reported 
by the two independent analyses of the NICER data\,\cite{Riley:2019yda,Miller:2019cac}. 
Although differing by 18 orders of magnitude in size and probing different regions of the
EOS, both PREX-2 and NICER favor a relatively stiff equation of state. However, some
tension emerges as we incorporate the tidal deformability of a $M\!=\!1.4\,M_{\odot}$ 
neutron star extracted from GW170817. Combining two analyses\,\cite{Abbott:PRL2017,
Abbott:2018exr}, the recommended upper limit for the tidal deformability was reported to
be 580 at the 90\% confidence level. As it stands, all models displayed in the figure will 
be inconsistent with such an upper limit. Such a softening of the EOS at intermediate 
densities---bracketed by a stiff EOS at both low densities and high densities as suggested 
by PREX-2 and massive neutron stars, respectively, may be indicative of a phase transition 
in the stellar interior\,\cite{Fattoyev:2017jql}. While enormously exciting, one must wait for 
the next generation of experimental and observational facilities to validate such a scenario.

\begin{figure}[h]
  \includegraphics[width=0.85\columnwidth]{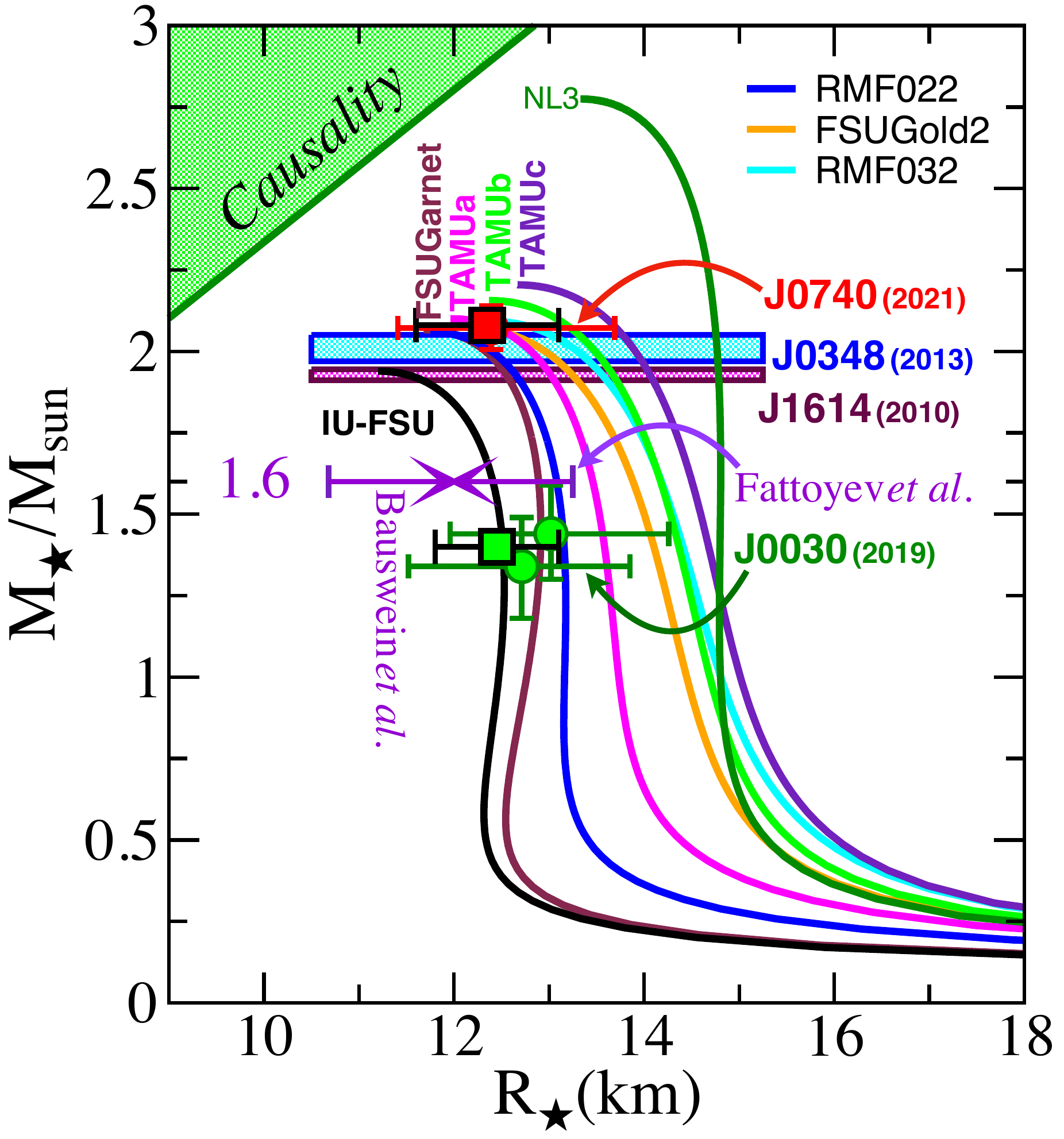}
 \caption{The ``Holy-Grail'' of nuclear-star structure: The Mass-Radius relation 
           	as predicted by a set of theoretical models. The historical detection 
	        of gravitational waves, the precise measurement of heavy neutron 
	        stars, the simultaneous determinations of the mass and radius of two 
	        neutron stars, and the extraction of the neutron-skin thickness of a
	        heavy nucleus are shaping our advances in determining the MR relation
	        and ultimately the equation of state of neutron rich matter.}
 \label{Fig6}
\end{figure}

Finally, Fig.\ref{Fig6} displays the holy grail of neutron-star structure, namely, the mass-radius 
(MR) relation informed by the many pioneering developments in the field. Mass constraints 
obtained from electromagnetic observations of the three most massive neutron stars to date
are shown with the horizontal bars\,\cite{Demorest:2010bx,Antoniadis:2013pzd,
Cromartie:2019kug,Fonseca:2021wxt}. In turn, the two horizontal arrows in the figure 
indicate constraints on stellar radii obtained exclusively from GW170817, suggesting
that the equation of state in the relevant density region is soft\,\cite{Fattoyev:2017jql,
Bauswein:2017vtn}. Such a scenario excludes several models that predict larger stellar radii.
By measuring relatively large radii for PSR J0030 and PSR J0740, NICER has relaxed 
some of the tight constraint imposed by GW170817. NICER results are denoted in the
figure by the horizontal error bars located at about $M\!=\!1.4\,M_{\odot}$ and 
$M\!=\!2\,M_{\odot}$. The excluded causality region was adopted from 
Ref.\cite{Lattimer:2006xb}. Relative to the status of the field five years ago, the 
progress in tightening the MR relation---and consequently the EOS---is simply 
remarkable. 
  
In summary, due a confluence of pioneering developments in a variety of 
interconnected disciplines, we have entered the golden era of neutron 
stars\,\cite{Baym:2019,Baym:2017whm}. Remarkable advances are
anticipated across all these disciplines during the next few years. Among
these, the operation of brand new terrestrial facilities that will measure the 
properties of atomic nuclei at the limits of existence, new telescopes operating 
across the entire electromagnetic spectrum that will determine with unprecedented 
precision the mass-radius relation, and future gravitational-wave observatories 
that will make countless detections of gravitational waves from compact binary 
coalescence. Such remarkable developments, combined with advances in our
theoretical understanding of dense, neutron-rich matter, will ultimately lead to 
a detailed understanding of the structure, dynamics, and composition of neutron 
stars, arguably, among the most fascinating objects in the universe. 

\begin{acknowledgments}\vspace{-10pt}
 I am grateful to all my colleagues---especially my graduate students---that contributed to this 
 work and to the community that has made the study of neutron stars one of the most exciting 
 endeavors in all of nuclear science. This material is based upon work supported by the U.S. 
 Department of Energy Office of Science, Office of Nuclear Physics under Award Number 
 DE-FG02-92ER40750. 
\end{acknowledgments}

\vfill\eject
\bibliography{./main.bbl}

\begin{thebibliography}{213}%
\makeatletter
\providecommand \@ifxundefined [1]{%
 \@ifx{#1\undefined}
}%
\providecommand \@ifnum [1]{%
 \ifnum #1\expandafter \@firstoftwo
 \else \expandafter \@secondoftwo
 \fi
}%
\providecommand \@ifx [1]{%
 \ifx #1\expandafter \@firstoftwo
 \else \expandafter \@secondoftwo
 \fi
}%
\providecommand \natexlab [1]{#1}%
\providecommand \enquote  [1]{``#1''}%
\providecommand \bibnamefont  [1]{#1}%
\providecommand \bibfnamefont [1]{#1}%
\providecommand \citenamefont [1]{#1}%
\providecommand \href@noop [0]{\@secondoftwo}%
\providecommand \href [0]{\begingroup \@sanitize@url \@href}%
\providecommand \@href[1]{\@@startlink{#1}\@@href}%
\providecommand \@@href[1]{\endgroup#1\@@endlink}%
\providecommand \@sanitize@url [0]{\catcode `\\12\catcode `\$12\catcode
  `\&12\catcode `\#12\catcode `\^12\catcode `\_12\catcode `\%12\relax}%
\providecommand \@@startlink[1]{}%
\providecommand \@@endlink[0]{}%
\providecommand \url  [0]{\begingroup\@sanitize@url \@url }%
\providecommand \@url [1]{\endgroup\@href {#1}{\urlprefix }}%
\providecommand \urlprefix  [0]{URL }%
\providecommand \Eprint [0]{\href }%
\providecommand \doibase [0]{http://dx.doi.org/}%
\providecommand \selectlanguage [0]{\@gobble}%
\providecommand \bibinfo  [0]{\@secondoftwo}%
\providecommand \bibfield  [0]{\@secondoftwo}%
\providecommand \translation [1]{[#1]}%
\providecommand \BibitemOpen [0]{}%
\providecommand \bibitemStop [0]{}%
\providecommand \bibitemNoStop [0]{.\EOS\space}%
\providecommand \EOS [0]{\spacefactor3000\relax}%
\providecommand \BibitemShut  [1]{\csname bibitem#1\endcsname}%
\let\auto@bib@innerbib\@empty
\bibitem [{\citenamefont {Abbott}\ \emph
  {et~al.}(2017{\natexlab{a}})\citenamefont {Abbott} \emph
  {et~al.}}]{Abbott:PRL2017}%
  \BibitemOpen
  \bibfield  {author} {\bibinfo {author} {\bibfnamefont {B.~P.}\ \bibnamefont
  {Abbott}} \emph {et~al.} (\bibinfo {collaboration} {Virgo, LIGO
  Scientific}),\ }\href {\doibase 10.1103/PhysRevLett.119.161101} {\bibfield
  {journal} {\bibinfo  {journal} {Phys. Rev. Lett.}\ }\textbf {\bibinfo
  {volume} {119}},\ \bibinfo {pages} {161101} (\bibinfo {year}
  {2017}{\natexlab{a}})}\BibitemShut {NoStop}%
\bibitem [{\citenamefont {Bauswein}\ \emph {et~al.}(2017)\citenamefont
  {Bauswein}, \citenamefont {Just}, \citenamefont {Janka},\ and\ \citenamefont
  {Stergioulas}}]{Bauswein:2017vtn}%
  \BibitemOpen
  \bibfield  {author} {\bibinfo {author} {\bibfnamefont {A.}~\bibnamefont
  {Bauswein}}, \bibinfo {author} {\bibfnamefont {O.}~\bibnamefont {Just}},
  \bibinfo {author} {\bibfnamefont {H.-T.}\ \bibnamefont {Janka}}, \ and\
  \bibinfo {author} {\bibfnamefont {N.}~\bibnamefont {Stergioulas}},\ }\href
  {\doibase 10.3847/2041-8213/aa9994} {\bibfield  {journal} {\bibinfo
  {journal} {Astrophys. J.}\ }\textbf {\bibinfo {volume} {850}},\ \bibinfo
  {pages} {L34} (\bibinfo {year} {2017})}\BibitemShut {NoStop}%
\bibitem [{\citenamefont {Fattoyev}\ \emph {et~al.}(2018)\citenamefont
  {Fattoyev}, \citenamefont {Piekarewicz},\ and\ \citenamefont
  {Horowitz}}]{Fattoyev:2017jql}%
  \BibitemOpen
  \bibfield  {author} {\bibinfo {author} {\bibfnamefont {F.~J.}\ \bibnamefont
  {Fattoyev}}, \bibinfo {author} {\bibfnamefont {J.}~\bibnamefont
  {Piekarewicz}}, \ and\ \bibinfo {author} {\bibfnamefont {C.~J.}\ \bibnamefont
  {Horowitz}},\ }\href {\doibase 10.1103/PhysRevLett.120.172702} {\bibfield
  {journal} {\bibinfo  {journal} {Phys. Rev. Lett.}\ }\textbf {\bibinfo
  {volume} {120}},\ \bibinfo {pages} {172702} (\bibinfo {year}
  {2018})}\BibitemShut {NoStop}%
\bibitem [{\citenamefont {Annala}\ \emph {et~al.}(2018)\citenamefont {Annala},
  \citenamefont {Gorda}, \citenamefont {Kurkela},\ and\ \citenamefont
  {Vuorinen}}]{Annala:2017llu}%
  \BibitemOpen
  \bibfield  {author} {\bibinfo {author} {\bibfnamefont {E.}~\bibnamefont
  {Annala}}, \bibinfo {author} {\bibfnamefont {T.}~\bibnamefont {Gorda}},
  \bibinfo {author} {\bibfnamefont {A.}~\bibnamefont {Kurkela}}, \ and\
  \bibinfo {author} {\bibfnamefont {A.}~\bibnamefont {Vuorinen}},\ }\href
  {\doibase 10.1103/PhysRevLett.120.172703} {\bibfield  {journal} {\bibinfo
  {journal} {Phys. Rev. Lett.}\ }\textbf {\bibinfo {volume} {120}},\ \bibinfo
  {pages} {172703} (\bibinfo {year} {2018})}\BibitemShut {NoStop}%
\bibitem [{\citenamefont {Abbott}\ \emph {et~al.}(2018)\citenamefont {Abbott}
  \emph {et~al.}}]{Abbott:2018exr}%
  \BibitemOpen
  \bibfield  {author} {\bibinfo {author} {\bibfnamefont {B.~P.}\ \bibnamefont
  {Abbott}} \emph {et~al.} (\bibinfo {collaboration} {Virgo, LIGO
  Scientific}),\ }\href {\doibase 10.1103/PhysRevLett.121.161101} {\bibfield
  {journal} {\bibinfo  {journal} {Phys. Rev. Lett.}\ }\textbf {\bibinfo
  {volume} {121}},\ \bibinfo {pages} {161101} (\bibinfo {year}
  {2018})}\BibitemShut {NoStop}%
\bibitem [{\citenamefont {Most}\ \emph {et~al.}(2018)\citenamefont {Most},
  \citenamefont {Weih}, \citenamefont {Rezzolla},\ and\ \citenamefont
  {Schaffner-Bielich}}]{Most:2018hfd}%
  \BibitemOpen
  \bibfield  {author} {\bibinfo {author} {\bibfnamefont {E.~R.}\ \bibnamefont
  {Most}}, \bibinfo {author} {\bibfnamefont {L.~R.}\ \bibnamefont {Weih}},
  \bibinfo {author} {\bibfnamefont {L.}~\bibnamefont {Rezzolla}}, \ and\
  \bibinfo {author} {\bibfnamefont {J.}~\bibnamefont {Schaffner-Bielich}},\
  }\href {\doibase 10.1103/PhysRevLett.120.261103} {\bibfield  {journal}
  {\bibinfo  {journal} {Phys. Rev. Lett.}\ }\textbf {\bibinfo {volume} {120}},\
  \bibinfo {pages} {261103} (\bibinfo {year} {2018})}\BibitemShut {NoStop}%
\bibitem [{\citenamefont {Tews}\ \emph {et~al.}(2018)\citenamefont {Tews},
  \citenamefont {Margueron},\ and\ \citenamefont {Reddy}}]{Tews:2018chv}%
  \BibitemOpen
  \bibfield  {author} {\bibinfo {author} {\bibfnamefont {I.}~\bibnamefont
  {Tews}}, \bibinfo {author} {\bibfnamefont {J.}~\bibnamefont {Margueron}}, \
  and\ \bibinfo {author} {\bibfnamefont {S.}~\bibnamefont {Reddy}},\ }\href
  {\doibase 10.1103/PhysRevC.98.045804} {\bibfield  {journal} {\bibinfo
  {journal} {Phys. Rev.}\ }\textbf {\bibinfo {volume} {C98}},\ \bibinfo {pages}
  {045804} (\bibinfo {year} {2018})}\BibitemShut {NoStop}%
\bibitem [{\citenamefont {Malik}\ \emph {et~al.}(2018)\citenamefont {Malik},
  \citenamefont {Alam}, \citenamefont {Fortin}, \citenamefont {Providencia},
  \citenamefont {Agrawal}, \citenamefont {Jha}, \citenamefont {Kumar},\ and\
  \citenamefont {Patra}}]{Malik:2018zcf}%
  \BibitemOpen
  \bibfield  {author} {\bibinfo {author} {\bibfnamefont {T.}~\bibnamefont
  {Malik}}, \bibinfo {author} {\bibfnamefont {N.}~\bibnamefont {Alam}},
  \bibinfo {author} {\bibfnamefont {M.}~\bibnamefont {Fortin}}, \bibinfo
  {author} {\bibfnamefont {C.}~\bibnamefont {Providencia}}, \bibinfo {author}
  {\bibfnamefont {B.~K.}\ \bibnamefont {Agrawal}}, \bibinfo {author}
  {\bibfnamefont {T.~K.}\ \bibnamefont {Jha}}, \bibinfo {author} {\bibfnamefont
  {B.}~\bibnamefont {Kumar}}, \ and\ \bibinfo {author} {\bibfnamefont {S.~K.}\
  \bibnamefont {Patra}},\ }\href {\doibase 10.1103/PhysRevC.98.035804}
  {\bibfield  {journal} {\bibinfo  {journal} {Phys. Rev.}\ }\textbf {\bibinfo
  {volume} {C98}},\ \bibinfo {pages} {035804} (\bibinfo {year}
  {2018})}\BibitemShut {NoStop}%
\bibitem [{\citenamefont {Radice}\ \emph {et~al.}(2018)\citenamefont {Radice},
  \citenamefont {Perego}, \citenamefont {Zappa},\ and\ \citenamefont
  {Bernuzzi}}]{Radice:2017lry}%
  \BibitemOpen
  \bibfield  {author} {\bibinfo {author} {\bibfnamefont {D.}~\bibnamefont
  {Radice}}, \bibinfo {author} {\bibfnamefont {A.}~\bibnamefont {Perego}},
  \bibinfo {author} {\bibfnamefont {F.}~\bibnamefont {Zappa}}, \ and\ \bibinfo
  {author} {\bibfnamefont {S.}~\bibnamefont {Bernuzzi}},\ }\href {\doibase
  10.3847/2041-8213/aaa402} {\bibfield  {journal} {\bibinfo  {journal}
  {Astrophys. J. Lett.}\ }\textbf {\bibinfo {volume} {852}},\ \bibinfo {pages}
  {L29} (\bibinfo {year} {2018})}\BibitemShut {NoStop}%
\bibitem [{\citenamefont {Radice}\ and\ \citenamefont
  {Dai}(2019)}]{Radice:2018ozg}%
  \BibitemOpen
  \bibfield  {author} {\bibinfo {author} {\bibfnamefont {D.}~\bibnamefont
  {Radice}}\ and\ \bibinfo {author} {\bibfnamefont {L.}~\bibnamefont {Dai}},\
  }\href {\doibase 10.1140/epja/i2019-12716-4} {\bibfield  {journal} {\bibinfo
  {journal} {Eur. Phys. J. A}\ }\textbf {\bibinfo {volume} {55}},\ \bibinfo
  {pages} {50} (\bibinfo {year} {2019})}\BibitemShut {NoStop}%
\bibitem [{\citenamefont {Tews}\ \emph {et~al.}(2019)\citenamefont {Tews},
  \citenamefont {Margueron},\ and\ \citenamefont {Reddy}}]{Tews:2019cap}%
  \BibitemOpen
  \bibfield  {author} {\bibinfo {author} {\bibfnamefont {I.}~\bibnamefont
  {Tews}}, \bibinfo {author} {\bibfnamefont {J.}~\bibnamefont {Margueron}}, \
  and\ \bibinfo {author} {\bibfnamefont {S.}~\bibnamefont {Reddy}},\ }\href
  {\doibase 10.1140/epja/i2019-12774-6} {\bibfield  {journal} {\bibinfo
  {journal} {Eur. Phys. J. A}\ }\textbf {\bibinfo {volume} {55}},\ \bibinfo
  {pages} {97} (\bibinfo {year} {2019})}\BibitemShut {NoStop}%
\bibitem [{\citenamefont {Capano}\ \emph {et~al.}(2019)\citenamefont {Capano},
  \citenamefont {Tews}, \citenamefont {Brown}, \citenamefont {Margalit},
  \citenamefont {De}, \citenamefont {Kumar}, \citenamefont {Brown},
  \citenamefont {Krishnan},\ and\ \citenamefont {Reddy}}]{Capano:2019eae}%
  \BibitemOpen
  \bibfield  {author} {\bibinfo {author} {\bibfnamefont {C.~D.}\ \bibnamefont
  {Capano}}, \bibinfo {author} {\bibfnamefont {I.}~\bibnamefont {Tews}},
  \bibinfo {author} {\bibfnamefont {S.~M.}\ \bibnamefont {Brown}}, \bibinfo
  {author} {\bibfnamefont {B.}~\bibnamefont {Margalit}}, \bibinfo {author}
  {\bibfnamefont {S.}~\bibnamefont {De}}, \bibinfo {author} {\bibfnamefont
  {S.}~\bibnamefont {Kumar}}, \bibinfo {author} {\bibfnamefont {D.~A.}\
  \bibnamefont {Brown}}, \bibinfo {author} {\bibfnamefont {B.}~\bibnamefont
  {Krishnan}}, \ and\ \bibinfo {author} {\bibfnamefont {S.}~\bibnamefont
  {Reddy}},\ }\href {\doibase 10.1038/s41550-020-1014-6} {\bibfield  {journal}
  {\bibinfo  {journal} {Nature Astronomy}\ }\textbf {\bibinfo {volume} {4}},\
  \bibinfo {pages} {625} (\bibinfo {year} {2019})}\BibitemShut {NoStop}%
\bibitem [{\citenamefont {Tsang}\ \emph {et~al.}(2019)\citenamefont {Tsang},
  \citenamefont {Lynch}, \citenamefont {Danielewicz},\ and\ \citenamefont
  {Tsang}}]{Tsang:2019mlz}%
  \BibitemOpen
  \bibfield  {author} {\bibinfo {author} {\bibfnamefont {M.}~\bibnamefont
  {Tsang}}, \bibinfo {author} {\bibfnamefont {W.}~\bibnamefont {Lynch}},
  \bibinfo {author} {\bibfnamefont {P.}~\bibnamefont {Danielewicz}}, \ and\
  \bibinfo {author} {\bibfnamefont {C.}~\bibnamefont {Tsang}},\ }\href
  {\doibase 10.1016/j.physletb.2019.06.059} {\bibfield  {journal} {\bibinfo
  {journal} {Phys. Lett. B}\ }\textbf {\bibinfo {volume} {795}},\ \bibinfo
  {pages} {533} (\bibinfo {year} {2019})}\BibitemShut {NoStop}%
\bibitem [{\citenamefont {Tsang}\ \emph {et~al.}(2020)\citenamefont {Tsang},
  \citenamefont {Tsang}, \citenamefont {Danielewicz}, \citenamefont {Lynch},\
  and\ \citenamefont {Fattoyev}}]{Tsang:2020lmb}%
  \BibitemOpen
  \bibfield  {author} {\bibinfo {author} {\bibfnamefont {C.~Y.}\ \bibnamefont
  {Tsang}}, \bibinfo {author} {\bibfnamefont {M.~B.}\ \bibnamefont {Tsang}},
  \bibinfo {author} {\bibfnamefont {P.}~\bibnamefont {Danielewicz}}, \bibinfo
  {author} {\bibfnamefont {W.~G.}\ \bibnamefont {Lynch}}, \ and\ \bibinfo
  {author} {\bibfnamefont {F.~J.}\ \bibnamefont {Fattoyev}},\ }\href {\doibase
  10.1103/PhysRevC.102.045808} {\bibfield  {journal} {\bibinfo  {journal}
  {Phys. Rev. C}\ }\textbf {\bibinfo {volume} {102}},\ \bibinfo {pages}
  {045808} (\bibinfo {year} {2020})}\BibitemShut {NoStop}%
\bibitem [{\citenamefont {Drischler}\ \emph {et~al.}(2020)\citenamefont
  {Drischler}, \citenamefont {Furnstahl}, \citenamefont {Melendez},\ and\
  \citenamefont {Phillips}}]{Drischler:2020hwi}%
  \BibitemOpen
  \bibfield  {author} {\bibinfo {author} {\bibfnamefont {C.}~\bibnamefont
  {Drischler}}, \bibinfo {author} {\bibfnamefont {R.}~\bibnamefont
  {Furnstahl}}, \bibinfo {author} {\bibfnamefont {J.}~\bibnamefont {Melendez}},
  \ and\ \bibinfo {author} {\bibfnamefont {D.}~\bibnamefont {Phillips}},\
  }\href {\doibase 10.1103/PhysRevLett.125.202702} {\bibfield  {journal}
  {\bibinfo  {journal} {Phys. Rev. Lett.}\ }\textbf {\bibinfo {volume} {125}},\
  \bibinfo {pages} {202702} (\bibinfo {year} {2020})}\BibitemShut {NoStop}%
\bibitem [{\citenamefont {Landry}\ \emph {et~al.}(2020)\citenamefont {Landry},
  \citenamefont {Essick},\ and\ \citenamefont
  {Chatziioannou}}]{Landry:2020vaw}%
  \BibitemOpen
  \bibfield  {author} {\bibinfo {author} {\bibfnamefont {P.}~\bibnamefont
  {Landry}}, \bibinfo {author} {\bibfnamefont {R.}~\bibnamefont {Essick}}, \
  and\ \bibinfo {author} {\bibfnamefont {K.}~\bibnamefont {Chatziioannou}},\
  }\href {\doibase 10.1103/PhysRevD.101.123007} {\bibfield  {journal} {\bibinfo
   {journal} {Phys. Rev. D}\ }\textbf {\bibinfo {volume} {101}},\ \bibinfo
  {pages} {123007} (\bibinfo {year} {2020})}\BibitemShut {NoStop}%
\bibitem [{\citenamefont {Xie}\ and\ \citenamefont {Li}(2021)}]{Xie:2020rwg}%
  \BibitemOpen
  \bibfield  {author} {\bibinfo {author} {\bibfnamefont {W.-J.}\ \bibnamefont
  {Xie}}\ and\ \bibinfo {author} {\bibfnamefont {B.-A.}\ \bibnamefont {Li}},\
  }\href {\doibase 10.1103/PhysRevC.103.035802} {\bibfield  {journal} {\bibinfo
   {journal} {Phys. Rev. C}\ }\textbf {\bibinfo {volume} {103}},\ \bibinfo
  {pages} {035802} (\bibinfo {year} {2021})}\BibitemShut {NoStop}%
\bibitem [{\citenamefont {Essick}\ \emph {et~al.}(2021)\citenamefont {Essick},
  \citenamefont {Tews}, \citenamefont {Landry},\ and\ \citenamefont
  {Schwenk}}]{Essick:2021kjb}%
  \BibitemOpen
  \bibfield  {author} {\bibinfo {author} {\bibfnamefont {R.}~\bibnamefont
  {Essick}}, \bibinfo {author} {\bibfnamefont {I.}~\bibnamefont {Tews}},
  \bibinfo {author} {\bibfnamefont {P.}~\bibnamefont {Landry}}, \ and\ \bibinfo
  {author} {\bibfnamefont {A.}~\bibnamefont {Schwenk}},\ }\href {\doibase
  10.1103/PhysRevLett.127.192701} {\bibfield  {journal} {\bibinfo  {journal}
  {Phys. Rev. Lett.}\ }\textbf {\bibinfo {volume} {127}},\ \bibinfo {pages}
  {192701} (\bibinfo {year} {2021})}\BibitemShut {NoStop}%
\bibitem [{\citenamefont {Chatziioannou}(2022)}]{Chatziioannou:2021tdi}%
  \BibitemOpen
  \bibfield  {author} {\bibinfo {author} {\bibfnamefont {K.}~\bibnamefont
  {Chatziioannou}},\ }\href {\doibase 10.1103/PhysRevD.105.084021} {\bibfield
  {journal} {\bibinfo  {journal} {Phys. Rev. D}\ }\textbf {\bibinfo {volume}
  {105}},\ \bibinfo {pages} {084021} (\bibinfo {year} {2022})}\BibitemShut
  {NoStop}%
\bibitem [{\citenamefont {Drout}\ \emph {et~al.}(2017)\citenamefont {Drout}
  \emph {et~al.}}]{Drout:2017ijr}%
  \BibitemOpen
  \bibfield  {author} {\bibinfo {author} {\bibfnamefont {M.~R.}\ \bibnamefont
  {Drout}} \emph {et~al.},\ }\href {\doibase 10.1126/science.aaq0049}
  {\bibfield  {journal} {\bibinfo  {journal} {Science}\ }\textbf {\bibinfo
  {volume} {358}},\ \bibinfo {pages} {1570} (\bibinfo {year}
  {2017})}\BibitemShut {NoStop}%
\bibitem [{\citenamefont {Cowperthwaite}\ \emph {et~al.}(2017)\citenamefont
  {Cowperthwaite} \emph {et~al.}}]{Cowperthwaite:2017dyu}%
  \BibitemOpen
  \bibfield  {author} {\bibinfo {author} {\bibfnamefont {P.~S.}\ \bibnamefont
  {Cowperthwaite}} \emph {et~al.},\ }\href {\doibase 10.3847/2041-8213/aa8fc7}
  {\bibfield  {journal} {\bibinfo  {journal} {Astrophys. J.}\ }\textbf
  {\bibinfo {volume} {848}},\ \bibinfo {pages} {L17} (\bibinfo {year}
  {2017})}\BibitemShut {NoStop}%
\bibitem [{\citenamefont {Chornock}\ \emph {et~al.}(2017)\citenamefont
  {Chornock} \emph {et~al.}}]{Chornock:2017sdf}%
  \BibitemOpen
  \bibfield  {author} {\bibinfo {author} {\bibfnamefont {R.}~\bibnamefont
  {Chornock}} \emph {et~al.},\ }\href {\doibase 10.3847/2041-8213/aa905c}
  {\bibfield  {journal} {\bibinfo  {journal} {Astrophys. J.}\ }\textbf
  {\bibinfo {volume} {848}},\ \bibinfo {pages} {L19} (\bibinfo {year}
  {2017})}\BibitemShut {NoStop}%
\bibitem [{\citenamefont {Nicholl}\ \emph {et~al.}(2017)\citenamefont {Nicholl}
  \emph {et~al.}}]{Nicholl:2017ahq}%
  \BibitemOpen
  \bibfield  {author} {\bibinfo {author} {\bibfnamefont {M.}~\bibnamefont
  {Nicholl}} \emph {et~al.},\ }\href@noop {} {\bibfield  {journal} {\bibinfo
  {journal} {Astrophys. J.}\ }\textbf {\bibinfo {volume} {848}},\ \bibinfo
  {pages} {L18} (\bibinfo {year} {2017})}\BibitemShut {NoStop}%
\bibitem [{\citenamefont {Baade}\ and\ \citenamefont
  {Zwicky}(1934)}]{Baade:1934}%
  \BibitemOpen
  \bibfield  {author} {\bibinfo {author} {\bibfnamefont {W.}~\bibnamefont
  {Baade}}\ and\ \bibinfo {author} {\bibfnamefont {F.}~\bibnamefont {Zwicky}},\
  }\href {\doibase 10.1103/PhysRev.45.130} {\bibfield  {journal} {\bibinfo
  {journal} {Phys. Rev.}\ }\textbf {\bibinfo {volume} {45}},\ \bibinfo {pages}
  {138} (\bibinfo {year} {1934})}\BibitemShut {NoStop}%
\bibitem [{\citenamefont {Yakovlev}\ \emph {et~al.}(2013)\citenamefont
  {Yakovlev}, \citenamefont {Haensel}, \citenamefont {Baym},\ and\
  \citenamefont {Pethick}}]{Yakovlev:2012rd}%
  \BibitemOpen
  \bibfield  {author} {\bibinfo {author} {\bibfnamefont {D.~G.}\ \bibnamefont
  {Yakovlev}}, \bibinfo {author} {\bibfnamefont {P.}~\bibnamefont {Haensel}},
  \bibinfo {author} {\bibfnamefont {G.}~\bibnamefont {Baym}}, \ and\ \bibinfo
  {author} {\bibfnamefont {C.~J.}\ \bibnamefont {Pethick}},\ }\href {\doibase
  10.3367/UFNe.0183.201303f.0307} {\bibfield  {journal} {\bibinfo  {journal}
  {Phys. Usp.}\ }\textbf {\bibinfo {volume} {56}},\ \bibinfo {pages} {289}
  (\bibinfo {year} {2013})},\ \Eprint {http://arxiv.org/abs/1210.0682}
  {arXiv:1210.0682 [physics.hist-ph]} \BibitemShut {NoStop}%
\bibitem [{\citenamefont {Oppenheimer}\ and\ \citenamefont
  {Volkoff}(1939)}]{Opp39_PR55}%
  \BibitemOpen
  \bibfield  {author} {\bibinfo {author} {\bibfnamefont {J.~R.}\ \bibnamefont
  {Oppenheimer}}\ and\ \bibinfo {author} {\bibfnamefont {G.~M.}\ \bibnamefont
  {Volkoff}},\ }\href@noop {} {\bibfield  {journal} {\bibinfo  {journal} {Phys.
  Rev.}\ }\textbf {\bibinfo {volume} {55}},\ \bibinfo {pages} {374} (\bibinfo
  {year} {1939})}\BibitemShut {NoStop}%
\bibitem [{\citenamefont {Tolman}(1939)}]{Tol39_PR55}%
  \BibitemOpen
  \bibfield  {author} {\bibinfo {author} {\bibfnamefont {R.~C.}\ \bibnamefont
  {Tolman}},\ }\href@noop {} {\bibfield  {journal} {\bibinfo  {journal} {Phys.
  Rev.}\ }\textbf {\bibinfo {volume} {55}},\ \bibinfo {pages} {364} (\bibinfo
  {year} {1939})}\BibitemShut {NoStop}%
\bibitem [{\citenamefont {Hewish}\ \emph {et~al.}(1968)\citenamefont {Hewish},
  \citenamefont {Bell}, \citenamefont {Pilkington}, \citenamefont {Scott},\
  and\ \citenamefont {Collins}}]{Hewish:1968}%
  \BibitemOpen
  \bibfield  {author} {\bibinfo {author} {\bibfnamefont {A.}~\bibnamefont
  {Hewish}}, \bibinfo {author} {\bibfnamefont {S.}~\bibnamefont {Bell}},
  \bibinfo {author} {\bibfnamefont {J.}~\bibnamefont {Pilkington}}, \bibinfo
  {author} {\bibfnamefont {P.}~\bibnamefont {Scott}}, \ and\ \bibinfo {author}
  {\bibfnamefont {R.}~\bibnamefont {Collins}},\ }\href {\doibase
  10.1038/217709a0} {\bibfield  {journal} {\bibinfo  {journal} {Nature}\
  }\textbf {\bibinfo {volume} {217}},\ \bibinfo {pages} {709} (\bibinfo {year}
  {1968})}\BibitemShut {NoStop}%
\bibitem [{\citenamefont {Lattimer}\ and\ \citenamefont
  {Prakash}(2004)}]{Lattimer:2004pg}%
  \BibitemOpen
  \bibfield  {author} {\bibinfo {author} {\bibfnamefont {J.~M.}\ \bibnamefont
  {Lattimer}}\ and\ \bibinfo {author} {\bibfnamefont {M.}~\bibnamefont
  {Prakash}},\ }\href@noop {} {\bibfield  {journal} {\bibinfo  {journal}
  {Science}\ }\textbf {\bibinfo {volume} {304}},\ \bibinfo {pages} {536}
  (\bibinfo {year} {2004})}\BibitemShut {NoStop}%
\bibitem [{\citenamefont {Demorest}\ \emph {et~al.}(2010)\citenamefont
  {Demorest}, \citenamefont {Pennucci}, \citenamefont {Ransom}, \citenamefont
  {Roberts},\ and\ \citenamefont {Hessels}}]{Demorest:2010bx}%
  \BibitemOpen
  \bibfield  {author} {\bibinfo {author} {\bibfnamefont {P.}~\bibnamefont
  {Demorest}}, \bibinfo {author} {\bibfnamefont {T.}~\bibnamefont {Pennucci}},
  \bibinfo {author} {\bibfnamefont {S.}~\bibnamefont {Ransom}}, \bibinfo
  {author} {\bibfnamefont {M.}~\bibnamefont {Roberts}}, \ and\ \bibinfo
  {author} {\bibfnamefont {J.}~\bibnamefont {Hessels}},\ }\href {\doibase
  10.1038/nature09466} {\bibfield  {journal} {\bibinfo  {journal} {Nature}\
  }\textbf {\bibinfo {volume} {467}},\ \bibinfo {pages} {1081} (\bibinfo {year}
  {2010})}\BibitemShut {NoStop}%
\bibitem [{\citenamefont {Antoniadis}\ \emph {et~al.}(2013)\citenamefont
  {Antoniadis}, \citenamefont {Freire}, \citenamefont {Wex}, \citenamefont
  {Tauris}, \citenamefont {Lynch} \emph {et~al.}}]{Antoniadis:2013pzd}%
  \BibitemOpen
  \bibfield  {author} {\bibinfo {author} {\bibfnamefont {J.}~\bibnamefont
  {Antoniadis}}, \bibinfo {author} {\bibfnamefont {P.~C.}\ \bibnamefont
  {Freire}}, \bibinfo {author} {\bibfnamefont {N.}~\bibnamefont {Wex}},
  \bibinfo {author} {\bibfnamefont {T.~M.}\ \bibnamefont {Tauris}}, \bibinfo
  {author} {\bibfnamefont {R.~S.}\ \bibnamefont {Lynch}},  \emph {et~al.},\
  }\href {\doibase 10.1126/science.1233232} {\bibfield  {journal} {\bibinfo
  {journal} {Science}\ }\textbf {\bibinfo {volume} {340}},\ \bibinfo {pages}
  {6131} (\bibinfo {year} {2013})}\BibitemShut {NoStop}%
\bibitem [{\citenamefont {Cromartie}\ \emph {et~al.}(2019)\citenamefont
  {Cromartie} \emph {et~al.}}]{Cromartie:2019kug}%
  \BibitemOpen
  \bibfield  {author} {\bibinfo {author} {\bibfnamefont {H.~T.}\ \bibnamefont
  {Cromartie}} \emph {et~al.},\ }\href@noop {} {\bibfield  {journal} {\bibinfo
  {journal} {Nat. Astron.}\ }\textbf {\bibinfo {volume} {4}},\ \bibinfo {pages}
  {72} (\bibinfo {year} {2019})}\BibitemShut {NoStop}%
\bibitem [{\citenamefont {Fonseca}\ \emph {et~al.}(2021)\citenamefont {Fonseca}
  \emph {et~al.}}]{Fonseca:2021wxt}%
  \BibitemOpen
  \bibfield  {author} {\bibinfo {author} {\bibfnamefont {E.}~\bibnamefont
  {Fonseca}} \emph {et~al.},\ }\href@noop {} {\  (\bibinfo {year} {2021})},\
  \Eprint {http://arxiv.org/abs/2104.00880} {arXiv:2104.00880 [astro-ph.HE]}
  \BibitemShut {NoStop}%
\bibitem [{\citenamefont {Romani}\ \emph {et~al.}(2022)\citenamefont {Romani},
  \citenamefont {Kandel}, \citenamefont {Filippenko}, \citenamefont {Brink},\
  and\ \citenamefont {Zheng}}]{Romani:2022jhd}%
  \BibitemOpen
  \bibfield  {author} {\bibinfo {author} {\bibfnamefont {R.~W.}\ \bibnamefont
  {Romani}}, \bibinfo {author} {\bibfnamefont {D.}~\bibnamefont {Kandel}},
  \bibinfo {author} {\bibfnamefont {A.~V.}\ \bibnamefont {Filippenko}},
  \bibinfo {author} {\bibfnamefont {T.~G.}\ \bibnamefont {Brink}}, \ and\
  \bibinfo {author} {\bibfnamefont {W.}~\bibnamefont {Zheng}},\ }\href
  {\doibase 10.3847/2041-8213/ac8007} {\bibfield  {journal} {\bibinfo
  {journal} {Astrophys. J. Lett.}\ }\textbf {\bibinfo {volume} {934}},\
  \bibinfo {pages} {L18} (\bibinfo {year} {2022})}\BibitemShut {NoStop}%
\bibitem [{\citenamefont {Page}\ \emph {et~al.}(2004)\citenamefont {Page},
  \citenamefont {Lattimer}, \citenamefont {Prakash},\ and\ \citenamefont
  {Steiner}}]{Page:2004fy}%
  \BibitemOpen
  \bibfield  {author} {\bibinfo {author} {\bibfnamefont {D.}~\bibnamefont
  {Page}}, \bibinfo {author} {\bibfnamefont {J.~M.}\ \bibnamefont {Lattimer}},
  \bibinfo {author} {\bibfnamefont {M.}~\bibnamefont {Prakash}}, \ and\
  \bibinfo {author} {\bibfnamefont {A.~W.}\ \bibnamefont {Steiner}},\ }\href
  {\doibase 10.1086/424844} {\bibfield  {journal} {\bibinfo  {journal}
  {Astrophys. J. Suppl.}\ }\textbf {\bibinfo {volume} {155}},\ \bibinfo {pages}
  {623} (\bibinfo {year} {2004})}\BibitemShut {NoStop}%
\bibitem [{\citenamefont {Page}\ and\ \citenamefont
  {Reddy}(2006)}]{Page:2006ud}%
  \BibitemOpen
  \bibfield  {author} {\bibinfo {author} {\bibfnamefont {D.}~\bibnamefont
  {Page}}\ and\ \bibinfo {author} {\bibfnamefont {S.}~\bibnamefont {Reddy}},\
  }\href {\doibase 10.1146/annurev.nucl.56.080805.140600} {\bibfield  {journal}
  {\bibinfo  {journal} {Ann. Rev. Nucl. Part. Sci.}\ }\textbf {\bibinfo
  {volume} {56}},\ \bibinfo {pages} {327} (\bibinfo {year} {2006})}\BibitemShut
  {NoStop}%
\bibitem [{\citenamefont {Page}\ \emph {et~al.}(2009)\citenamefont {Page},
  \citenamefont {Lattimer}, \citenamefont {Prakash},\ and\ \citenamefont
  {Steiner}}]{Page:2009fu}%
  \BibitemOpen
  \bibfield  {author} {\bibinfo {author} {\bibfnamefont {D.}~\bibnamefont
  {Page}}, \bibinfo {author} {\bibfnamefont {J.~M.}\ \bibnamefont {Lattimer}},
  \bibinfo {author} {\bibfnamefont {M.}~\bibnamefont {Prakash}}, \ and\
  \bibinfo {author} {\bibfnamefont {A.~W.}\ \bibnamefont {Steiner}},\ }\href
  {\doibase 10.1088/0004-637X/707/2/1131} {\bibfield  {journal} {\bibinfo
  {journal} {Astrophys. J.}\ }\textbf {\bibinfo {volume} {707}},\ \bibinfo
  {pages} {1131} (\bibinfo {year} {2009})}\BibitemShut {NoStop}%
\bibitem [{\citenamefont {Ozel}\ \emph {et~al.}(2010)\citenamefont {Ozel},
  \citenamefont {Baym},\ and\ \citenamefont {Guver}}]{Ozel:2010fw}%
  \BibitemOpen
  \bibfield  {author} {\bibinfo {author} {\bibfnamefont {F.}~\bibnamefont
  {Ozel}}, \bibinfo {author} {\bibfnamefont {G.}~\bibnamefont {Baym}}, \ and\
  \bibinfo {author} {\bibfnamefont {T.}~\bibnamefont {Guver}},\ }\href
  {\doibase 10.1103/PhysRevD.82.101301} {\bibfield  {journal} {\bibinfo
  {journal} {Phys. Rev.}\ }\textbf {\bibinfo {volume} {D82}},\ \bibinfo {pages}
  {101301} (\bibinfo {year} {2010})}\BibitemShut {NoStop}%
\bibitem [{\citenamefont {Steiner}\ \emph {et~al.}(2010)\citenamefont
  {Steiner}, \citenamefont {Lattimer},\ and\ \citenamefont
  {Brown}}]{Steiner:2010fz}%
  \BibitemOpen
  \bibfield  {author} {\bibinfo {author} {\bibfnamefont {A.~W.}\ \bibnamefont
  {Steiner}}, \bibinfo {author} {\bibfnamefont {J.~M.}\ \bibnamefont
  {Lattimer}}, \ and\ \bibinfo {author} {\bibfnamefont {E.~F.}\ \bibnamefont
  {Brown}},\ }\href@noop {} {\bibfield  {journal} {\bibinfo  {journal}
  {Astrophys. J.}\ }\textbf {\bibinfo {volume} {722}},\ \bibinfo {pages} {33}
  (\bibinfo {year} {2010})}\BibitemShut {NoStop}%
\bibitem [{\citenamefont {Suleimanov}\ \emph {et~al.}(2011)\citenamefont
  {Suleimanov}, \citenamefont {Poutanen}, \citenamefont {Revnivtsev},\ and\
  \citenamefont {Werner}}]{Suleimanov:2010th}%
  \BibitemOpen
  \bibfield  {author} {\bibinfo {author} {\bibfnamefont {V.}~\bibnamefont
  {Suleimanov}}, \bibinfo {author} {\bibfnamefont {J.}~\bibnamefont
  {Poutanen}}, \bibinfo {author} {\bibfnamefont {M.}~\bibnamefont
  {Revnivtsev}}, \ and\ \bibinfo {author} {\bibfnamefont {K.}~\bibnamefont
  {Werner}},\ }\href {\doibase 10.1088/0004-637X/742/2/122} {\bibfield
  {journal} {\bibinfo  {journal} {Astrophys. J.}\ }\textbf {\bibinfo {volume}
  {742}},\ \bibinfo {pages} {122} (\bibinfo {year} {2011})}\BibitemShut
  {NoStop}%
\bibitem [{\citenamefont {Guillot}\ \emph {et~al.}(2013)\citenamefont
  {Guillot}, \citenamefont {Servillat}, \citenamefont {Webb},\ and\
  \citenamefont {Rutledge}}]{Guillot:2013wu}%
  \BibitemOpen
  \bibfield  {author} {\bibinfo {author} {\bibfnamefont {S.}~\bibnamefont
  {Guillot}}, \bibinfo {author} {\bibfnamefont {M.}~\bibnamefont {Servillat}},
  \bibinfo {author} {\bibfnamefont {N.~A.}\ \bibnamefont {Webb}}, \ and\
  \bibinfo {author} {\bibfnamefont {R.~E.}\ \bibnamefont {Rutledge}},\ }\href
  {\doibase 10.1088/0004-637X/772/1/7} {\bibfield  {journal} {\bibinfo
  {journal} {Astrophys. J.}\ }\textbf {\bibinfo {volume} {772}},\ \bibinfo
  {pages} {7} (\bibinfo {year} {2013})}\BibitemShut {NoStop}%
\bibitem [{\citenamefont {Lattimer}\ and\ \citenamefont
  {Steiner}(2014)}]{Lattimer:2013hma}%
  \BibitemOpen
  \bibfield  {author} {\bibinfo {author} {\bibfnamefont {J.~M.}\ \bibnamefont
  {Lattimer}}\ and\ \bibinfo {author} {\bibfnamefont {A.~W.}\ \bibnamefont
  {Steiner}},\ }\href {\doibase 10.1088/0004-637X/784/2/123} {\bibfield
  {journal} {\bibinfo  {journal} {Astrophys. J.}\ }\textbf {\bibinfo {volume}
  {784}},\ \bibinfo {pages} {123} (\bibinfo {year} {2014})}\BibitemShut
  {NoStop}%
\bibitem [{\citenamefont {Heinke}\ \emph {et~al.}(2014)\citenamefont {Heinke},
  \citenamefont {Cohn}, \citenamefont {Lugger}, \citenamefont {Webb},
  \citenamefont {Ho} \emph {et~al.}}]{Heinke:2014xaa}%
  \BibitemOpen
  \bibfield  {author} {\bibinfo {author} {\bibfnamefont {C.~O.}\ \bibnamefont
  {Heinke}}, \bibinfo {author} {\bibfnamefont {H.~N.}\ \bibnamefont {Cohn}},
  \bibinfo {author} {\bibfnamefont {P.~M.}\ \bibnamefont {Lugger}}, \bibinfo
  {author} {\bibfnamefont {N.~A.}\ \bibnamefont {Webb}}, \bibinfo {author}
  {\bibfnamefont {W.}~\bibnamefont {Ho}},  \emph {et~al.},\ }\href {\doibase
  10.1093/mnras/stu1449} {\bibfield  {journal} {\bibinfo  {journal} {Mon. Not.
  Roy. Astron. Soc.}\ }\textbf {\bibinfo {volume} {444}},\ \bibinfo {pages}
  {443} (\bibinfo {year} {2014})}\BibitemShut {NoStop}%
\bibitem [{\citenamefont {Guillot}\ and\ \citenamefont
  {Rutledge}(2014)}]{Guillot:2014lla}%
  \BibitemOpen
  \bibfield  {author} {\bibinfo {author} {\bibfnamefont {S.}~\bibnamefont
  {Guillot}}\ and\ \bibinfo {author} {\bibfnamefont {R.~E.}\ \bibnamefont
  {Rutledge}},\ }\href {\doibase 10.1088/2041-8205/796/1/L3} {\bibfield
  {journal} {\bibinfo  {journal} {Astrophys. J.}\ }\textbf {\bibinfo {volume}
  {796}},\ \bibinfo {pages} {L3} (\bibinfo {year} {2014})}\BibitemShut
  {NoStop}%
\bibitem [{\citenamefont {Ozel}\ \emph {et~al.}(2016)\citenamefont {Ozel},
  \citenamefont {Psaltis}, \citenamefont {Guver}, \citenamefont {Baym},
  \citenamefont {Heinke},\ and\ \citenamefont {Guillot}}]{Ozel:2015fia}%
  \BibitemOpen
  \bibfield  {author} {\bibinfo {author} {\bibfnamefont {F.}~\bibnamefont
  {Ozel}}, \bibinfo {author} {\bibfnamefont {D.}~\bibnamefont {Psaltis}},
  \bibinfo {author} {\bibfnamefont {T.}~\bibnamefont {Guver}}, \bibinfo
  {author} {\bibfnamefont {G.}~\bibnamefont {Baym}}, \bibinfo {author}
  {\bibfnamefont {C.}~\bibnamefont {Heinke}}, \ and\ \bibinfo {author}
  {\bibfnamefont {S.}~\bibnamefont {Guillot}},\ }\href {\doibase
  10.3847/0004-637X/820/1/28} {\bibfield  {journal} {\bibinfo  {journal}
  {Astrophys. J.}\ }\textbf {\bibinfo {volume} {820}},\ \bibinfo {pages} {28}
  (\bibinfo {year} {2016})}\BibitemShut {NoStop}%
\bibitem [{\citenamefont {Watts}\ \emph {et~al.}(2016)\citenamefont {Watts}
  \emph {et~al.}}]{Watts:2016uzu}%
  \BibitemOpen
  \bibfield  {author} {\bibinfo {author} {\bibfnamefont {A.~L.}\ \bibnamefont
  {Watts}} \emph {et~al.},\ }\href {\doibase 10.1103/RevModPhys.88.021001}
  {\bibfield  {journal} {\bibinfo  {journal} {Rev. Mod. Phys.}\ }\textbf
  {\bibinfo {volume} {88}},\ \bibinfo {pages} {021001} (\bibinfo {year}
  {2016})}\BibitemShut {NoStop}%
\bibitem [{\citenamefont {Steiner}\ \emph {et~al.}(2018)\citenamefont
  {Steiner}, \citenamefont {Heinke}, \citenamefont {Bogdanov}, \citenamefont
  {Li}, \citenamefont {Ho}, \citenamefont {Bahramian},\ and\ \citenamefont
  {Han}}]{Steiner:2017vmg}%
  \BibitemOpen
  \bibfield  {author} {\bibinfo {author} {\bibfnamefont {A.~W.}\ \bibnamefont
  {Steiner}}, \bibinfo {author} {\bibfnamefont {C.~O.}\ \bibnamefont {Heinke}},
  \bibinfo {author} {\bibfnamefont {S.}~\bibnamefont {Bogdanov}}, \bibinfo
  {author} {\bibfnamefont {C.}~\bibnamefont {Li}}, \bibinfo {author}
  {\bibfnamefont {W.~C.~G.}\ \bibnamefont {Ho}}, \bibinfo {author}
  {\bibfnamefont {A.}~\bibnamefont {Bahramian}}, \ and\ \bibinfo {author}
  {\bibfnamefont {S.}~\bibnamefont {Han}},\ }\href {\doibase
  10.1093/mnras/sty215} {\bibfield  {journal} {\bibinfo  {journal} {Mon. Not.
  Roy. Astron. Soc.}\ }\textbf {\bibinfo {volume} {476}},\ \bibinfo {pages}
  {421} (\bibinfo {year} {2018})}\BibitemShut {NoStop}%
\bibitem [{\citenamefont {Nattila}\ \emph {et~al.}(2017)\citenamefont
  {Nattila}, \citenamefont {Miller}, \citenamefont {Steiner}, \citenamefont
  {Kajava}, \citenamefont {Suleimanov},\ and\ \citenamefont
  {Poutanen}}]{Nattila:2017wtj}%
  \BibitemOpen
  \bibfield  {author} {\bibinfo {author} {\bibfnamefont {J.}~\bibnamefont
  {Nattila}}, \bibinfo {author} {\bibfnamefont {M.~C.}\ \bibnamefont {Miller}},
  \bibinfo {author} {\bibfnamefont {A.~W.}\ \bibnamefont {Steiner}}, \bibinfo
  {author} {\bibfnamefont {J.~J.~E.}\ \bibnamefont {Kajava}}, \bibinfo {author}
  {\bibfnamefont {V.~F.}\ \bibnamefont {Suleimanov}}, \ and\ \bibinfo {author}
  {\bibfnamefont {J.}~\bibnamefont {Poutanen}},\ }\href {\doibase
  10.1051/0004-6361/201731082} {\bibfield  {journal} {\bibinfo  {journal}
  {Astron. Astrophys.}\ }\textbf {\bibinfo {volume} {608}},\ \bibinfo {pages}
  {A31} (\bibinfo {year} {2017})}\BibitemShut {NoStop}%
\bibitem [{\citenamefont {Riley}\ \emph {et~al.}(2019)\citenamefont {Riley}
  \emph {et~al.}}]{Riley:2019yda}%
  \BibitemOpen
  \bibfield  {author} {\bibinfo {author} {\bibfnamefont {T.~E.}\ \bibnamefont
  {Riley}} \emph {et~al.},\ }\href {\doibase 10.3847/2041-8213/ab481c}
  {\bibfield  {journal} {\bibinfo  {journal} {Astrophys. J. Lett.}\ }\textbf
  {\bibinfo {volume} {887}},\ \bibinfo {pages} {L21} (\bibinfo {year}
  {2019})}\BibitemShut {NoStop}%
\bibitem [{\citenamefont {Miller}\ \emph {et~al.}(2019)\citenamefont {Miller}
  \emph {et~al.}}]{Miller:2019cac}%
  \BibitemOpen
  \bibfield  {author} {\bibinfo {author} {\bibfnamefont {M.~C.}\ \bibnamefont
  {Miller}} \emph {et~al.},\ }\href@noop {} {\bibfield  {journal} {\bibinfo
  {journal} {Astrophys. J. Lett.}\ }\textbf {\bibinfo {volume} {887}},\
  \bibinfo {pages} {L24} (\bibinfo {year} {2019})}\BibitemShut {NoStop}%
\bibitem [{\citenamefont {Miller}\ \emph {et~al.}(2021)\citenamefont {Miller}
  \emph {et~al.}}]{Miller:2021qha}%
  \BibitemOpen
  \bibfield  {author} {\bibinfo {author} {\bibfnamefont {M.~C.}\ \bibnamefont
  {Miller}} \emph {et~al.},\ }\href {\doibase 10.3847/2041-8213/ac089b}
  {\bibfield  {journal} {\bibinfo  {journal} {Astrophys. J. Lett.}\ }\textbf
  {\bibinfo {volume} {918}},\ \bibinfo {pages} {L28} (\bibinfo {year}
  {2021})}\BibitemShut {NoStop}%
\bibitem [{\citenamefont {Riley}\ \emph {et~al.}(2021)\citenamefont {Riley}
  \emph {et~al.}}]{Riley:2021pdl}%
  \BibitemOpen
  \bibfield  {author} {\bibinfo {author} {\bibfnamefont {T.~E.}\ \bibnamefont
  {Riley}} \emph {et~al.},\ }\href {\doibase 10.3847/2041-8213/ac0a81}
  {\bibfield  {journal} {\bibinfo  {journal} {Astrophys. J. Lett.}\ }\textbf
  {\bibinfo {volume} {918}},\ \bibinfo {pages} {L27} (\bibinfo {year}
  {2021})}\BibitemShut {NoStop}%
\bibitem [{\citenamefont {Raaijmakers}\ \emph {et~al.}(2021)\citenamefont
  {Raaijmakers}, \citenamefont {Greif}, \citenamefont {Hebeler}, \citenamefont
  {Hinderer}, \citenamefont {Nissanke}, \citenamefont {Schwenk}, \citenamefont
  {Riley}, \citenamefont {Watts}, \citenamefont {Lattimer},\ and\ \citenamefont
  {Ho}}]{Raaijmakers:2021uju}%
  \BibitemOpen
  \bibfield  {author} {\bibinfo {author} {\bibfnamefont {G.}~\bibnamefont
  {Raaijmakers}}, \bibinfo {author} {\bibfnamefont {S.~K.}\ \bibnamefont
  {Greif}}, \bibinfo {author} {\bibfnamefont {K.}~\bibnamefont {Hebeler}},
  \bibinfo {author} {\bibfnamefont {T.}~\bibnamefont {Hinderer}}, \bibinfo
  {author} {\bibfnamefont {S.}~\bibnamefont {Nissanke}}, \bibinfo {author}
  {\bibfnamefont {A.}~\bibnamefont {Schwenk}}, \bibinfo {author} {\bibfnamefont
  {T.~E.}\ \bibnamefont {Riley}}, \bibinfo {author} {\bibfnamefont {A.~L.}\
  \bibnamefont {Watts}}, \bibinfo {author} {\bibfnamefont {J.~M.}\ \bibnamefont
  {Lattimer}}, \ and\ \bibinfo {author} {\bibfnamefont {W.~C.~G.}\ \bibnamefont
  {Ho}},\ }\href@noop {} {\  (\bibinfo {year} {2021})},\ \Eprint
  {http://arxiv.org/abs/2105.06981} {arXiv:2105.06981 [astro-ph.HE]}
  \BibitemShut {NoStop}%
\bibitem [{\citenamefont {Chamel}\ and\ \citenamefont
  {Haensel}(2008)}]{Chamel:2008ca}%
  \BibitemOpen
  \bibfield  {author} {\bibinfo {author} {\bibfnamefont {N.}~\bibnamefont
  {Chamel}}\ and\ \bibinfo {author} {\bibfnamefont {P.}~\bibnamefont
  {Haensel}},\ }\href@noop {} {\bibfield  {journal} {\bibinfo  {journal}
  {Living Rev. Rel.}\ }\textbf {\bibinfo {volume} {11}},\ \bibinfo {pages} {10}
  (\bibinfo {year} {2008})}\BibitemShut {NoStop}%
\bibitem [{\citenamefont {Bertulani}\ and\ \citenamefont
  {Piekarewicz}(2012)}]{Bertulani:2012}%
  \BibitemOpen
  \bibfield  {author} {\bibinfo {author} {\bibfnamefont {C.}~\bibnamefont
  {Bertulani}}\ and\ \bibinfo {author} {\bibfnamefont {J.}~\bibnamefont
  {Piekarewicz}},\ }\enquote {\bibinfo {title} {Neutron star crust.}}\ \
  (\bibinfo  {publisher} {Nova Science Publishers, Hauppauge New York},\
  \bibinfo {year} {2012})\BibitemShut {NoStop}%
\bibitem [{\citenamefont {Ravenhall}\ \emph {et~al.}(1983)\citenamefont
  {Ravenhall}, \citenamefont {Pethick},\ and\ \citenamefont
  {Wilson}}]{Ravenhall:1983uh}%
  \BibitemOpen
  \bibfield  {author} {\bibinfo {author} {\bibfnamefont {D.~G.}\ \bibnamefont
  {Ravenhall}}, \bibinfo {author} {\bibfnamefont {C.~J.}\ \bibnamefont
  {Pethick}}, \ and\ \bibinfo {author} {\bibfnamefont {J.~R.}\ \bibnamefont
  {Wilson}},\ }\href@noop {} {\bibfield  {journal} {\bibinfo  {journal} {Phys.
  Rev. Lett.}\ }\textbf {\bibinfo {volume} {50}},\ \bibinfo {pages} {2066}
  (\bibinfo {year} {1983})}\BibitemShut {NoStop}%
\bibitem [{\citenamefont {Hashimoto}\ \emph {et~al.}(1984)\citenamefont
  {Hashimoto}, \citenamefont {Seki},\ and\ \citenamefont
  {Yamada}}]{Hashimoto:1984}%
  \BibitemOpen
  \bibfield  {author} {\bibinfo {author} {\bibfnamefont {M.}~\bibnamefont
  {Hashimoto}}, \bibinfo {author} {\bibfnamefont {H.}~\bibnamefont {Seki}}, \
  and\ \bibinfo {author} {\bibfnamefont {M.}~\bibnamefont {Yamada}},\
  }\href@noop {} {\bibfield  {journal} {\bibinfo  {journal} {Prog. Theor.
  Phys.}\ }\textbf {\bibinfo {volume} {71}},\ \bibinfo {pages} {320} (\bibinfo
  {year} {1984})}\BibitemShut {NoStop}%
\bibitem [{\citenamefont {Baym}\ \emph {et~al.}(1971)\citenamefont {Baym},
  \citenamefont {Pethick},\ and\ \citenamefont {Sutherland}}]{Baym:1971pw}%
  \BibitemOpen
  \bibfield  {author} {\bibinfo {author} {\bibfnamefont {G.}~\bibnamefont
  {Baym}}, \bibinfo {author} {\bibfnamefont {C.}~\bibnamefont {Pethick}}, \
  and\ \bibinfo {author} {\bibfnamefont {P.}~\bibnamefont {Sutherland}},\
  }\href@noop {} {\bibfield  {journal} {\bibinfo  {journal} {Astrophys. J.}\
  }\textbf {\bibinfo {volume} {170}},\ \bibinfo {pages} {299} (\bibinfo {year}
  {1971})}\BibitemShut {NoStop}%
\bibitem [{\citenamefont {Coldwell-Horsfall}\ and\ \citenamefont
  {Maradudin}(1960)}]{Coldwell:1960}%
  \BibitemOpen
  \bibfield  {author} {\bibinfo {author} {\bibfnamefont {R.~A.}\ \bibnamefont
  {Coldwell-Horsfall}}\ and\ \bibinfo {author} {\bibfnamefont {A.~A.}\
  \bibnamefont {Maradudin}},\ }\href {\doibase 10.1063/1.1703670} {\bibfield
  {journal} {\bibinfo  {journal} {Journal of Mathematical Physics}\ }\textbf
  {\bibinfo {volume} {1}},\ \bibinfo {pages} {395} (\bibinfo {year}
  {1960})}\BibitemShut {NoStop}%
\bibitem [{\citenamefont {Sholl}(1967)}]{Sholl:1967}%
  \BibitemOpen
  \bibfield  {author} {\bibinfo {author} {\bibfnamefont {C.~A.}\ \bibnamefont
  {Sholl}},\ }\href@noop {} {\bibfield  {journal} {\bibinfo  {journal}
  {Proceedings of the Physical Society}\ }\textbf {\bibinfo {volume} {92}},\
  \bibinfo {pages} {434} (\bibinfo {year} {1967})}\BibitemShut {NoStop}%
\bibitem [{\citenamefont {Ruester}\ \emph {et~al.}(2006)\citenamefont
  {Ruester}, \citenamefont {Hempel},\ and\ \citenamefont
  {Schaffner-Bielich}}]{Ruester:2005fm}%
  \BibitemOpen
  \bibfield  {author} {\bibinfo {author} {\bibfnamefont {S.~B.}\ \bibnamefont
  {Ruester}}, \bibinfo {author} {\bibfnamefont {M.}~\bibnamefont {Hempel}}, \
  and\ \bibinfo {author} {\bibfnamefont {J.}~\bibnamefont
  {Schaffner-Bielich}},\ }\href@noop {} {\bibfield  {journal} {\bibinfo
  {journal} {Phys. Rev.}\ }\textbf {\bibinfo {volume} {C73}},\ \bibinfo {pages}
  {035804} (\bibinfo {year} {2006})}\BibitemShut {NoStop}%
\bibitem [{\citenamefont {Roca-Maza}\ and\ \citenamefont
  {Piekarewicz}(2008)}]{RocaMaza:2008ja}%
  \BibitemOpen
  \bibfield  {author} {\bibinfo {author} {\bibfnamefont {X.}~\bibnamefont
  {Roca-Maza}}\ and\ \bibinfo {author} {\bibfnamefont {J.}~\bibnamefont
  {Piekarewicz}},\ }\href@noop {} {\bibfield  {journal} {\bibinfo  {journal}
  {Phys. Rev.}\ }\textbf {\bibinfo {volume} {C78}},\ \bibinfo {pages} {025807}
  (\bibinfo {year} {2008})}\BibitemShut {NoStop}%
\bibitem [{\citenamefont {Bethe}\ and\ \citenamefont
  {Bacher}(1936)}]{Bethe:1936}%
  \BibitemOpen
  \bibfield  {author} {\bibinfo {author} {\bibfnamefont {H.~A.}\ \bibnamefont
  {Bethe}}\ and\ \bibinfo {author} {\bibfnamefont {R.~F.}\ \bibnamefont
  {Bacher}},\ }\href {\doibase 10.1103/RevModPhys.8.82} {\bibfield  {journal}
  {\bibinfo  {journal} {Rev. Mod. Phys.}\ }\textbf {\bibinfo {volume} {8}},\
  \bibinfo {pages} {82} (\bibinfo {year} {1936})}\BibitemShut {NoStop}%
\bibitem [{\citenamefont {von Weizs{\"a}cker}(1935)}]{Weizsacker:1935}%
  \BibitemOpen
  \bibfield  {author} {\bibinfo {author} {\bibfnamefont {C.~F.}\ \bibnamefont
  {von Weizs{\"a}cker}},\ }\href@noop {} {\bibfield  {journal} {\bibinfo
  {journal} {Z. Physik}\ }\textbf {\bibinfo {volume} {96}},\ \bibinfo {pages}
  {431} (\bibinfo {year} {1935})}\BibitemShut {NoStop}%
\bibitem [{\citenamefont {Duflo}(1994)}]{Duflo:1994}%
  \BibitemOpen
  \bibfield  {author} {\bibinfo {author} {\bibfnamefont {J.}~\bibnamefont
  {Duflo}},\ }\href {\doibase 10.1016/0375-9474(94)90737-4} {\bibfield
  {journal} {\bibinfo  {journal} {Nucl. Phys.}\ }\textbf {\bibinfo {volume}
  {A576}},\ \bibinfo {pages} {29} (\bibinfo {year} {1994})}\BibitemShut
  {NoStop}%
\bibitem [{\citenamefont {Duflo}\ and\ \citenamefont
  {Zuker}(1995)}]{Duflo:1995}%
  \BibitemOpen
  \bibfield  {author} {\bibinfo {author} {\bibfnamefont {J.}~\bibnamefont
  {Duflo}}\ and\ \bibinfo {author} {\bibfnamefont {A.}~\bibnamefont {Zuker}},\
  }\href {\doibase 10.1103/PhysRevC.52.R23} {\bibfield  {journal} {\bibinfo
  {journal} {Phys. Rev.}\ }\textbf {\bibinfo {volume} {C52}},\ \bibinfo {pages}
  {R23} (\bibinfo {year} {1995})}\BibitemShut {NoStop}%
\bibitem [{\citenamefont {M\"oller}\ \emph {et~al.}(1996)\citenamefont
  {M\"oller}, \citenamefont {Nix},\ and\ \citenamefont
  {Kratz}}]{Moller:1997bz}%
  \BibitemOpen
  \bibfield  {author} {\bibinfo {author} {\bibfnamefont {P.}~\bibnamefont
  {M\"oller}}, \bibinfo {author} {\bibfnamefont {J.~R.}\ \bibnamefont {Nix}}, \
  and\ \bibinfo {author} {\bibfnamefont {K.~L.}\ \bibnamefont {Kratz}},\
  }\href@noop {} {\bibfield  {journal} {\bibinfo  {journal} {Atom. Data Nucl.
  Data Tabl.}\ }\textbf {\bibinfo {volume} {66}},\ \bibinfo {pages} {131}
  (\bibinfo {year} {1996})}\BibitemShut {NoStop}%
\bibitem [{\citenamefont {M\"oller}\ \emph {et~al.}(2012)\citenamefont
  {M\"oller}, \citenamefont {Myers}, \citenamefont {Sagawa},\ and\
  \citenamefont {Yoshida}}]{Moller:2012}%
  \BibitemOpen
  \bibfield  {author} {\bibinfo {author} {\bibfnamefont {P.}~\bibnamefont
  {M\"oller}}, \bibinfo {author} {\bibfnamefont {W.~D.}\ \bibnamefont {Myers}},
  \bibinfo {author} {\bibfnamefont {H.}~\bibnamefont {Sagawa}}, \ and\ \bibinfo
  {author} {\bibfnamefont {S.}~\bibnamefont {Yoshida}},\ }\href@noop {}
  {\bibfield  {journal} {\bibinfo  {journal} {Phys. Rev. Lett.}\ }\textbf
  {\bibinfo {volume} {108}},\ \bibinfo {pages} {052501} (\bibinfo {year}
  {2012})}\BibitemShut {NoStop}%
\bibitem [{\citenamefont {Goriely}\ \emph {et~al.}(2009)\citenamefont
  {Goriely}, \citenamefont {Chamel},\ and\ \citenamefont
  {Pearson}}]{Goriely:2009mw}%
  \BibitemOpen
  \bibfield  {author} {\bibinfo {author} {\bibfnamefont {S.}~\bibnamefont
  {Goriely}}, \bibinfo {author} {\bibfnamefont {N.}~\bibnamefont {Chamel}}, \
  and\ \bibinfo {author} {\bibfnamefont {J.~M.}\ \bibnamefont {Pearson}},\
  }\href@noop {} {\bibfield  {journal} {\bibinfo  {journal} {Phys. Rev. Lett.}\
  }\textbf {\bibinfo {volume} {102}},\ \bibinfo {pages} {152503} (\bibinfo
  {year} {2009})}\BibitemShut {NoStop}%
\bibitem [{\citenamefont {Goriely}\ \emph {et~al.}(2013)\citenamefont
  {Goriely}, \citenamefont {Chamel},\ and\ \citenamefont
  {Pearson}}]{Goriely:2013nxa}%
  \BibitemOpen
  \bibfield  {author} {\bibinfo {author} {\bibfnamefont {S.}~\bibnamefont
  {Goriely}}, \bibinfo {author} {\bibfnamefont {N.}~\bibnamefont {Chamel}}, \
  and\ \bibinfo {author} {\bibfnamefont {J.~M.}\ \bibnamefont {Pearson}},\
  }\href {\doibase 10.1103/PhysRevC.88.061302} {\bibfield  {journal} {\bibinfo
  {journal} {Phys. Rev.}\ }\textbf {\bibinfo {volume} {C88}},\ \bibinfo {pages}
  {061302} (\bibinfo {year} {2013})}\BibitemShut {NoStop}%
\bibitem [{\citenamefont {Kortelainen}\ \emph {et~al.}(2010)\citenamefont
  {Kortelainen}, \citenamefont {Lesinski}, \citenamefont {More}, \citenamefont
  {Nazarewicz}, \citenamefont {Sarich} \emph {et~al.}}]{Kortelainen:2010hv}%
  \BibitemOpen
  \bibfield  {author} {\bibinfo {author} {\bibfnamefont {M.}~\bibnamefont
  {Kortelainen}}, \bibinfo {author} {\bibfnamefont {T.}~\bibnamefont
  {Lesinski}}, \bibinfo {author} {\bibfnamefont {J.}~\bibnamefont {More}},
  \bibinfo {author} {\bibfnamefont {W.}~\bibnamefont {Nazarewicz}}, \bibinfo
  {author} {\bibfnamefont {J.}~\bibnamefont {Sarich}},  \emph {et~al.},\
  }\href@noop {} {\bibfield  {journal} {\bibinfo  {journal} {Phys. Rev.}\
  }\textbf {\bibinfo {volume} {C82}},\ \bibinfo {pages} {024313} (\bibinfo
  {year} {2010})}\BibitemShut {NoStop}%
\bibitem [{\citenamefont {Gernoth}\ \emph {et~al.}(1993)\citenamefont
  {Gernoth}, \citenamefont {Clark}, \citenamefont {Prater},\ and\ \citenamefont
  {Bohr}}]{Gernoth:1993}%
  \BibitemOpen
  \bibfield  {author} {\bibinfo {author} {\bibfnamefont {K.}~\bibnamefont
  {Gernoth}}, \bibinfo {author} {\bibfnamefont {J.}~\bibnamefont {Clark}},
  \bibinfo {author} {\bibfnamefont {J.}~\bibnamefont {Prater}}, \ and\ \bibinfo
  {author} {\bibfnamefont {H.}~\bibnamefont {Bohr}},\ }\href {\doibase
  10.1016/0370-2693(93)90738-4} {\bibfield  {journal} {\bibinfo  {journal}
  {Phys. Lett. B}\ }\textbf {\bibinfo {volume} {300}},\ \bibinfo {pages} {1}
  (\bibinfo {year} {1993})}\BibitemShut {NoStop}%
\bibitem [{\citenamefont {Clark}\ and\ \citenamefont
  {Li}(2006)}]{Clark:2006ua}%
  \BibitemOpen
  \bibfield  {author} {\bibinfo {author} {\bibfnamefont {J.~W.}\ \bibnamefont
  {Clark}}\ and\ \bibinfo {author} {\bibfnamefont {H.}~\bibnamefont {Li}},\
  }\href@noop {} {\bibfield  {journal} {\bibinfo  {journal} {Int. J. Mod.
  Phys.}\ }\textbf {\bibinfo {volume} {B20}},\ \bibinfo {pages} {5015}
  (\bibinfo {year} {2006})}\BibitemShut {NoStop}%
\bibitem [{\citenamefont {Athanassopoulos}\ \emph {et~al.}(2004)\citenamefont
  {Athanassopoulos}, \citenamefont {Mavrommatis}, \citenamefont {Gernoth},\
  and\ \citenamefont {Clark}}]{Athanassopoulos:2003qe}%
  \BibitemOpen
  \bibfield  {author} {\bibinfo {author} {\bibfnamefont {S.}~\bibnamefont
  {Athanassopoulos}}, \bibinfo {author} {\bibfnamefont {E.}~\bibnamefont
  {Mavrommatis}}, \bibinfo {author} {\bibfnamefont {K.~A.}\ \bibnamefont
  {Gernoth}}, \ and\ \bibinfo {author} {\bibfnamefont {J.~W.}\ \bibnamefont
  {Clark}},\ }\href@noop {} {\bibfield  {journal} {\bibinfo  {journal} {Nucl.
  Phys.}\ }\textbf {\bibinfo {volume} {A743}},\ \bibinfo {pages} {222}
  (\bibinfo {year} {2004})}\BibitemShut {NoStop}%
\bibitem [{\citenamefont {Utama}\ \emph {et~al.}(2016)\citenamefont {Utama},
  \citenamefont {Piekarewicz},\ and\ \citenamefont {Prosper}}]{Utama:2015hva}%
  \BibitemOpen
  \bibfield  {author} {\bibinfo {author} {\bibfnamefont {R.}~\bibnamefont
  {Utama}}, \bibinfo {author} {\bibfnamefont {J.}~\bibnamefont {Piekarewicz}},
  \ and\ \bibinfo {author} {\bibfnamefont {H.~B.}\ \bibnamefont {Prosper}},\
  }\href {\doibase 10.1103/PhysRevC.93.014311} {\bibfield  {journal} {\bibinfo
  {journal} {Phys. Rev.}\ }\textbf {\bibinfo {volume} {C93}},\ \bibinfo {pages}
  {014311} (\bibinfo {year} {2016})}\BibitemShut {NoStop}%
\bibitem [{\citenamefont {Utama}\ and\ \citenamefont
  {Piekarewicz}(2017)}]{Utama:2017wqe}%
  \BibitemOpen
  \bibfield  {author} {\bibinfo {author} {\bibfnamefont {R.}~\bibnamefont
  {Utama}}\ and\ \bibinfo {author} {\bibfnamefont {J.}~\bibnamefont
  {Piekarewicz}},\ }\href {\doibase 10.1103/PhysRevC.96.044308} {\bibfield
  {journal} {\bibinfo  {journal} {Phys. Rev.}\ }\textbf {\bibinfo {volume}
  {C96}},\ \bibinfo {pages} {044308} (\bibinfo {year} {2017})}\BibitemShut
  {NoStop}%
\bibitem [{\citenamefont {Utama}\ and\ \citenamefont
  {Piekarewicz}(2018)}]{Utama:2017ytc}%
  \BibitemOpen
  \bibfield  {author} {\bibinfo {author} {\bibfnamefont {R.}~\bibnamefont
  {Utama}}\ and\ \bibinfo {author} {\bibfnamefont {J.}~\bibnamefont
  {Piekarewicz}},\ }\href {\doibase 10.1103/PhysRevC.97.014306} {\bibfield
  {journal} {\bibinfo  {journal} {Phys. Rev.}\ }\textbf {\bibinfo {volume}
  {C97}},\ \bibinfo {pages} {014306} (\bibinfo {year} {2018})}\BibitemShut
  {NoStop}%
\bibitem [{\citenamefont {Neufcourt}\ \emph {et~al.}(2018)\citenamefont
  {Neufcourt}, \citenamefont {Cao}, \citenamefont {Nazarewicz},\ and\
  \citenamefont {Viens}}]{Neufcourt:2018syo}%
  \BibitemOpen
  \bibfield  {author} {\bibinfo {author} {\bibfnamefont {L.}~\bibnamefont
  {Neufcourt}}, \bibinfo {author} {\bibfnamefont {Y.}~\bibnamefont {Cao}},
  \bibinfo {author} {\bibfnamefont {W.}~\bibnamefont {Nazarewicz}}, \ and\
  \bibinfo {author} {\bibfnamefont {F.}~\bibnamefont {Viens}},\ }\href
  {\doibase 10.1103/PhysRevC.98.034318} {\bibfield  {journal} {\bibinfo
  {journal} {Phys. Rev.}\ }\textbf {\bibinfo {volume} {C98}},\ \bibinfo {pages}
  {034318} (\bibinfo {year} {2018})}\BibitemShut {NoStop}%
\bibitem [{\citenamefont {Neufcourt}\ \emph {et~al.}(2019)\citenamefont
  {Neufcourt}, \citenamefont {Cao}, \citenamefont {Nazarewicz}, \citenamefont
  {Olsen},\ and\ \citenamefont {Viens}}]{Neufcourt:2019qvd}%
  \BibitemOpen
  \bibfield  {author} {\bibinfo {author} {\bibfnamefont {L.}~\bibnamefont
  {Neufcourt}}, \bibinfo {author} {\bibfnamefont {Y.}~\bibnamefont {Cao}},
  \bibinfo {author} {\bibfnamefont {W.}~\bibnamefont {Nazarewicz}}, \bibinfo
  {author} {\bibfnamefont {E.}~\bibnamefont {Olsen}}, \ and\ \bibinfo {author}
  {\bibfnamefont {F.}~\bibnamefont {Viens}},\ }\href {\doibase
  10.1103/PhysRevLett.122.062502} {\bibfield  {journal} {\bibinfo  {journal}
  {Phys. Rev. Lett.}\ }\textbf {\bibinfo {volume} {122}},\ \bibinfo {pages}
  {062502} (\bibinfo {year} {2019})}\BibitemShut {NoStop}%
\bibitem [{\citenamefont {Neufcourt}\ \emph {et~al.}(2020)\citenamefont
  {Neufcourt}, \citenamefont {Cao}, \citenamefont {Giuliani}, \citenamefont
  {Nazarewicz}, \citenamefont {Olsen},\ and\ \citenamefont
  {Tarasov}}]{Neufcourt:2020nme}%
  \BibitemOpen
  \bibfield  {author} {\bibinfo {author} {\bibfnamefont {L.}~\bibnamefont
  {Neufcourt}}, \bibinfo {author} {\bibfnamefont {Y.}~\bibnamefont {Cao}},
  \bibinfo {author} {\bibfnamefont {S.~A.}\ \bibnamefont {Giuliani}}, \bibinfo
  {author} {\bibfnamefont {W.}~\bibnamefont {Nazarewicz}}, \bibinfo {author}
  {\bibfnamefont {E.}~\bibnamefont {Olsen}}, \ and\ \bibinfo {author}
  {\bibfnamefont {O.~B.}\ \bibnamefont {Tarasov}},\ }\href {\doibase
  10.1103/PhysRevC.101.044307} {\bibfield  {journal} {\bibinfo  {journal}
  {Phys. Rev. C}\ }\textbf {\bibinfo {volume} {101}},\ \bibinfo {pages}
  {044307} (\bibinfo {year} {2020})}\BibitemShut {NoStop}%
\bibitem [{\citenamefont {Niu}\ and\ \citenamefont
  {Liang}(2018)}]{Niu:2018csp}%
  \BibitemOpen
  \bibfield  {author} {\bibinfo {author} {\bibfnamefont {Z.~M.}\ \bibnamefont
  {Niu}}\ and\ \bibinfo {author} {\bibfnamefont {H.~Z.}\ \bibnamefont
  {Liang}},\ }\href {\doibase 10.1016/j.physletb.2018.01.002} {\bibfield
  {journal} {\bibinfo  {journal} {Phys. Lett. B}\ }\textbf {\bibinfo {volume}
  {778}},\ \bibinfo {pages} {48} (\bibinfo {year} {2018})}\BibitemShut
  {NoStop}%
\bibitem [{\citenamefont {Lovell}\ \emph {et~al.}(2022)\citenamefont {Lovell},
  \citenamefont {Mohan}, \citenamefont {Sprouse},\ and\ \citenamefont
  {Mumpower}}]{Lovell:2022pkw}%
  \BibitemOpen
  \bibfield  {author} {\bibinfo {author} {\bibfnamefont {A.~E.}\ \bibnamefont
  {Lovell}}, \bibinfo {author} {\bibfnamefont {A.~T.}\ \bibnamefont {Mohan}},
  \bibinfo {author} {\bibfnamefont {T.~M.}\ \bibnamefont {Sprouse}}, \ and\
  \bibinfo {author} {\bibfnamefont {M.~R.}\ \bibnamefont {Mumpower}},\ }\href
  {\doibase 10.1103/PhysRevC.106.014305} {\bibfield  {journal} {\bibinfo
  {journal} {Phys. Rev. C}\ }\textbf {\bibinfo {volume} {106}},\ \bibinfo
  {pages} {014305} (\bibinfo {year} {2022})}\BibitemShut {NoStop}%
\bibitem [{\citenamefont {Giraud}\ \emph {et~al.}(2022)\citenamefont {Giraud}
  \emph {et~al.}}]{Giraud:2022cgb}%
  \BibitemOpen
  \bibfield  {author} {\bibinfo {author} {\bibfnamefont {S.}~\bibnamefont
  {Giraud}} \emph {et~al.},\ }\href {\doibase 10.1016/j.physletb.2022.137309}
  {\bibfield  {journal} {\bibinfo  {journal} {Phys. Lett. B}\ }\textbf
  {\bibinfo {volume} {833}},\ \bibinfo {pages} {137309} (\bibinfo {year}
  {2022})}\BibitemShut {NoStop}%
\bibitem [{\citenamefont {Wolf}\ \emph {et~al.}(2013)\citenamefont {Wolf} \emph
  {et~al.}}]{Wolf:2013ge}%
  \BibitemOpen
  \bibfield  {author} {\bibinfo {author} {\bibfnamefont {R.}~\bibnamefont
  {Wolf}} \emph {et~al.},\ }\href {\doibase 10.1103/PhysRevLett.110.041101}
  {\bibfield  {journal} {\bibinfo  {journal} {Phys. Rev. Lett.}\ }\textbf
  {\bibinfo {volume} {110}},\ \bibinfo {pages} {041101} (\bibinfo {year}
  {2013})}\BibitemShut {NoStop}%
\bibitem [{\citenamefont {Pearson}\ \emph {et~al.}(2011)\citenamefont
  {Pearson}, \citenamefont {Goriely},\ and\ \citenamefont
  {Chamel}}]{Pearson:2011zz}%
  \BibitemOpen
  \bibfield  {author} {\bibinfo {author} {\bibfnamefont {J.}~\bibnamefont
  {Pearson}}, \bibinfo {author} {\bibfnamefont {S.}~\bibnamefont {Goriely}}, \
  and\ \bibinfo {author} {\bibfnamefont {N.}~\bibnamefont {Chamel}},\ }\href
  {\doibase 10.1103/PhysRevC.83.065810} {\bibfield  {journal} {\bibinfo
  {journal} {Phys. Rev.}\ }\textbf {\bibinfo {volume} {C83}},\ \bibinfo {pages}
  {065810} (\bibinfo {year} {2011})}\BibitemShut {NoStop}%
\bibitem [{\citenamefont {Horowitz}\ \emph {et~al.}(2020)\citenamefont
  {Horowitz}, \citenamefont {Piekarewicz},\ and\ \citenamefont
  {Reed}}]{Horowitz:2020evx}%
  \BibitemOpen
  \bibfield  {author} {\bibinfo {author} {\bibfnamefont {C.~J.}\ \bibnamefont
  {Horowitz}}, \bibinfo {author} {\bibfnamefont {J.}~\bibnamefont
  {Piekarewicz}}, \ and\ \bibinfo {author} {\bibfnamefont {B.}~\bibnamefont
  {Reed}},\ }\href {\doibase 10.1103/PhysRevC.102.044321} {\bibfield  {journal}
  {\bibinfo  {journal} {Phys. Rev. C}\ }\textbf {\bibinfo {volume} {102}},\
  \bibinfo {pages} {044321} (\bibinfo {year} {2020})}\BibitemShut {NoStop}%
\bibitem [{JBP()}]{JBPGC}%
  \BibitemOpen
  \href@noop {} {\enquote {\bibinfo {title} {{\rm Jodrell Bank Centre for
  Astrophysics} (glitch catalogue) {\tt
  http://www.jb.man.ac.uk/pulsar/glitches.html}},}\ }\BibitemShut {NoStop}%
\bibitem [{\citenamefont {Andersson}\ \emph {et~al.}(2012)\citenamefont
  {Andersson}, \citenamefont {Glampedakis}, \citenamefont {Ho},\ and\
  \citenamefont {Espinoza}}]{Andersson:2012iu}%
  \BibitemOpen
  \bibfield  {author} {\bibinfo {author} {\bibfnamefont {N.}~\bibnamefont
  {Andersson}}, \bibinfo {author} {\bibfnamefont {K.}~\bibnamefont
  {Glampedakis}}, \bibinfo {author} {\bibfnamefont {W.}~\bibnamefont {Ho}}, \
  and\ \bibinfo {author} {\bibfnamefont {C.}~\bibnamefont {Espinoza}},\ }\href
  {\doibase 10.1103/PhysRevLett.109.241103} {\bibfield  {journal} {\bibinfo
  {journal} {Phys. Rev. Lett.}\ }\textbf {\bibinfo {volume} {109}},\ \bibinfo
  {pages} {241103} (\bibinfo {year} {2012})}\BibitemShut {NoStop}%
\bibitem [{\citenamefont {Chamel}(2013)}]{Chamel:2012ae}%
  \BibitemOpen
  \bibfield  {author} {\bibinfo {author} {\bibfnamefont {N.}~\bibnamefont
  {Chamel}},\ }\href {\doibase 10.1103/PhysRevLett.110.011101} {\bibfield
  {journal} {\bibinfo  {journal} {Phys. Rev. Lett.}\ }\textbf {\bibinfo
  {volume} {110}},\ \bibinfo {pages} {011101} (\bibinfo {year}
  {2013})}\BibitemShut {NoStop}%
\bibitem [{\citenamefont {Anderson}\ and\ \citenamefont
  {Itoh}(1975)}]{Anderson:1975zze}%
  \BibitemOpen
  \bibfield  {author} {\bibinfo {author} {\bibfnamefont {P.}~\bibnamefont
  {Anderson}}\ and\ \bibinfo {author} {\bibfnamefont {N.}~\bibnamefont
  {Itoh}},\ }\href {\doibase 10.1038/256025a0} {\bibfield  {journal} {\bibinfo
  {journal} {Nature}\ }\textbf {\bibinfo {volume} {256}},\ \bibinfo {pages}
  {25} (\bibinfo {year} {1975})}\BibitemShut {NoStop}%
\bibitem [{\citenamefont {Alpar}\ \emph {et~al.}(1984)\citenamefont {Alpar},
  \citenamefont {Langer},\ and\ \citenamefont {Sauls}}]{Alpar:1984}%
  \BibitemOpen
  \bibfield  {author} {\bibinfo {author} {\bibfnamefont {M.}~\bibnamefont
  {Alpar}}, \bibinfo {author} {\bibfnamefont {S.}~\bibnamefont {Langer}}, \
  and\ \bibinfo {author} {\bibfnamefont {J.}~\bibnamefont {Sauls}},\
  }\href@noop {} {\bibfield  {journal} {\bibinfo  {journal} {Astrophys. J.}\
  }\textbf {\bibinfo {volume} {282}},\ \bibinfo {pages} {533} (\bibinfo {year}
  {1984})}\BibitemShut {NoStop}%
\bibitem [{\citenamefont {Pines}\ and\ \citenamefont
  {Alpar}(1985)}]{Pines:1985kz}%
  \BibitemOpen
  \bibfield  {author} {\bibinfo {author} {\bibfnamefont {D.}~\bibnamefont
  {Pines}}\ and\ \bibinfo {author} {\bibfnamefont {M.}~\bibnamefont {Alpar}},\
  }\href@noop {} {\bibfield  {journal} {\bibinfo  {journal} {Nature}\ }\textbf
  {\bibinfo {volume} {316}},\ \bibinfo {pages} {27} (\bibinfo {year}
  {1985})}\BibitemShut {NoStop}%
\bibitem [{\citenamefont {Horowitz}\ \emph
  {et~al.}(2004{\natexlab{a}})\citenamefont {Horowitz}, \citenamefont
  {Perez-Garcia},\ and\ \citenamefont {Piekarewicz}}]{Horowitz:2004yf}%
  \BibitemOpen
  \bibfield  {author} {\bibinfo {author} {\bibfnamefont {C.~J.}\ \bibnamefont
  {Horowitz}}, \bibinfo {author} {\bibfnamefont {M.~A.}\ \bibnamefont
  {Perez-Garcia}}, \ and\ \bibinfo {author} {\bibfnamefont {J.}~\bibnamefont
  {Piekarewicz}},\ }\href@noop {} {\bibfield  {journal} {\bibinfo  {journal}
  {Phys. Rev.}\ }\textbf {\bibinfo {volume} {C69}},\ \bibinfo {pages} {045804}
  (\bibinfo {year} {2004}{\natexlab{a}})}\BibitemShut {NoStop}%
\bibitem [{\citenamefont {Maruyama}\ \emph {et~al.}(2005)\citenamefont
  {Maruyama}, \citenamefont {Tatsumi}, \citenamefont {Voskresensky},
  \citenamefont {Tanigawa},\ and\ \citenamefont {Chiba}}]{Maruyama:2005vb}%
  \BibitemOpen
  \bibfield  {author} {\bibinfo {author} {\bibfnamefont {T.}~\bibnamefont
  {Maruyama}}, \bibinfo {author} {\bibfnamefont {T.}~\bibnamefont {Tatsumi}},
  \bibinfo {author} {\bibfnamefont {D.~N.}\ \bibnamefont {Voskresensky}},
  \bibinfo {author} {\bibfnamefont {T.}~\bibnamefont {Tanigawa}}, \ and\
  \bibinfo {author} {\bibfnamefont {S.}~\bibnamefont {Chiba}},\ }\href
  {\doibase 10.1103/PhysRevC.72.015802} {\bibfield  {journal} {\bibinfo
  {journal} {Phys. Rev.}\ }\textbf {\bibinfo {volume} {C72}},\ \bibinfo {pages}
  {015802} (\bibinfo {year} {2005})}\BibitemShut {NoStop}%
\bibitem [{\citenamefont {Avancini}\ \emph {et~al.}(2008)\citenamefont
  {Avancini}, \citenamefont {Menezes}, \citenamefont {Alloy}, \citenamefont
  {Marinelli}, \citenamefont {Moraes} \emph {et~al.}}]{Avancini:2008zz}%
  \BibitemOpen
  \bibfield  {author} {\bibinfo {author} {\bibfnamefont {S.}~\bibnamefont
  {Avancini}}, \bibinfo {author} {\bibfnamefont {D.}~\bibnamefont {Menezes}},
  \bibinfo {author} {\bibfnamefont {M.}~\bibnamefont {Alloy}}, \bibinfo
  {author} {\bibfnamefont {J.}~\bibnamefont {Marinelli}}, \bibinfo {author}
  {\bibfnamefont {M.}~\bibnamefont {Moraes}},  \emph {et~al.},\ }\href
  {\doibase 10.1103/PhysRevC.78.015802} {\bibfield  {journal} {\bibinfo
  {journal} {Phys. Rev.}\ }\textbf {\bibinfo {volume} {C78}},\ \bibinfo {pages}
  {015802} (\bibinfo {year} {2008})}\BibitemShut {NoStop}%
\bibitem [{\citenamefont {Newton}\ and\ \citenamefont
  {Stone}(2009)}]{Newton:2009zz}%
  \BibitemOpen
  \bibfield  {author} {\bibinfo {author} {\bibfnamefont {W.}~\bibnamefont
  {Newton}}\ and\ \bibinfo {author} {\bibfnamefont {J.}~\bibnamefont {Stone}},\
  }\href {\doibase 10.1103/PhysRevC.79.055801} {\bibfield  {journal} {\bibinfo
  {journal} {Phys. Rev.}\ }\textbf {\bibinfo {volume} {C79}},\ \bibinfo {pages}
  {055801} (\bibinfo {year} {2009})}\BibitemShut {NoStop}%
\bibitem [{\citenamefont {Avancini}\ \emph {et~al.}(2012)\citenamefont
  {Avancini}, \citenamefont {Barros}, \citenamefont {Brito}, \citenamefont
  {Chiacchiera}, \citenamefont {Menezes},\ and\ \citenamefont
  {Providencia}}]{Avancini:2012bj}%
  \BibitemOpen
  \bibfield  {author} {\bibinfo {author} {\bibfnamefont {S.~S.}\ \bibnamefont
  {Avancini}}, \bibinfo {author} {\bibfnamefont {C.~C.}\ \bibnamefont {Barros},
  \bibfnamefont {Jr}}, \bibinfo {author} {\bibfnamefont {L.}~\bibnamefont
  {Brito}}, \bibinfo {author} {\bibfnamefont {S.}~\bibnamefont {Chiacchiera}},
  \bibinfo {author} {\bibfnamefont {D.~P.}\ \bibnamefont {Menezes}}, \ and\
  \bibinfo {author} {\bibfnamefont {C.}~\bibnamefont {Providencia}},\ }\href
  {\doibase 10.1103/PhysRevC.85.035806} {\bibfield  {journal} {\bibinfo
  {journal} {Phys. Rev.}\ }\textbf {\bibinfo {volume} {C85}},\ \bibinfo {pages}
  {035806} (\bibinfo {year} {2012})}\BibitemShut {NoStop}%
\bibitem [{\citenamefont {Schuetrumpf}\ and\ \citenamefont
  {Nazarewicz}(2015)}]{Schuetrumpf:2015nza}%
  \BibitemOpen
  \bibfield  {author} {\bibinfo {author} {\bibfnamefont {B.}~\bibnamefont
  {Schuetrumpf}}\ and\ \bibinfo {author} {\bibfnamefont {W.}~\bibnamefont
  {Nazarewicz}},\ }\href {\doibase 10.1103/PhysRevC.92.045806} {\bibfield
  {journal} {\bibinfo  {journal} {Phys. Rev.}\ }\textbf {\bibinfo {volume}
  {C92}},\ \bibinfo {pages} {045806} (\bibinfo {year} {2015})}\BibitemShut
  {NoStop}%
\bibitem [{\citenamefont {Fattoyev}\ \emph {et~al.}(2017)\citenamefont
  {Fattoyev}, \citenamefont {Horowitz},\ and\ \citenamefont
  {Schuetrumpf}}]{Fattoyev:2017zhb}%
  \BibitemOpen
  \bibfield  {author} {\bibinfo {author} {\bibfnamefont {F.}~\bibnamefont
  {Fattoyev}}, \bibinfo {author} {\bibfnamefont {C.}~\bibnamefont {Horowitz}},
  \ and\ \bibinfo {author} {\bibfnamefont {B.}~\bibnamefont {Schuetrumpf}},\
  }\href {\doibase 10.1103/PhysRevC.95.055804} {\bibfield  {journal} {\bibinfo
  {journal} {Phys. Rev. C}\ }\textbf {\bibinfo {volume} {95}},\ \bibinfo
  {pages} {055804} (\bibinfo {year} {2017})}\BibitemShut {NoStop}%
\bibitem [{\citenamefont {Newton}\ \emph {et~al.}(2022)\citenamefont {Newton},
  \citenamefont {Kaltenborn}, \citenamefont {Cantu}, \citenamefont {Wang},
  \citenamefont {Stinson},\ and\ \citenamefont
  {Rikovska~Stone}}]{Newton:2021vyd}%
  \BibitemOpen
  \bibfield  {author} {\bibinfo {author} {\bibfnamefont {W.~G.}\ \bibnamefont
  {Newton}}, \bibinfo {author} {\bibfnamefont {M.~A.}\ \bibnamefont
  {Kaltenborn}}, \bibinfo {author} {\bibfnamefont {S.}~\bibnamefont {Cantu}},
  \bibinfo {author} {\bibfnamefont {S.}~\bibnamefont {Wang}}, \bibinfo {author}
  {\bibfnamefont {A.}~\bibnamefont {Stinson}}, \ and\ \bibinfo {author}
  {\bibfnamefont {J.}~\bibnamefont {Rikovska~Stone}},\ }\href {\doibase
  10.1103/PhysRevC.105.025806} {\bibfield  {journal} {\bibinfo  {journal}
  {Phys. Rev. C}\ }\textbf {\bibinfo {volume} {105}},\ \bibinfo {pages}
  {025806} (\bibinfo {year} {2022})}\BibitemShut {NoStop}%
\bibitem [{\citenamefont {Watanabe}\ \emph {et~al.}(2003)\citenamefont
  {Watanabe}, \citenamefont {Sato}, \citenamefont {Yasuoka},\ and\
  \citenamefont {Ebisuzaki}}]{Watanabe:2003xu}%
  \BibitemOpen
  \bibfield  {author} {\bibinfo {author} {\bibfnamefont {G.}~\bibnamefont
  {Watanabe}}, \bibinfo {author} {\bibfnamefont {K.}~\bibnamefont {Sato}},
  \bibinfo {author} {\bibfnamefont {K.}~\bibnamefont {Yasuoka}}, \ and\
  \bibinfo {author} {\bibfnamefont {T.}~\bibnamefont {Ebisuzaki}},\ }\href
  {\doibase 10.1103/PhysRevC.68.035806} {\bibfield  {journal} {\bibinfo
  {journal} {Phys. Rev.}\ }\textbf {\bibinfo {volume} {C68}},\ \bibinfo {pages}
  {035806} (\bibinfo {year} {2003})}\BibitemShut {NoStop}%
\bibitem [{\citenamefont {Watanabe}\ \emph {et~al.}(2005)\citenamefont
  {Watanabe}, \citenamefont {Maruyama}, \citenamefont {Sato}, \citenamefont
  {Yasuoka},\ and\ \citenamefont {Ebisuzaki}}]{Watanabe:2004tr}%
  \BibitemOpen
  \bibfield  {author} {\bibinfo {author} {\bibfnamefont {G.}~\bibnamefont
  {Watanabe}}, \bibinfo {author} {\bibfnamefont {T.}~\bibnamefont {Maruyama}},
  \bibinfo {author} {\bibfnamefont {K.}~\bibnamefont {Sato}}, \bibinfo {author}
  {\bibfnamefont {K.}~\bibnamefont {Yasuoka}}, \ and\ \bibinfo {author}
  {\bibfnamefont {T.}~\bibnamefont {Ebisuzaki}},\ }\href {\doibase
  10.1103/PhysRevLett.94.031101} {\bibfield  {journal} {\bibinfo  {journal}
  {Phys. Rev. Lett.}\ }\textbf {\bibinfo {volume} {94}},\ \bibinfo {pages}
  {031101} (\bibinfo {year} {2005})}\BibitemShut {NoStop}%
\bibitem [{\citenamefont {Watanabe}\ \emph {et~al.}(2009)\citenamefont
  {Watanabe}, \citenamefont {Sonoda}, \citenamefont {Maruyama}, \citenamefont
  {Sato}, \citenamefont {Yasuoka} \emph {et~al.}}]{Watanabe:2009vi}%
  \BibitemOpen
  \bibfield  {author} {\bibinfo {author} {\bibfnamefont {G.}~\bibnamefont
  {Watanabe}}, \bibinfo {author} {\bibfnamefont {H.}~\bibnamefont {Sonoda}},
  \bibinfo {author} {\bibfnamefont {T.}~\bibnamefont {Maruyama}}, \bibinfo
  {author} {\bibfnamefont {K.}~\bibnamefont {Sato}}, \bibinfo {author}
  {\bibfnamefont {K.}~\bibnamefont {Yasuoka}},  \emph {et~al.},\ }\href
  {\doibase 10.1103/PhysRevLett.103.121101} {\bibfield  {journal} {\bibinfo
  {journal} {Phys. Rev. Lett.}\ }\textbf {\bibinfo {volume} {103}},\ \bibinfo
  {pages} {121101} (\bibinfo {year} {2009})}\BibitemShut {NoStop}%
\bibitem [{\citenamefont {Piekarewicz}\ and\ \citenamefont
  {Toledo~Sanchez}(2012)}]{Piekarewicz:2011qc}%
  \BibitemOpen
  \bibfield  {author} {\bibinfo {author} {\bibfnamefont {J.}~\bibnamefont
  {Piekarewicz}}\ and\ \bibinfo {author} {\bibfnamefont {G.}~\bibnamefont
  {Toledo~Sanchez}},\ }\href@noop {} {\bibfield  {journal} {\bibinfo  {journal}
  {Phys. Rev.}\ }\textbf {\bibinfo {volume} {C85}},\ \bibinfo {pages} {015807}
  (\bibinfo {year} {2012})}\BibitemShut {NoStop}%
\bibitem [{\citenamefont {Caplan}\ \emph {et~al.}(2015)\citenamefont {Caplan},
  \citenamefont {Schneider}, \citenamefont {Horowitz},\ and\ \citenamefont
  {Berry}}]{Caplan:2014gaa}%
  \BibitemOpen
  \bibfield  {author} {\bibinfo {author} {\bibfnamefont {M.~E.}\ \bibnamefont
  {Caplan}}, \bibinfo {author} {\bibfnamefont {A.~S.}\ \bibnamefont
  {Schneider}}, \bibinfo {author} {\bibfnamefont {C.~J.}\ \bibnamefont
  {Horowitz}}, \ and\ \bibinfo {author} {\bibfnamefont {D.~K.}\ \bibnamefont
  {Berry}},\ }\href {\doibase 10.1103/PhysRevC.91.065802} {\bibfield  {journal}
  {\bibinfo  {journal} {Phys. Rev.}\ }\textbf {\bibinfo {volume} {C91}},\
  \bibinfo {pages} {065802} (\bibinfo {year} {2015})}\BibitemShut {NoStop}%
\bibitem [{\citenamefont {da~Silva~Schneider}\ \emph
  {et~al.}(2018)\citenamefont {da~Silva~Schneider}, \citenamefont {Caplan},
  \citenamefont {Berry},\ and\ \citenamefont {Horowitz}}]{Schneider:2018yby}%
  \BibitemOpen
  \bibfield  {author} {\bibinfo {author} {\bibfnamefont {A.}~\bibnamefont
  {da~Silva~Schneider}}, \bibinfo {author} {\bibfnamefont {M.~E.}\ \bibnamefont
  {Caplan}}, \bibinfo {author} {\bibfnamefont {D.~K.}\ \bibnamefont {Berry}}, \
  and\ \bibinfo {author} {\bibfnamefont {C.~J.}\ \bibnamefont {Horowitz}},\
  }\href {\doibase 10.1103/PhysRevC.98.055801} {\bibfield  {journal} {\bibinfo
  {journal} {Phys. Rev. C}\ }\textbf {\bibinfo {volume} {98}},\ \bibinfo
  {pages} {055801} (\bibinfo {year} {2018})}\BibitemShut {NoStop}%
\bibitem [{\citenamefont {Lopez}\ \emph {et~al.}(2021)\citenamefont {Lopez},
  \citenamefont {Dorso},\ and\ \citenamefont {Frank}}]{Lopez:2020zne}%
  \BibitemOpen
  \bibfield  {author} {\bibinfo {author} {\bibfnamefont {J.~A.}\ \bibnamefont
  {Lopez}}, \bibinfo {author} {\bibfnamefont {C.~O.}\ \bibnamefont {Dorso}}, \
  and\ \bibinfo {author} {\bibfnamefont {G.~A.}\ \bibnamefont {Frank}},\ }\href
  {\doibase 10.1007/s11467-020-1004-2} {\bibfield  {journal} {\bibinfo
  {journal} {Front. Phys. (Beijing)}\ }\textbf {\bibinfo {volume} {16}},\
  \bibinfo {pages} {24301} (\bibinfo {year} {2021})}\BibitemShut {NoStop}%
\bibitem [{\citenamefont {Shafieepour}\ \emph {et~al.}(2022)\citenamefont
  {Shafieepour}, \citenamefont {Moshfegh},\ and\ \citenamefont
  {Piekarewicz}}]{Shafieepour:2022meo}%
  \BibitemOpen
  \bibfield  {author} {\bibinfo {author} {\bibfnamefont {R.}~\bibnamefont
  {Shafieepour}}, \bibinfo {author} {\bibfnamefont {H.~R.}\ \bibnamefont
  {Moshfegh}}, \ and\ \bibinfo {author} {\bibfnamefont {J.}~\bibnamefont
  {Piekarewicz}},\ }\href@noop {} {\bibfield  {journal} {\bibinfo  {journal}
  {Phys. Rev. C}\ }\textbf {\bibinfo {volume} {105}},\ \bibinfo {pages}
  {055809} (\bibinfo {year} {2022})}\BibitemShut {NoStop}%
\bibitem [{\citenamefont {Horowitz}\ \emph
  {et~al.}(2004{\natexlab{b}})\citenamefont {Horowitz}, \citenamefont
  {Perez-Garcia}, \citenamefont {Carriere}, \citenamefont {Berry},\ and\
  \citenamefont {Piekarewicz}}]{Horowitz:2004pv}%
  \BibitemOpen
  \bibfield  {author} {\bibinfo {author} {\bibfnamefont {C.~J.}\ \bibnamefont
  {Horowitz}}, \bibinfo {author} {\bibfnamefont {M.~A.}\ \bibnamefont
  {Perez-Garcia}}, \bibinfo {author} {\bibfnamefont {J.}~\bibnamefont
  {Carriere}}, \bibinfo {author} {\bibfnamefont {D.~K.}\ \bibnamefont {Berry}},
  \ and\ \bibinfo {author} {\bibfnamefont {J.}~\bibnamefont {Piekarewicz}},\
  }\href@noop {} {\bibfield  {journal} {\bibinfo  {journal} {Phys. Rev.}\
  }\textbf {\bibinfo {volume} {C70}},\ \bibinfo {pages} {065806} (\bibinfo
  {year} {2004}{\natexlab{b}})}\BibitemShut {NoStop}%
\bibitem [{\citenamefont {Kycia}\ \emph {et~al.}(2017)\citenamefont {Kycia},
  \citenamefont {Kubis},\ and\ \citenamefont {W\'ojcik}}]{Kycia:2017ibr}%
  \BibitemOpen
  \bibfield  {author} {\bibinfo {author} {\bibfnamefont {R.~A.}\ \bibnamefont
  {Kycia}}, \bibinfo {author} {\bibfnamefont {S.}~\bibnamefont {Kubis}}, \ and\
  \bibinfo {author} {\bibfnamefont {W.}~\bibnamefont {W\'ojcik}},\ }\href
  {\doibase 10.1103/PhysRevC.96.025803} {\bibfield  {journal} {\bibinfo
  {journal} {Phys. Rev. C}\ }\textbf {\bibinfo {volume} {96}},\ \bibinfo
  {pages} {025803} (\bibinfo {year} {2017})}\BibitemShut {NoStop}%
\bibitem [{\citenamefont {Horowitz}\ \emph {et~al.}(2005)\citenamefont
  {Horowitz}, \citenamefont {Perez-Garcia}, \citenamefont {Berry},\ and\
  \citenamefont {Piekarewicz}}]{Horowitz:2005zb}%
  \BibitemOpen
  \bibfield  {author} {\bibinfo {author} {\bibfnamefont {C.~J.}\ \bibnamefont
  {Horowitz}}, \bibinfo {author} {\bibfnamefont {M.~A.}\ \bibnamefont
  {Perez-Garcia}}, \bibinfo {author} {\bibfnamefont {D.~K.}\ \bibnamefont
  {Berry}}, \ and\ \bibinfo {author} {\bibfnamefont {J.}~\bibnamefont
  {Piekarewicz}},\ }\href@noop {} {\bibfield  {journal} {\bibinfo  {journal}
  {Phys. Rev.}\ }\textbf {\bibinfo {volume} {C72}},\ \bibinfo {pages} {035801}
  (\bibinfo {year} {2005})}\BibitemShut {NoStop}%
\bibitem [{\citenamefont {Horowitz}\ and\ \citenamefont
  {Berry}(2008)}]{Horowitz:2008vf}%
  \BibitemOpen
  \bibfield  {author} {\bibinfo {author} {\bibfnamefont {C.~J.}\ \bibnamefont
  {Horowitz}}\ and\ \bibinfo {author} {\bibfnamefont {D.~K.}\ \bibnamefont
  {Berry}},\ }\href {\doibase 10.1103/PhysRevC.78.035806} {\bibfield  {journal}
  {\bibinfo  {journal} {Phys. Rev.}\ }\textbf {\bibinfo {volume} {C78}},\
  \bibinfo {pages} {035806} (\bibinfo {year} {2008})}\BibitemShut {NoStop}%
\bibitem [{\citenamefont {Horowitz}\ and\ \citenamefont
  {Kadau}(2009)}]{Horowitz:2009ya}%
  \BibitemOpen
  \bibfield  {author} {\bibinfo {author} {\bibfnamefont {C.~J.}\ \bibnamefont
  {Horowitz}}\ and\ \bibinfo {author} {\bibfnamefont {K.}~\bibnamefont
  {Kadau}},\ }\href {\doibase 10.1103/PhysRevLett.102.191102} {\bibfield
  {journal} {\bibinfo  {journal} {Phys. Rev. Lett.}\ }\textbf {\bibinfo
  {volume} {102}},\ \bibinfo {pages} {191102} (\bibinfo {year}
  {2009})}\BibitemShut {NoStop}%
\bibitem [{\citenamefont {Grygorov}\ \emph {et~al.}(2010)\citenamefont
  {Grygorov}, \citenamefont {Gogelein},\ and\ \citenamefont
  {Muther}}]{Grygorov:2010zz}%
  \BibitemOpen
  \bibfield  {author} {\bibinfo {author} {\bibfnamefont {P.}~\bibnamefont
  {Grygorov}}, \bibinfo {author} {\bibfnamefont {P.}~\bibnamefont {Gogelein}},
  \ and\ \bibinfo {author} {\bibfnamefont {H.}~\bibnamefont {Muther}},\ }\href
  {\doibase 10.1088/0954-3899/37/7/075203} {\bibfield  {journal} {\bibinfo
  {journal} {J.Phys.G}\ }\textbf {\bibinfo {volume} {G37}},\ \bibinfo {pages}
  {075203} (\bibinfo {year} {2010})}\BibitemShut {NoStop}%
\bibitem [{\citenamefont {Chamel}(2012)}]{Chamel:2012zn}%
  \BibitemOpen
  \bibfield  {author} {\bibinfo {author} {\bibfnamefont {N.}~\bibnamefont
  {Chamel}},\ }\href {\doibase 10.1103/PhysRevC.85.035801,
  10.1103/PhysRevC.85.039902} {\bibfield  {journal} {\bibinfo  {journal} {Phys.
  Rev.}\ }\textbf {\bibinfo {volume} {C85}},\ \bibinfo {pages} {035801}
  (\bibinfo {year} {2012})}\BibitemShut {NoStop}%
\bibitem [{\citenamefont {Horowitz}\ \emph {et~al.}(2015)\citenamefont
  {Horowitz}, \citenamefont {Berry}, \citenamefont {Briggs}, \citenamefont
  {Caplan}, \citenamefont {Cumming},\ and\ \citenamefont
  {Schneider}}]{Horowitz:2014xca}%
  \BibitemOpen
  \bibfield  {author} {\bibinfo {author} {\bibfnamefont {C.~J.}\ \bibnamefont
  {Horowitz}}, \bibinfo {author} {\bibfnamefont {D.~K.}\ \bibnamefont {Berry}},
  \bibinfo {author} {\bibfnamefont {C.~M.}\ \bibnamefont {Briggs}}, \bibinfo
  {author} {\bibfnamefont {M.~E.}\ \bibnamefont {Caplan}}, \bibinfo {author}
  {\bibfnamefont {A.}~\bibnamefont {Cumming}}, \ and\ \bibinfo {author}
  {\bibfnamefont {A.~S.}\ \bibnamefont {Schneider}},\ }\href {\doibase
  10.1103/PhysRevLett.114.031102} {\bibfield  {journal} {\bibinfo  {journal}
  {Phys. Rev. Lett.}\ }\textbf {\bibinfo {volume} {114}},\ \bibinfo {pages}
  {031102} (\bibinfo {year} {2015})}\BibitemShut {NoStop}%
\bibitem [{\citenamefont {Caplan}\ and\ \citenamefont
  {Horowitz}(2017)}]{Caplan:2016uvu}%
  \BibitemOpen
  \bibfield  {author} {\bibinfo {author} {\bibfnamefont {M.~E.}\ \bibnamefont
  {Caplan}}\ and\ \bibinfo {author} {\bibfnamefont {C.~J.}\ \bibnamefont
  {Horowitz}},\ }\href {\doibase 10.1103/RevModPhys.89.041002} {\bibfield
  {journal} {\bibinfo  {journal} {Rev. Mod. Phys.}\ }\textbf {\bibinfo {volume}
  {89}},\ \bibinfo {pages} {041002} (\bibinfo {year} {2017})}\BibitemShut
  {NoStop}%
\bibitem [{\citenamefont {Nandi}\ and\ \citenamefont
  {Schramm}(2018)}]{Nandi:2017aqq}%
  \BibitemOpen
  \bibfield  {author} {\bibinfo {author} {\bibfnamefont {R.}~\bibnamefont
  {Nandi}}\ and\ \bibinfo {author} {\bibfnamefont {S.}~\bibnamefont
  {Schramm}},\ }\href {\doibase 10.3847/1538-4357/aa9f12} {\bibfield  {journal}
  {\bibinfo  {journal} {Astrophys. J.}\ }\textbf {\bibinfo {volume} {852}},\
  \bibinfo {pages} {135} (\bibinfo {year} {2018})}\BibitemShut {NoStop}%
\bibitem [{\citenamefont {Lin}\ \emph {et~al.}(2020)\citenamefont {Lin},
  \citenamefont {Caplan}, \citenamefont {Horowitz},\ and\ \citenamefont
  {Lunardini}}]{Lin:2020nxy}%
  \BibitemOpen
  \bibfield  {author} {\bibinfo {author} {\bibfnamefont {Z.}~\bibnamefont
  {Lin}}, \bibinfo {author} {\bibfnamefont {M.~E.}\ \bibnamefont {Caplan}},
  \bibinfo {author} {\bibfnamefont {C.~J.}\ \bibnamefont {Horowitz}}, \ and\
  \bibinfo {author} {\bibfnamefont {C.}~\bibnamefont {Lunardini}},\ }\href
  {\doibase 10.1103/PhysRevC.102.045801} {\bibfield  {journal} {\bibinfo
  {journal} {Phys. Rev. C}\ }\textbf {\bibinfo {volume} {102}},\ \bibinfo
  {pages} {045801} (\bibinfo {year} {2020})}\BibitemShut {NoStop}%
\bibitem [{\citenamefont {Pons}\ \emph {et~al.}(2013)\citenamefont {Pons},
  \citenamefont {Vigan\'o},\ and\ \citenamefont {Rea}}]{Pons:2013nea}%
  \BibitemOpen
  \bibfield  {author} {\bibinfo {author} {\bibfnamefont {J.~A.}\ \bibnamefont
  {Pons}}, \bibinfo {author} {\bibfnamefont {D.}~\bibnamefont {Vigan\'o}}, \
  and\ \bibinfo {author} {\bibfnamefont {N.}~\bibnamefont {Rea}},\ }\href@noop
  {} {\bibfield  {journal} {\bibinfo  {journal} {Nature Physics, 9,}\ }\textbf
  {\bibinfo {volume} {431-434}} (\bibinfo {year} {2013})}\BibitemShut {NoStop}%
\bibitem [{\citenamefont {Brown}\ \emph {et~al.}(2018)\citenamefont {Brown},
  \citenamefont {Cumming}, \citenamefont {Fattoyev}, \citenamefont {Horowitz},
  \citenamefont {Page},\ and\ \citenamefont {Reddy}}]{Brown:2017gxd}%
  \BibitemOpen
  \bibfield  {author} {\bibinfo {author} {\bibfnamefont {E.~F.}\ \bibnamefont
  {Brown}}, \bibinfo {author} {\bibfnamefont {A.}~\bibnamefont {Cumming}},
  \bibinfo {author} {\bibfnamefont {F.~J.}\ \bibnamefont {Fattoyev}}, \bibinfo
  {author} {\bibfnamefont {C.}~\bibnamefont {Horowitz}}, \bibinfo {author}
  {\bibfnamefont {D.}~\bibnamefont {Page}}, \ and\ \bibinfo {author}
  {\bibfnamefont {S.}~\bibnamefont {Reddy}},\ }\href {\doibase
  10.1103/PhysRevLett.120.182701} {\bibfield  {journal} {\bibinfo  {journal}
  {Phys. Rev. Lett.}\ }\textbf {\bibinfo {volume} {120}},\ \bibinfo {pages}
  {182701} (\bibinfo {year} {2018})}\BibitemShut {NoStop}%
\bibitem [{\citenamefont {Horowitz}\ and\ \citenamefont
  {Piekarewicz}(2001{\natexlab{a}})}]{Horowitz:2000xj}%
  \BibitemOpen
  \bibfield  {author} {\bibinfo {author} {\bibfnamefont {C.~J.}\ \bibnamefont
  {Horowitz}}\ and\ \bibinfo {author} {\bibfnamefont {J.}~\bibnamefont
  {Piekarewicz}},\ }\href@noop {} {\bibfield  {journal} {\bibinfo  {journal}
  {Phys. Rev. Lett.}\ }\textbf {\bibinfo {volume} {86}},\ \bibinfo {pages}
  {5647} (\bibinfo {year} {2001}{\natexlab{a}})}\BibitemShut {NoStop}%
\bibitem [{\citenamefont {Horowitz}\ and\ \citenamefont
  {Piekarewicz}(2002)}]{Horowitz:2002mb}%
  \BibitemOpen
  \bibfield  {author} {\bibinfo {author} {\bibfnamefont {C.~J.}\ \bibnamefont
  {Horowitz}}\ and\ \bibinfo {author} {\bibfnamefont {J.}~\bibnamefont
  {Piekarewicz}},\ }\href@noop {} {\bibfield  {journal} {\bibinfo  {journal}
  {Phys. Rev.}\ }\textbf {\bibinfo {volume} {C66}},\ \bibinfo {pages} {055803}
  (\bibinfo {year} {2002})}\BibitemShut {NoStop}%
\bibitem [{\citenamefont {Lattimer}\ \emph {et~al.}(1991)\citenamefont
  {Lattimer}, \citenamefont {Prakash}, \citenamefont {Pethick},\ and\
  \citenamefont {Haensel}}]{Lattimer:1991ib}%
  \BibitemOpen
  \bibfield  {author} {\bibinfo {author} {\bibfnamefont {J.~M.}\ \bibnamefont
  {Lattimer}}, \bibinfo {author} {\bibfnamefont {M.}~\bibnamefont {Prakash}},
  \bibinfo {author} {\bibfnamefont {C.~J.}\ \bibnamefont {Pethick}}, \ and\
  \bibinfo {author} {\bibfnamefont {P.}~\bibnamefont {Haensel}},\ }\href@noop
  {} {\bibfield  {journal} {\bibinfo  {journal} {Phys. Rev. Lett.}\ }\textbf
  {\bibinfo {volume} {66}},\ \bibinfo {pages} {2701} (\bibinfo {year}
  {1991})}\BibitemShut {NoStop}%
\bibitem [{\citenamefont {Adhikari}\ \emph {et~al.}(2021)\citenamefont
  {Adhikari} \emph {et~al.}}]{Adhikari:2021phr}%
  \BibitemOpen
  \bibfield  {author} {\bibinfo {author} {\bibfnamefont {D.}~\bibnamefont
  {Adhikari}} \emph {et~al.} (\bibinfo {collaboration} {PREX}),\ }\href
  {\doibase 10.1103/PhysRevLett.126.172502} {\bibfield  {journal} {\bibinfo
  {journal} {Phys. Rev. Lett.}\ }\textbf {\bibinfo {volume} {126}},\ \bibinfo
  {pages} {172502} (\bibinfo {year} {2021})}\BibitemShut {NoStop}%
\bibitem [{\citenamefont {Reed}\ \emph {et~al.}(2021)\citenamefont {Reed},
  \citenamefont {Fattoyev}, \citenamefont {Horowitz},\ and\ \citenamefont
  {Piekarewicz}}]{Reed:2021nqk}%
  \BibitemOpen
  \bibfield  {author} {\bibinfo {author} {\bibfnamefont {B.~T.}\ \bibnamefont
  {Reed}}, \bibinfo {author} {\bibfnamefont {F.~J.}\ \bibnamefont {Fattoyev}},
  \bibinfo {author} {\bibfnamefont {C.~J.}\ \bibnamefont {Horowitz}}, \ and\
  \bibinfo {author} {\bibfnamefont {J.}~\bibnamefont {Piekarewicz}},\ }\href
  {\doibase 10.1103/PhysRevLett.126.172503} {\bibfield  {journal} {\bibinfo
  {journal} {Phys. Rev. Lett.}\ }\textbf {\bibinfo {volume} {126}},\ \bibinfo
  {pages} {172503} (\bibinfo {year} {2021})}\BibitemShut {NoStop}%
\bibitem [{\citenamefont {Vidana}\ \emph {et~al.}(2000)\citenamefont {Vidana},
  \citenamefont {Polls}, \citenamefont {Ramos}, \citenamefont {Engvik},\ and\
  \citenamefont {Hjorth-Jensen}}]{Vidana:2000ew}%
  \BibitemOpen
  \bibfield  {author} {\bibinfo {author} {\bibfnamefont {I.}~\bibnamefont
  {Vidana}}, \bibinfo {author} {\bibfnamefont {A.}~\bibnamefont {Polls}},
  \bibinfo {author} {\bibfnamefont {A.}~\bibnamefont {Ramos}}, \bibinfo
  {author} {\bibfnamefont {L.}~\bibnamefont {Engvik}}, \ and\ \bibinfo {author}
  {\bibfnamefont {M.}~\bibnamefont {Hjorth-Jensen}},\ }\href {\doibase
  10.1103/PhysRevC.62.035801} {\bibfield  {journal} {\bibinfo  {journal} {Phys.
  Rev. C}\ }\textbf {\bibinfo {volume} {62}},\ \bibinfo {pages} {035801}
  (\bibinfo {year} {2000})}\BibitemShut {NoStop}%
\bibitem [{\citenamefont {Sammarruca}(2009)}]{Sammarruca:2009wn}%
  \BibitemOpen
  \bibfield  {author} {\bibinfo {author} {\bibfnamefont {F.}~\bibnamefont
  {Sammarruca}},\ }\href {\doibase 10.1103/PhysRevC.79.034301} {\bibfield
  {journal} {\bibinfo  {journal} {Phys. Rev. C}\ }\textbf {\bibinfo {volume}
  {79}},\ \bibinfo {pages} {034301} (\bibinfo {year} {2009})}\BibitemShut
  {NoStop}%
\bibitem [{\citenamefont {Weissenborn}\ \emph {et~al.}(2012)\citenamefont
  {Weissenborn}, \citenamefont {Chatterjee},\ and\ \citenamefont
  {Schaffner-Bielich}}]{Weissenborn:2011ut}%
  \BibitemOpen
  \bibfield  {author} {\bibinfo {author} {\bibfnamefont {S.}~\bibnamefont
  {Weissenborn}}, \bibinfo {author} {\bibfnamefont {D.}~\bibnamefont
  {Chatterjee}}, \ and\ \bibinfo {author} {\bibfnamefont {J.}~\bibnamefont
  {Schaffner-Bielich}},\ }\href {\doibase 10.1103/PhysRevC.85.065802}
  {\bibfield  {journal} {\bibinfo  {journal} {Phys. Rev. C}\ }\textbf {\bibinfo
  {volume} {85}},\ \bibinfo {pages} {065802} (\bibinfo {year} {2012})},\
  \bibinfo {note} {[Erratum: Phys.Rev.C 90, 019904 (2014)]}\BibitemShut
  {NoStop}%
\bibitem [{\citenamefont {Massot}\ \emph {et~al.}(2012)\citenamefont {Massot},
  \citenamefont {Margueron},\ and\ \citenamefont {Chanfray}}]{Massot:2012pf}%
  \BibitemOpen
  \bibfield  {author} {\bibinfo {author} {\bibfnamefont {E.}~\bibnamefont
  {Massot}}, \bibinfo {author} {\bibfnamefont {J.}~\bibnamefont {Margueron}}, \
  and\ \bibinfo {author} {\bibfnamefont {G.}~\bibnamefont {Chanfray}},\ }\href
  {\doibase 10.1209/0295-5075/97/39002} {\bibfield  {journal} {\bibinfo
  {journal} {EPL}\ }\textbf {\bibinfo {volume} {97}},\ \bibinfo {pages} {39002}
  (\bibinfo {year} {2012})}\BibitemShut {NoStop}%
\bibitem [{\citenamefont {Lonardoni}\ \emph {et~al.}(2015)\citenamefont
  {Lonardoni}, \citenamefont {Lovato}, \citenamefont {Gandolfi},\ and\
  \citenamefont {Pederiva}}]{Lonardoni:2014bwa}%
  \BibitemOpen
  \bibfield  {author} {\bibinfo {author} {\bibfnamefont {D.}~\bibnamefont
  {Lonardoni}}, \bibinfo {author} {\bibfnamefont {A.}~\bibnamefont {Lovato}},
  \bibinfo {author} {\bibfnamefont {S.}~\bibnamefont {Gandolfi}}, \ and\
  \bibinfo {author} {\bibfnamefont {F.}~\bibnamefont {Pederiva}},\ }\href
  {\doibase 10.1103/PhysRevLett.114.092301} {\bibfield  {journal} {\bibinfo
  {journal} {Phys. Rev. Lett.}\ }\textbf {\bibinfo {volume} {114}},\ \bibinfo
  {pages} {092301} (\bibinfo {year} {2015})}\BibitemShut {NoStop}%
\bibitem [{\citenamefont {Oertel}\ \emph {et~al.}(2016)\citenamefont {Oertel},
  \citenamefont {Gulminelli}, \citenamefont {Provid\^encia},\ and\
  \citenamefont {Raduta}}]{Oertel:2016xsn}%
  \BibitemOpen
  \bibfield  {author} {\bibinfo {author} {\bibfnamefont {M.}~\bibnamefont
  {Oertel}}, \bibinfo {author} {\bibfnamefont {F.}~\bibnamefont {Gulminelli}},
  \bibinfo {author} {\bibfnamefont {C.}~\bibnamefont {Provid\^encia}}, \ and\
  \bibinfo {author} {\bibfnamefont {A.~R.}\ \bibnamefont {Raduta}},\ }\href
  {\doibase 10.1140/epja/i2016-16050-1} {\bibfield  {journal} {\bibinfo
  {journal} {Eur. Phys. J. A}\ }\textbf {\bibinfo {volume} {52}},\ \bibinfo
  {pages} {50} (\bibinfo {year} {2016})}\BibitemShut {NoStop}%
\bibitem [{\citenamefont {Oertel}\ \emph {et~al.}(2017)\citenamefont {Oertel},
  \citenamefont {Hempel}, \citenamefont {Kl\"ahn},\ and\ \citenamefont
  {Typel}}]{Oertel:2016bki}%
  \BibitemOpen
  \bibfield  {author} {\bibinfo {author} {\bibfnamefont {M.}~\bibnamefont
  {Oertel}}, \bibinfo {author} {\bibfnamefont {M.}~\bibnamefont {Hempel}},
  \bibinfo {author} {\bibfnamefont {T.}~\bibnamefont {Kl\"ahn}}, \ and\
  \bibinfo {author} {\bibfnamefont {S.}~\bibnamefont {Typel}},\ }\href
  {\doibase 10.1103/RevModPhys.89.015007} {\bibfield  {journal} {\bibinfo
  {journal} {Rev. Mod. Phys.}\ }\textbf {\bibinfo {volume} {89}},\ \bibinfo
  {pages} {015007} (\bibinfo {year} {2017})}\BibitemShut {NoStop}%
\bibitem [{\citenamefont {Tolos}\ \emph {et~al.}(2017)\citenamefont {Tolos},
  \citenamefont {Centelles},\ and\ \citenamefont {Ramos}}]{Tolos:2016hhl}%
  \BibitemOpen
  \bibfield  {author} {\bibinfo {author} {\bibfnamefont {L.}~\bibnamefont
  {Tolos}}, \bibinfo {author} {\bibfnamefont {M.}~\bibnamefont {Centelles}}, \
  and\ \bibinfo {author} {\bibfnamefont {A.}~\bibnamefont {Ramos}},\ }\href
  {\doibase 10.3847/1538-4357/834/1/3} {\bibfield  {journal} {\bibinfo
  {journal} {Astrophys. J.}\ }\textbf {\bibinfo {volume} {834}},\ \bibinfo
  {pages} {3} (\bibinfo {year} {2017})}\BibitemShut {NoStop}%
\bibitem [{\citenamefont {Fortin}\ \emph {et~al.}(2017)\citenamefont {Fortin},
  \citenamefont {Avancini}, \citenamefont {Provid\^encia},\ and\ \citenamefont
  {Vida\~na}}]{Fortin:2017cvt}%
  \BibitemOpen
  \bibfield  {author} {\bibinfo {author} {\bibfnamefont {M.}~\bibnamefont
  {Fortin}}, \bibinfo {author} {\bibfnamefont {S.~S.}\ \bibnamefont
  {Avancini}}, \bibinfo {author} {\bibfnamefont {C.}~\bibnamefont
  {Provid\^encia}}, \ and\ \bibinfo {author} {\bibfnamefont {I.}~\bibnamefont
  {Vida\~na}},\ }\href {\doibase 10.1103/PhysRevC.95.065803} {\bibfield
  {journal} {\bibinfo  {journal} {Phys. Rev. C}\ }\textbf {\bibinfo {volume}
  {95}},\ \bibinfo {pages} {065803} (\bibinfo {year} {2017})}\BibitemShut
  {NoStop}%
\bibitem [{\citenamefont {Negreiros}\ \emph {et~al.}(2018)\citenamefont
  {Negreiros}, \citenamefont {Tolos}, \citenamefont {Centelles}, \citenamefont
  {Ramos},\ and\ \citenamefont {Dexheimer}}]{Negreiros:2018cho}%
  \BibitemOpen
  \bibfield  {author} {\bibinfo {author} {\bibfnamefont {R.}~\bibnamefont
  {Negreiros}}, \bibinfo {author} {\bibfnamefont {L.}~\bibnamefont {Tolos}},
  \bibinfo {author} {\bibfnamefont {M.}~\bibnamefont {Centelles}}, \bibinfo
  {author} {\bibfnamefont {A.}~\bibnamefont {Ramos}}, \ and\ \bibinfo {author}
  {\bibfnamefont {V.}~\bibnamefont {Dexheimer}},\ }\href {\doibase
  10.3847/1538-4357/aad049} {\bibfield  {journal} {\bibinfo  {journal}
  {Astrophys. J.}\ }\textbf {\bibinfo {volume} {863}},\ \bibinfo {pages} {104}
  (\bibinfo {year} {2018})}\BibitemShut {NoStop}%
\bibitem [{\citenamefont {Logoteta}(2021)}]{Logoteta:2021iuy}%
  \BibitemOpen
  \bibfield  {author} {\bibinfo {author} {\bibfnamefont {D.}~\bibnamefont
  {Logoteta}},\ }\href {\doibase 10.3390/universe7110408} {\bibfield  {journal}
  {\bibinfo  {journal} {Universe}\ }\textbf {\bibinfo {volume} {7}},\ \bibinfo
  {pages} {408} (\bibinfo {year} {2021})}\BibitemShut {NoStop}%
\bibitem [{\citenamefont {Sawyer}(1972)}]{Sawyer:1972}%
  \BibitemOpen
  \bibfield  {author} {\bibinfo {author} {\bibfnamefont {R.~F.}\ \bibnamefont
  {Sawyer}},\ }\href {\doibase 10.1103/PhysRevLett.29.382} {\bibfield
  {journal} {\bibinfo  {journal} {Phys. Rev. Lett.}\ }\textbf {\bibinfo
  {volume} {29}},\ \bibinfo {pages} {382} (\bibinfo {year} {1972})}\BibitemShut
  {NoStop}%
\bibitem [{\citenamefont {Migdal}(1973)}]{Migdal:1973jkf}%
  \BibitemOpen
  \bibfield  {author} {\bibinfo {author} {\bibfnamefont {A.~B.}\ \bibnamefont
  {Migdal}},\ }\href {\doibase 10.1016/0370-2693(73)90640-0} {\bibfield
  {journal} {\bibinfo  {journal} {Phys. Lett. B}\ }\textbf {\bibinfo {volume}
  {45}},\ \bibinfo {pages} {448} (\bibinfo {year} {1973})}\BibitemShut
  {NoStop}%
\bibitem [{\citenamefont {Baym}\ and\ \citenamefont
  {Flowers}(1974)}]{Baym:1974vzp}%
  \BibitemOpen
  \bibfield  {author} {\bibinfo {author} {\bibfnamefont {G.}~\bibnamefont
  {Baym}}\ and\ \bibinfo {author} {\bibfnamefont {E.}~\bibnamefont {Flowers}},\
  }\href {\doibase 10.1016/0375-9474(74)90583-1} {\bibfield  {journal}
  {\bibinfo  {journal} {Nucl. Phys. A}\ }\textbf {\bibinfo {volume} {222}},\
  \bibinfo {pages} {29} (\bibinfo {year} {1974})}\BibitemShut {NoStop}%
\bibitem [{\citenamefont {Weise}\ and\ \citenamefont
  {Brown}(1975)}]{Weise:1975tk}%
  \BibitemOpen
  \bibfield  {author} {\bibinfo {author} {\bibfnamefont {W.}~\bibnamefont
  {Weise}}\ and\ \bibinfo {author} {\bibfnamefont {G.~E.}\ \bibnamefont
  {Brown}},\ }\href {\doibase 10.1016/0370-2693(75)90658-9} {\bibfield
  {journal} {\bibinfo  {journal} {Phys. Lett. B}\ }\textbf {\bibinfo {volume}
  {58}},\ \bibinfo {pages} {300} (\bibinfo {year} {1975})}\BibitemShut
  {NoStop}%
\bibitem [{\citenamefont {Migdal}(1978)}]{Migdal:1978}%
  \BibitemOpen
  \bibfield  {author} {\bibinfo {author} {\bibfnamefont {A.~B.}\ \bibnamefont
  {Migdal}},\ }\href {\doibase 10.1103/RevModPhys.50.107} {\bibfield  {journal}
  {\bibinfo  {journal} {Rev. Mod. Phys.}\ }\textbf {\bibinfo {volume} {50}},\
  \bibinfo {pages} {107} (\bibinfo {year} {1978})}\BibitemShut {NoStop}%
\bibitem [{\citenamefont {Kaplan}\ and\ \citenamefont
  {Nelson}(1986)}]{Kaplan:1986yq}%
  \BibitemOpen
  \bibfield  {author} {\bibinfo {author} {\bibfnamefont {D.~B.}\ \bibnamefont
  {Kaplan}}\ and\ \bibinfo {author} {\bibfnamefont {A.~E.}\ \bibnamefont
  {Nelson}},\ }\href {\doibase 10.1016/0370-2693(86)90331-X} {\bibfield
  {journal} {\bibinfo  {journal} {Phys. Lett. B}\ }\textbf {\bibinfo {volume}
  {175}},\ \bibinfo {pages} {57} (\bibinfo {year} {1986})}\BibitemShut
  {NoStop}%
\bibitem [{\citenamefont {Brown}\ \emph {et~al.}(1992)\citenamefont {Brown},
  \citenamefont {Thorsson}, \citenamefont {Kubodera},\ and\ \citenamefont
  {Rho}}]{Brown:1992ib}%
  \BibitemOpen
  \bibfield  {author} {\bibinfo {author} {\bibfnamefont {G.~E.}\ \bibnamefont
  {Brown}}, \bibinfo {author} {\bibfnamefont {V.}~\bibnamefont {Thorsson}},
  \bibinfo {author} {\bibfnamefont {K.}~\bibnamefont {Kubodera}}, \ and\
  \bibinfo {author} {\bibfnamefont {M.}~\bibnamefont {Rho}},\ }\href {\doibase
  10.1016/0370-2693(92)91386-N} {\bibfield  {journal} {\bibinfo  {journal}
  {Phys. Lett. B}\ }\textbf {\bibinfo {volume} {291}},\ \bibinfo {pages} {355}
  (\bibinfo {year} {1992})}\BibitemShut {NoStop}%
\bibitem [{\citenamefont {Thorsson}\ \emph {et~al.}(1994)\citenamefont
  {Thorsson}, \citenamefont {Prakash},\ and\ \citenamefont
  {Lattimer}}]{Thorsson:1993bu}%
  \BibitemOpen
  \bibfield  {author} {\bibinfo {author} {\bibfnamefont {V.}~\bibnamefont
  {Thorsson}}, \bibinfo {author} {\bibfnamefont {M.}~\bibnamefont {Prakash}}, \
  and\ \bibinfo {author} {\bibfnamefont {J.~M.}\ \bibnamefont {Lattimer}},\
  }\href {\doibase 10.1016/0375-9474(94)90407-3} {\bibfield  {journal}
  {\bibinfo  {journal} {Nucl. Phys. A}\ }\textbf {\bibinfo {volume} {572}},\
  \bibinfo {pages} {693} (\bibinfo {year} {1994})},\ \bibinfo {note} {[Erratum:
  Nucl.Phys.A 574, 851 (1994)]}\BibitemShut {NoStop}%
\bibitem [{\citenamefont {Glendenning}\ and\ \citenamefont
  {Schaffner-Bielich}(1998)}]{Glendenning:1998zx}%
  \BibitemOpen
  \bibfield  {author} {\bibinfo {author} {\bibfnamefont {N.~K.}\ \bibnamefont
  {Glendenning}}\ and\ \bibinfo {author} {\bibfnamefont {J.}~\bibnamefont
  {Schaffner-Bielich}},\ }\href {\doibase 10.1103/PhysRevLett.81.4564}
  {\bibfield  {journal} {\bibinfo  {journal} {Phys. Rev. Lett.}\ }\textbf
  {\bibinfo {volume} {81}},\ \bibinfo {pages} {4564} (\bibinfo {year}
  {1998})}\BibitemShut {NoStop}%
\bibitem [{\citenamefont {Ramos}\ \emph {et~al.}(2001)\citenamefont {Ramos},
  \citenamefont {Schaffner-Bielich},\ and\ \citenamefont
  {Wambach}}]{Ramos:2000dq}%
  \BibitemOpen
  \bibfield  {author} {\bibinfo {author} {\bibfnamefont {A.}~\bibnamefont
  {Ramos}}, \bibinfo {author} {\bibfnamefont {J.}~\bibnamefont
  {Schaffner-Bielich}}, \ and\ \bibinfo {author} {\bibfnamefont
  {J.}~\bibnamefont {Wambach}},\ }\href@noop {} {\bibfield  {journal} {\bibinfo
   {journal} {Lect. Notes Phys.}\ }\textbf {\bibinfo {volume} {578}},\ \bibinfo
  {pages} {175} (\bibinfo {year} {2001})}\BibitemShut {NoStop}%
\bibitem [{\citenamefont {Toki}\ and\ \citenamefont
  {Weise}(1980)}]{Toki:1980zp}%
  \BibitemOpen
  \bibfield  {author} {\bibinfo {author} {\bibfnamefont {H.}~\bibnamefont
  {Toki}}\ and\ \bibinfo {author} {\bibfnamefont {W.}~\bibnamefont {Weise}},\
  }\href {\doibase 10.1016/0370-2693(80)90260-9} {\bibfield  {journal}
  {\bibinfo  {journal} {Phys. Lett. B}\ }\textbf {\bibinfo {volume} {92}},\
  \bibinfo {pages} {265} (\bibinfo {year} {1980})}\BibitemShut {NoStop}%
\bibitem [{\citenamefont {Alberico}\ \emph {et~al.}(1980)\citenamefont
  {Alberico}, \citenamefont {Ericson},\ and\ \citenamefont
  {Molinari}}]{Alberico:1980vb}%
  \BibitemOpen
  \bibfield  {author} {\bibinfo {author} {\bibfnamefont {W.~M.}\ \bibnamefont
  {Alberico}}, \bibinfo {author} {\bibfnamefont {M.}~\bibnamefont {Ericson}}, \
  and\ \bibinfo {author} {\bibfnamefont {A.}~\bibnamefont {Molinari}},\ }\href
  {\doibase 10.1016/0370-2693(80)90326-3} {\bibfield  {journal} {\bibinfo
  {journal} {Phys. Lett. B}\ }\textbf {\bibinfo {volume} {92}},\ \bibinfo
  {pages} {153} (\bibinfo {year} {1980})}\BibitemShut {NoStop}%
\bibitem [{\citenamefont {Alberico}\ \emph {et~al.}(1982)\citenamefont
  {Alberico}, \citenamefont {Ericson},\ and\ \citenamefont
  {Molinari}}]{Alberico:1981sz}%
  \BibitemOpen
  \bibfield  {author} {\bibinfo {author} {\bibfnamefont {W.~M.}\ \bibnamefont
  {Alberico}}, \bibinfo {author} {\bibfnamefont {M.}~\bibnamefont {Ericson}}, \
  and\ \bibinfo {author} {\bibfnamefont {A.}~\bibnamefont {Molinari}},\ }\href
  {\doibase 10.1016/0375-9474(82)90007-0} {\bibfield  {journal} {\bibinfo
  {journal} {Nucl. Phys. A}\ }\textbf {\bibinfo {volume} {379}},\ \bibinfo
  {pages} {429} (\bibinfo {year} {1982})}\BibitemShut {NoStop}%
\bibitem [{\citenamefont {McClelland}\ \emph {et~al.}(1992)\citenamefont
  {McClelland} \emph {et~al.}}]{McClelland:1992yb}%
  \BibitemOpen
  \bibfield  {author} {\bibinfo {author} {\bibfnamefont {J.~B.}\ \bibnamefont
  {McClelland}} \emph {et~al.},\ }\href {\doibase 10.1103/PhysRevLett.69.582}
  {\bibfield  {journal} {\bibinfo  {journal} {Phys. Rev. Lett.}\ }\textbf
  {\bibinfo {volume} {69}},\ \bibinfo {pages} {582} (\bibinfo {year}
  {1992})}\BibitemShut {NoStop}%
\bibitem [{\citenamefont {Chen}\ \emph {et~al.}(1993)\citenamefont {Chen} \emph
  {et~al.}}]{Chen:1993fs}%
  \BibitemOpen
  \bibfield  {author} {\bibinfo {author} {\bibfnamefont {X.~Y.}\ \bibnamefont
  {Chen}} \emph {et~al.},\ }\href {\doibase 10.1103/PhysRevC.47.2159}
  {\bibfield  {journal} {\bibinfo  {journal} {Phys. Rev. C}\ }\textbf {\bibinfo
  {volume} {47}},\ \bibinfo {pages} {2159} (\bibinfo {year}
  {1993})}\BibitemShut {NoStop}%
\bibitem [{\citenamefont {Wang}\ \emph {et~al.}(1994)\citenamefont {Wang} \emph
  {et~al.}}]{Wang:1994zzd}%
  \BibitemOpen
  \bibfield  {author} {\bibinfo {author} {\bibfnamefont {L.}~\bibnamefont
  {Wang}} \emph {et~al.},\ }\href {\doibase 10.1103/PhysRevC.50.2438}
  {\bibfield  {journal} {\bibinfo  {journal} {Phys. Rev. C}\ }\textbf {\bibinfo
  {volume} {50}},\ \bibinfo {pages} {2438} (\bibinfo {year}
  {1994})}\BibitemShut {NoStop}%
\bibitem [{\citenamefont {Pandharipande}\ \emph {et~al.}(1994)\citenamefont
  {Pandharipande}, \citenamefont {Carlson}, \citenamefont {Pieper},
  \citenamefont {Wiringa},\ and\ \citenamefont
  {Schiavilla}}]{Pandharipande:1994jc}%
  \BibitemOpen
  \bibfield  {author} {\bibinfo {author} {\bibfnamefont {V.~R.}\ \bibnamefont
  {Pandharipande}}, \bibinfo {author} {\bibfnamefont {J.}~\bibnamefont
  {Carlson}}, \bibinfo {author} {\bibfnamefont {S.~C.}\ \bibnamefont {Pieper}},
  \bibinfo {author} {\bibfnamefont {R.~B.}\ \bibnamefont {Wiringa}}, \ and\
  \bibinfo {author} {\bibfnamefont {R.}~\bibnamefont {Schiavilla}},\ }\href
  {\doibase 10.1103/PhysRevC.49.789} {\bibfield  {journal} {\bibinfo  {journal}
  {Phys. Rev. C}\ }\textbf {\bibinfo {volume} {49}},\ \bibinfo {pages} {789}
  (\bibinfo {year} {1994})}\BibitemShut {NoStop}%
\bibitem [{\citenamefont {Wilczek}(2000)}]{Wilczek:2000}%
  \BibitemOpen
  \bibfield  {author} {\bibinfo {author} {\bibfnamefont {F.}~\bibnamefont
  {Wilczek}},\ }\href@noop {} {\bibfield  {journal} {\bibinfo  {journal}
  {Physics Today}\ }\textbf {\bibinfo {volume} {53}},\ \bibinfo {pages} {22}
  (\bibinfo {year} {2000})}\BibitemShut {NoStop}%
\bibitem [{\citenamefont {Alford}\ \emph {et~al.}(1999)\citenamefont {Alford},
  \citenamefont {Rajagopal},\ and\ \citenamefont {Wilczek}}]{Alford:1998mk}%
  \BibitemOpen
  \bibfield  {author} {\bibinfo {author} {\bibfnamefont {M.~G.}\ \bibnamefont
  {Alford}}, \bibinfo {author} {\bibfnamefont {K.}~\bibnamefont {Rajagopal}}, \
  and\ \bibinfo {author} {\bibfnamefont {F.}~\bibnamefont {Wilczek}},\ }\href
  {\doibase 10.1016/S0550-3213(98)00668-3} {\bibfield  {journal} {\bibinfo
  {journal} {Nucl. Phys.}\ }\textbf {\bibinfo {volume} {B537}} (\bibinfo {year}
  {1999}),\ 10.1016/S0550-3213(98)00668-3}\BibitemShut {NoStop}%
\bibitem [{\citenamefont {Alford}\ \emph {et~al.}(1998)\citenamefont {Alford},
  \citenamefont {Rajagopal},\ and\ \citenamefont {Wilczek}}]{Alford:1997zt}%
  \BibitemOpen
  \bibfield  {author} {\bibinfo {author} {\bibfnamefont {M.~G.}\ \bibnamefont
  {Alford}}, \bibinfo {author} {\bibfnamefont {K.}~\bibnamefont {Rajagopal}}, \
  and\ \bibinfo {author} {\bibfnamefont {F.}~\bibnamefont {Wilczek}},\ }\href
  {\doibase 10.1016/S0370-2693(98)00051-3} {\bibfield  {journal} {\bibinfo
  {journal} {Phys. Lett. B}\ }\textbf {\bibinfo {volume} {422}},\ \bibinfo
  {pages} {247} (\bibinfo {year} {1998})}\BibitemShut {NoStop}%
\bibitem [{\citenamefont {Rajagopal}\ and\ \citenamefont
  {Wilczek}(2001)}]{Rajagopal:2000ff}%
  \BibitemOpen
  \bibfield  {author} {\bibinfo {author} {\bibfnamefont {K.}~\bibnamefont
  {Rajagopal}}\ and\ \bibinfo {author} {\bibfnamefont {F.}~\bibnamefont
  {Wilczek}},\ }\href {\doibase 10.1103/PhysRevLett.86.3492} {\bibfield
  {journal} {\bibinfo  {journal} {Phys. Rev. Lett.}\ }\textbf {\bibinfo
  {volume} {86}},\ \bibinfo {pages} {3492} (\bibinfo {year}
  {2001})}\BibitemShut {NoStop}%
\bibitem [{\citenamefont {Alford}\ \emph {et~al.}(2008)\citenamefont {Alford},
  \citenamefont {Schmitt}, \citenamefont {Rajagopal},\ and\ \citenamefont
  {Schafer}}]{Alford:2007xm}%
  \BibitemOpen
  \bibfield  {author} {\bibinfo {author} {\bibfnamefont {M.~G.}\ \bibnamefont
  {Alford}}, \bibinfo {author} {\bibfnamefont {A.}~\bibnamefont {Schmitt}},
  \bibinfo {author} {\bibfnamefont {K.}~\bibnamefont {Rajagopal}}, \ and\
  \bibinfo {author} {\bibfnamefont {T.}~\bibnamefont {Schafer}},\ }\href
  {\doibase 10.1103/RevModPhys.80.1455} {\bibfield  {journal} {\bibinfo
  {journal} {Rev. Mod. Phys.}\ }\textbf {\bibinfo {volume} {80}},\ \bibinfo
  {pages} {1455} (\bibinfo {year} {2008})}\BibitemShut {NoStop}%
\bibitem [{\citenamefont {Annala}\ \emph {et~al.}(2020)\citenamefont {Annala},
  \citenamefont {Gorda}, \citenamefont {Kurkela}, \citenamefont {N\"attil\"a},\
  and\ \citenamefont {Vuorinen}}]{Annala:2019puf}%
  \BibitemOpen
  \bibfield  {author} {\bibinfo {author} {\bibfnamefont {E.}~\bibnamefont
  {Annala}}, \bibinfo {author} {\bibfnamefont {T.}~\bibnamefont {Gorda}},
  \bibinfo {author} {\bibfnamefont {A.}~\bibnamefont {Kurkela}}, \bibinfo
  {author} {\bibfnamefont {J.}~\bibnamefont {N\"attil\"a}}, \ and\ \bibinfo
  {author} {\bibfnamefont {A.}~\bibnamefont {Vuorinen}},\ }\href {\doibase
  10.1038/s41567-020-0914-9} {\bibfield  {journal} {\bibinfo  {journal} {Nature
  Phys.}\ }\textbf {\bibinfo {volume} {16}},\ \bibinfo {pages} {907} (\bibinfo
  {year} {2020})}\BibitemShut {NoStop}%
\bibitem [{\citenamefont {Phillips}(1998)}]{Phillips1998}%
  \BibitemOpen
  \bibfield  {author} {\bibinfo {author} {\bibfnamefont {A.~C.}\ \bibnamefont
  {Phillips}},\ }\enquote {\bibinfo {title} {The physics of stars},}\ \
  (\bibinfo  {publisher} {John Wiley \& Sons, Chichester},\ \bibinfo {year}
  {1998})\ \bibinfo {edition} {2nd}\ ed.\BibitemShut {Stop}%
\bibitem [{\citenamefont {Weber}(1999)}]{Weber:1999}%
  \BibitemOpen
  \bibfield  {author} {\bibinfo {author} {\bibfnamefont {F.}~\bibnamefont
  {Weber}},\ }\enquote {\bibinfo {title} {Pulsars as astrophysical laboratories
  for nuclear and particle physics},}\ \ (\bibinfo  {publisher} {Institute of
  Physics Publishing},\ \bibinfo {address} {Bristol, UK},\ \bibinfo {year}
  {1999})\BibitemShut {NoStop}%
\bibitem [{\citenamefont {Glendenning}(2000)}]{Glendenning:2000}%
  \BibitemOpen
  \bibfield  {author} {\bibinfo {author} {\bibfnamefont {N.~K.}\ \bibnamefont
  {Glendenning}},\ }\enquote {\bibinfo {title} {Compact stars},}\ \ (\bibinfo
  {publisher} {Springer-Verlag New York},\ \bibinfo {year} {2000})\BibitemShut
  {NoStop}%
\bibitem [{\citenamefont {Piekarewicz}(2016)}]{Piekarewicz2016}%
  \BibitemOpen
  \bibfield  {author} {\bibinfo {author} {\bibfnamefont {J.}~\bibnamefont
  {Piekarewicz}},\ }\enquote {\bibinfo {title} {Neutron star matter equation of
  state},}\ in\ \href {\doibase 10.1007/978-3-319-20794-0_54-1} {\emph
  {\bibinfo {booktitle} {Handbook of Supernovae}}},\ \bibinfo {editor} {edited
  by\ \bibinfo {editor} {\bibfnamefont {A.~W.}\ \bibnamefont {Alsabti}}\ and\
  \bibinfo {editor} {\bibfnamefont {P.}~\bibnamefont {Murdin}}}\ (\bibinfo
  {publisher} {Springer International Publishing},\ \bibinfo {address} {Cham},\
  \bibinfo {year} {2016})\ pp.\ \bibinfo {pages} {1--20}\BibitemShut {NoStop}%
\bibitem [{\citenamefont {Press}\ \emph {et~al.}(1989)\citenamefont {Press},
  \citenamefont {Flannery}, \citenamefont {Teukolsky},\ and\ \citenamefont
  {Vetterling}}]{NumericalRecipes}%
  \BibitemOpen
  \bibfield  {author} {\bibinfo {author} {\bibfnamefont {W.~H.}\ \bibnamefont
  {Press}}, \bibinfo {author} {\bibfnamefont {B.~P.}\ \bibnamefont {Flannery}},
  \bibinfo {author} {\bibfnamefont {S.~A.}\ \bibnamefont {Teukolsky}}, \ and\
  \bibinfo {author} {\bibfnamefont {W.~T.}\ \bibnamefont {Vetterling}},\
  }\enquote {\bibinfo {title} {Numerical recipes: The art of scientific
  computing},}\ \ (\bibinfo  {publisher} {Cambridge University Press},\
  \bibinfo {year} {1989})\BibitemShut {NoStop}%
\bibitem [{\citenamefont {Weinberg}(1990)}]{Weinberg:1990rz}%
  \BibitemOpen
  \bibfield  {author} {\bibinfo {author} {\bibfnamefont {S.}~\bibnamefont
  {Weinberg}},\ }\href {\doibase 10.1016/0370-2693(90)90938-3} {\bibfield
  {journal} {\bibinfo  {journal} {Phys. Lett. B}\ }\textbf {\bibinfo {volume}
  {251}},\ \bibinfo {pages} {288} (\bibinfo {year} {1990})}\BibitemShut
  {NoStop}%
\bibitem [{\citenamefont {Gezerlis}\ and\ \citenamefont
  {Carlson}(2010)}]{Gezerlis:2009iw}%
  \BibitemOpen
  \bibfield  {author} {\bibinfo {author} {\bibfnamefont {A.}~\bibnamefont
  {Gezerlis}}\ and\ \bibinfo {author} {\bibfnamefont {J.}~\bibnamefont
  {Carlson}},\ }\href {\doibase 10.1103/PhysRevC.81.025803} {\bibfield
  {journal} {\bibinfo  {journal} {Phys. Rev.}\ }\textbf {\bibinfo {volume}
  {C81}},\ \bibinfo {pages} {025803} (\bibinfo {year} {2010})}\BibitemShut
  {NoStop}%
\bibitem [{\citenamefont {van Kolck}(1994)}]{vanKolck:1994yi}%
  \BibitemOpen
  \bibfield  {author} {\bibinfo {author} {\bibfnamefont {U.}~\bibnamefont {van
  Kolck}},\ }\href {\doibase 10.1103/PhysRevC.49.2932} {\bibfield  {journal}
  {\bibinfo  {journal} {Phys. Rev. C}\ }\textbf {\bibinfo {volume} {49}},\
  \bibinfo {pages} {2932} (\bibinfo {year} {1994})}\BibitemShut {NoStop}%
\bibitem [{\citenamefont {Ordonez}\ \emph {et~al.}(1996)\citenamefont
  {Ordonez}, \citenamefont {Ray},\ and\ \citenamefont {van
  Kolck}}]{Ordonez:1995rz}%
  \BibitemOpen
  \bibfield  {author} {\bibinfo {author} {\bibfnamefont {C.}~\bibnamefont
  {Ordonez}}, \bibinfo {author} {\bibfnamefont {L.}~\bibnamefont {Ray}}, \ and\
  \bibinfo {author} {\bibfnamefont {U.}~\bibnamefont {van Kolck}},\ }\href
  {\doibase 10.1103/PhysRevC.53.2086} {\bibfield  {journal} {\bibinfo
  {journal} {Phys. Rev. C}\ }\textbf {\bibinfo {volume} {53}},\ \bibinfo
  {pages} {2086} (\bibinfo {year} {1996})}\BibitemShut {NoStop}%
\bibitem [{\citenamefont {Rodriguez~Entem}\ \emph {et~al.}(2020)\citenamefont
  {Rodriguez~Entem}, \citenamefont {Machleidt},\ and\ \citenamefont
  {Nosyk}}]{RodriguezEntem:2020jgp}%
  \BibitemOpen
  \bibfield  {author} {\bibinfo {author} {\bibfnamefont {D.}~\bibnamefont
  {Rodriguez~Entem}}, \bibinfo {author} {\bibfnamefont {R.}~\bibnamefont
  {Machleidt}}, \ and\ \bibinfo {author} {\bibfnamefont {Y.}~\bibnamefont
  {Nosyk}},\ }\href {\doibase 10.3389/fphy.2020.00057} {\bibfield  {journal}
  {\bibinfo  {journal} {Front. in Phys.}\ }\textbf {\bibinfo {volume} {8}},\
  \bibinfo {pages} {57} (\bibinfo {year} {2020})}\BibitemShut {NoStop}%
\bibitem [{\citenamefont {Hebeler}\ and\ \citenamefont
  {Schwenk}(2010)}]{Hebeler:2009iv}%
  \BibitemOpen
  \bibfield  {author} {\bibinfo {author} {\bibfnamefont {K.}~\bibnamefont
  {Hebeler}}\ and\ \bibinfo {author} {\bibfnamefont {A.}~\bibnamefont
  {Schwenk}},\ }\href {\doibase 10.1103/PhysRevC.82.014314} {\bibfield
  {journal} {\bibinfo  {journal} {Phys. Rev.}\ }\textbf {\bibinfo {volume}
  {C82}},\ \bibinfo {pages} {014314} (\bibinfo {year} {2010})}\BibitemShut
  {NoStop}%
\bibitem [{\citenamefont {Tews}\ \emph {et~al.}(2013)\citenamefont {Tews},
  \citenamefont {Kruger}, \citenamefont {Hebeler},\ and\ \citenamefont
  {Schwenk}}]{Tews:2012fj}%
  \BibitemOpen
  \bibfield  {author} {\bibinfo {author} {\bibfnamefont {I.}~\bibnamefont
  {Tews}}, \bibinfo {author} {\bibfnamefont {T.}~\bibnamefont {Kruger}},
  \bibinfo {author} {\bibfnamefont {K.}~\bibnamefont {Hebeler}}, \ and\
  \bibinfo {author} {\bibfnamefont {A.}~\bibnamefont {Schwenk}},\ }\href
  {\doibase 10.1103/PhysRevLett.110.032504} {\bibfield  {journal} {\bibinfo
  {journal} {Phys. Rev. Lett.}\ }\textbf {\bibinfo {volume} {110}},\ \bibinfo
  {pages} {032504} (\bibinfo {year} {2013})}\BibitemShut {NoStop}%
\bibitem [{\citenamefont {Kruger}\ \emph {et~al.}(2013)\citenamefont {Kruger},
  \citenamefont {Tews}, \citenamefont {Hebeler},\ and\ \citenamefont
  {Schwenk}}]{Kruger:2013kua}%
  \BibitemOpen
  \bibfield  {author} {\bibinfo {author} {\bibfnamefont {T.}~\bibnamefont
  {Kruger}}, \bibinfo {author} {\bibfnamefont {I.}~\bibnamefont {Tews}},
  \bibinfo {author} {\bibfnamefont {K.}~\bibnamefont {Hebeler}}, \ and\
  \bibinfo {author} {\bibfnamefont {A.}~\bibnamefont {Schwenk}},\ }\href
  {\doibase 10.1103/PhysRevC.88.025802} {\bibfield  {journal} {\bibinfo
  {journal} {Phys. Rev.}\ }\textbf {\bibinfo {volume} {C88}},\ \bibinfo {pages}
  {025802} (\bibinfo {year} {2013})}\BibitemShut {NoStop}%
\bibitem [{\citenamefont {Lonardoni}\ \emph {et~al.}(2020)\citenamefont
  {Lonardoni}, \citenamefont {Tews}, \citenamefont {Gandolfi},\ and\
  \citenamefont {Carlson}}]{Lonardoni:2019ypg}%
  \BibitemOpen
  \bibfield  {author} {\bibinfo {author} {\bibfnamefont {D.}~\bibnamefont
  {Lonardoni}}, \bibinfo {author} {\bibfnamefont {I.}~\bibnamefont {Tews}},
  \bibinfo {author} {\bibfnamefont {S.}~\bibnamefont {Gandolfi}}, \ and\
  \bibinfo {author} {\bibfnamefont {J.}~\bibnamefont {Carlson}},\ }\href
  {\doibase 10.1103/PhysRevResearch.2.022033} {\bibfield  {journal} {\bibinfo
  {journal} {Phys. Rev. Res.}\ }\textbf {\bibinfo {volume} {2}},\ \bibinfo
  {pages} {022033} (\bibinfo {year} {2020})}\BibitemShut {NoStop}%
\bibitem [{\citenamefont {Drischler}\ \emph {et~al.}(2021)\citenamefont
  {Drischler}, \citenamefont {Holt},\ and\ \citenamefont
  {Wellenhofer}}]{Drischler:2021kxf}%
  \BibitemOpen
  \bibfield  {author} {\bibinfo {author} {\bibfnamefont {C.}~\bibnamefont
  {Drischler}}, \bibinfo {author} {\bibfnamefont {J.~W.}\ \bibnamefont {Holt}},
  \ and\ \bibinfo {author} {\bibfnamefont {C.}~\bibnamefont {Wellenhofer}},\
  }\href {\doibase 10.1146/annurev-nucl-102419-041903} {\bibfield  {journal}
  {\bibinfo  {journal} {Ann. Rev. Nucl. Part. Sci.}\ }\textbf {\bibinfo
  {volume} {71}},\ \bibinfo {pages} {403} (\bibinfo {year} {2021})}\BibitemShut
  {NoStop}%
\bibitem [{\citenamefont {Sammarruca}\ and\ \citenamefont
  {Millerson}(2021)}]{Sammarruca:2021mhv}%
  \BibitemOpen
  \bibfield  {author} {\bibinfo {author} {\bibfnamefont {F.}~\bibnamefont
  {Sammarruca}}\ and\ \bibinfo {author} {\bibfnamefont {R.}~\bibnamefont
  {Millerson}},\ }\href {\doibase 10.1103/PhysRevC.104.034308} {\bibfield
  {journal} {\bibinfo  {journal} {Phys. Rev. C}\ }\textbf {\bibinfo {volume}
  {104}},\ \bibinfo {pages} {034308} (\bibinfo {year} {2021})}\BibitemShut
  {NoStop}%
\bibitem [{\citenamefont {Sammarruca}\ and\ \citenamefont
  {Millerson}(2022)}]{Sammarruca:2022ser}%
  \BibitemOpen
  \bibfield  {author} {\bibinfo {author} {\bibfnamefont {F.}~\bibnamefont
  {Sammarruca}}\ and\ \bibinfo {author} {\bibfnamefont {R.}~\bibnamefont
  {Millerson}},\ }\href {\doibase 10.3390/universe8020133} {\bibfield
  {journal} {\bibinfo  {journal} {Universe}\ }\textbf {\bibinfo {volume} {8}},\
  \bibinfo {pages} {133} (\bibinfo {year} {2022})}\BibitemShut {NoStop}%
\bibitem [{\citenamefont {Alford}\ \emph {et~al.}(2022)\citenamefont {Alford},
  \citenamefont {Brodie}, \citenamefont {Haber},\ and\ \citenamefont
  {Tews}}]{Alford:2022bpp}%
  \BibitemOpen
  \bibfield  {author} {\bibinfo {author} {\bibfnamefont {M.~G.}\ \bibnamefont
  {Alford}}, \bibinfo {author} {\bibfnamefont {L.}~\bibnamefont {Brodie}},
  \bibinfo {author} {\bibfnamefont {A.}~\bibnamefont {Haber}}, \ and\ \bibinfo
  {author} {\bibfnamefont {I.}~\bibnamefont {Tews}},\ }\href@noop {} {\
  (\bibinfo {year} {2022})},\ \Eprint {http://arxiv.org/abs/2205.10283}
  {arXiv:2205.10283 [nucl-th]} \BibitemShut {NoStop}%
\bibitem [{\citenamefont {Brown}(2000)}]{Brown:2000}%
  \BibitemOpen
  \bibfield  {author} {\bibinfo {author} {\bibfnamefont {B.~A.}\ \bibnamefont
  {Brown}},\ }\href@noop {} {\bibfield  {journal} {\bibinfo  {journal} {Phys.
  Rev. Lett.}\ }\textbf {\bibinfo {volume} {85}},\ \bibinfo {pages} {5296}
  (\bibinfo {year} {2000})}\BibitemShut {NoStop}%
\bibitem [{\citenamefont {Furnstahl}(2002)}]{Furnstahl:2001un}%
  \BibitemOpen
  \bibfield  {author} {\bibinfo {author} {\bibfnamefont {R.~J.}\ \bibnamefont
  {Furnstahl}},\ }\href@noop {} {\bibfield  {journal} {\bibinfo  {journal}
  {Nucl. Phys.}\ }\textbf {\bibinfo {volume} {A706}},\ \bibinfo {pages} {85}
  (\bibinfo {year} {2002})}\BibitemShut {NoStop}%
\bibitem [{\citenamefont {Roca-Maza}\ \emph {et~al.}(2011)\citenamefont
  {Roca-Maza}, \citenamefont {Centelles}, \citenamefont {Vi\~nas},\ and\
  \citenamefont {Warda}}]{RocaMaza:2011pm}%
  \BibitemOpen
  \bibfield  {author} {\bibinfo {author} {\bibfnamefont {X.}~\bibnamefont
  {Roca-Maza}}, \bibinfo {author} {\bibfnamefont {M.}~\bibnamefont
  {Centelles}}, \bibinfo {author} {\bibfnamefont {X.}~\bibnamefont {Vi\~nas}},
  \ and\ \bibinfo {author} {\bibfnamefont {M.}~\bibnamefont {Warda}},\ }\href
  {\doibase 10.1103/PhysRevLett.106.252501} {\bibfield  {journal} {\bibinfo
  {journal} {Phys. Rev. Lett.}\ }\textbf {\bibinfo {volume} {106}},\ \bibinfo
  {pages} {252501} (\bibinfo {year} {2011})}\BibitemShut {NoStop}%
\bibitem [{\citenamefont {Piekarewicz}\ and\ \citenamefont
  {Fattoyev}(2019{\natexlab{a}})}]{Piekarewicz:2019ahf}%
  \BibitemOpen
  \bibfield  {author} {\bibinfo {author} {\bibfnamefont {J.}~\bibnamefont
  {Piekarewicz}}\ and\ \bibinfo {author} {\bibfnamefont {F.~J.}\ \bibnamefont
  {Fattoyev}},\ }\href {\doibase 10.1063/PT.3.4247} {\bibfield  {journal}
  {\bibinfo  {journal} {Physics Today}\ }\textbf {\bibinfo {volume} {72}},\
  \bibinfo {pages} {30} (\bibinfo {year} {2019}{\natexlab{a}})}\BibitemShut
  {NoStop}%
\bibitem [{\citenamefont {Horowitz}\ and\ \citenamefont
  {Piekarewicz}(2001{\natexlab{b}})}]{Horowitz:2001ya}%
  \BibitemOpen
  \bibfield  {author} {\bibinfo {author} {\bibfnamefont {C.~J.}\ \bibnamefont
  {Horowitz}}\ and\ \bibinfo {author} {\bibfnamefont {J.}~\bibnamefont
  {Piekarewicz}},\ }\href@noop {} {\bibfield  {journal} {\bibinfo  {journal}
  {Phys. Rev.}\ }\textbf {\bibinfo {volume} {C64}},\ \bibinfo {pages} {062802}
  (\bibinfo {year} {2001}{\natexlab{b}})}\BibitemShut {NoStop}%
\bibitem [{\citenamefont {Abrahamyan}\ \emph {et~al.}(2012)\citenamefont
  {Abrahamyan}, \citenamefont {Ahmed}, \citenamefont {Albataineh},
  \citenamefont {Aniol}, \citenamefont {Armstrong} \emph
  {et~al.}}]{Abrahamyan:2012gp}%
  \BibitemOpen
  \bibfield  {author} {\bibinfo {author} {\bibfnamefont {S.}~\bibnamefont
  {Abrahamyan}}, \bibinfo {author} {\bibfnamefont {Z.}~\bibnamefont {Ahmed}},
  \bibinfo {author} {\bibfnamefont {H.}~\bibnamefont {Albataineh}}, \bibinfo
  {author} {\bibfnamefont {K.}~\bibnamefont {Aniol}}, \bibinfo {author}
  {\bibfnamefont {D.~S.}\ \bibnamefont {Armstrong}},  \emph {et~al.},\
  }\href@noop {} {\bibfield  {journal} {\bibinfo  {journal} {Phys. Rev. Lett.}\
  }\textbf {\bibinfo {volume} {108}},\ \bibinfo {pages} {112502} (\bibinfo
  {year} {2012})}\BibitemShut {NoStop}%
\bibitem [{\citenamefont {Horowitz}\ \emph {et~al.}(2012)\citenamefont
  {Horowitz}, \citenamefont {Ahmed}, \citenamefont {Jen}, \citenamefont
  {Rakhman}, \citenamefont {Souder} \emph {et~al.}}]{Horowitz:2012tj}%
  \BibitemOpen
  \bibfield  {author} {\bibinfo {author} {\bibfnamefont {C.~J.}\ \bibnamefont
  {Horowitz}}, \bibinfo {author} {\bibfnamefont {Z.}~\bibnamefont {Ahmed}},
  \bibinfo {author} {\bibfnamefont {C.~M.}\ \bibnamefont {Jen}}, \bibinfo
  {author} {\bibfnamefont {A.}~\bibnamefont {Rakhman}}, \bibinfo {author}
  {\bibfnamefont {P.~A.}\ \bibnamefont {Souder}},  \emph {et~al.},\ }\href
  {\doibase 10.1103/PhysRevC.85.032501} {\bibfield  {journal} {\bibinfo
  {journal} {Phys. Rev.}\ }\textbf {\bibinfo {volume} {C85}},\ \bibinfo {pages}
  {032501} (\bibinfo {year} {2012})}\BibitemShut {NoStop}%
\bibitem [{\citenamefont {Lattimer}\ and\ \citenamefont
  {Prakash}(2007)}]{Lattimer:2006xb}%
  \BibitemOpen
  \bibfield  {author} {\bibinfo {author} {\bibfnamefont {J.~M.}\ \bibnamefont
  {Lattimer}}\ and\ \bibinfo {author} {\bibfnamefont {M.}~\bibnamefont
  {Prakash}},\ }\href@noop {} {\bibfield  {journal} {\bibinfo  {journal} {Phys.
  Rept.}\ }\textbf {\bibinfo {volume} {442}},\ \bibinfo {pages} {109} (\bibinfo
  {year} {2007})}\BibitemShut {NoStop}%
\bibitem [{\citenamefont {Carriere}\ \emph {et~al.}(2003)\citenamefont
  {Carriere}, \citenamefont {Horowitz},\ and\ \citenamefont
  {Piekarewicz}}]{Carriere:2002bx}%
  \BibitemOpen
  \bibfield  {author} {\bibinfo {author} {\bibfnamefont {J.}~\bibnamefont
  {Carriere}}, \bibinfo {author} {\bibfnamefont {C.~J.}\ \bibnamefont
  {Horowitz}}, \ and\ \bibinfo {author} {\bibfnamefont {J.}~\bibnamefont
  {Piekarewicz}},\ }\href@noop {} {\bibfield  {journal} {\bibinfo  {journal}
  {Astrophys. J.}\ }\textbf {\bibinfo {volume} {593}},\ \bibinfo {pages} {463}
  (\bibinfo {year} {2003})}\BibitemShut {NoStop}%
\bibitem [{\citenamefont {Yang}\ and\ \citenamefont
  {Piekarewicz}(2020)}]{Yang:2019fvs}%
  \BibitemOpen
  \bibfield  {author} {\bibinfo {author} {\bibfnamefont {J.}~\bibnamefont
  {Yang}}\ and\ \bibinfo {author} {\bibfnamefont {J.}~\bibnamefont
  {Piekarewicz}},\ }\href {\doibase 10.1146/annurev-nucl-101918-023608}
  {\bibfield  {journal} {\bibinfo  {journal} {Ann. Rev. Nucl. Part. Sci.}\
  }\textbf {\bibinfo {volume} {70}},\ \bibinfo {pages} {21} (\bibinfo {year}
  {2020})}\BibitemShut {NoStop}%
\bibitem [{\citenamefont {Tsang}\ \emph {et~al.}(2012)\citenamefont {Tsang},
  \citenamefont {Stone}, \citenamefont {Camera}, \citenamefont {Danielewicz},
  \citenamefont {Gandolfi} \emph {et~al.}}]{Tsang:2012se}%
  \BibitemOpen
  \bibfield  {author} {\bibinfo {author} {\bibfnamefont {M.}~\bibnamefont
  {Tsang}}, \bibinfo {author} {\bibfnamefont {J.}~\bibnamefont {Stone}},
  \bibinfo {author} {\bibfnamefont {F.}~\bibnamefont {Camera}}, \bibinfo
  {author} {\bibfnamefont {P.}~\bibnamefont {Danielewicz}}, \bibinfo {author}
  {\bibfnamefont {S.}~\bibnamefont {Gandolfi}},  \emph {et~al.},\ }\href
  {\doibase 10.1103/PhysRevC.86.015803} {\bibfield  {journal} {\bibinfo
  {journal} {Phys.Rev.}\ }\textbf {\bibinfo {volume} {C86}},\ \bibinfo {pages}
  {015803} (\bibinfo {year} {2012})}\BibitemShut {NoStop}%
\bibitem [{\citenamefont {Horowitz}\ \emph {et~al.}(2014)\citenamefont
  {Horowitz}, \citenamefont {Brown}, \citenamefont {Kim}, \citenamefont
  {Lynch}, \citenamefont {Michaels} \emph {et~al.}}]{Horowitz:2014bja}%
  \BibitemOpen
  \bibfield  {author} {\bibinfo {author} {\bibfnamefont {C.~J.}\ \bibnamefont
  {Horowitz}}, \bibinfo {author} {\bibfnamefont {E.~F.}\ \bibnamefont {Brown}},
  \bibinfo {author} {\bibfnamefont {Y.}~\bibnamefont {Kim}}, \bibinfo {author}
  {\bibfnamefont {W.~G.}\ \bibnamefont {Lynch}}, \bibinfo {author}
  {\bibfnamefont {R.}~\bibnamefont {Michaels}},  \emph {et~al.},\ }\href@noop
  {} {\bibfield  {journal} {\bibinfo  {journal} {J. Phys.}\ }\textbf {\bibinfo
  {volume} {G41}},\ \bibinfo {pages} {093001} (\bibinfo {year}
  {2014})}\BibitemShut {NoStop}%
\bibitem [{\citenamefont {Thiel}\ \emph {et~al.}(2019)\citenamefont {Thiel},
  \citenamefont {Sfienti}, \citenamefont {Piekarewicz}, \citenamefont
  {Horowitz},\ and\ \citenamefont {Vanderhaeghen}}]{Thiel:2019tkm}%
  \BibitemOpen
  \bibfield  {author} {\bibinfo {author} {\bibfnamefont {M.}~\bibnamefont
  {Thiel}}, \bibinfo {author} {\bibfnamefont {C.}~\bibnamefont {Sfienti}},
  \bibinfo {author} {\bibfnamefont {J.}~\bibnamefont {Piekarewicz}}, \bibinfo
  {author} {\bibfnamefont {C.~J.}\ \bibnamefont {Horowitz}}, \ and\ \bibinfo
  {author} {\bibfnamefont {M.}~\bibnamefont {Vanderhaeghen}},\ }\href {\doibase
  10.1088/1361-6471/ab2c6d} {\bibfield  {journal} {\bibinfo  {journal} {J.
  Phys.}\ }\textbf {\bibinfo {volume} {G46}},\ \bibinfo {pages} {093003}
  (\bibinfo {year} {2019})}\BibitemShut {NoStop}%
\bibitem [{\citenamefont {Psaltis}\ \emph {et~al.}(2014)\citenamefont
  {Psaltis}, \citenamefont {\"Ozel},\ and\ \citenamefont
  {Chakrabarty}}]{Psaltis:2013fha}%
  \BibitemOpen
  \bibfield  {author} {\bibinfo {author} {\bibfnamefont {D.}~\bibnamefont
  {Psaltis}}, \bibinfo {author} {\bibfnamefont {F.}~\bibnamefont {\"Ozel}}, \
  and\ \bibinfo {author} {\bibfnamefont {D.}~\bibnamefont {Chakrabarty}},\
  }\href {\doibase 10.1088/0004-637X/787/2/136} {\bibfield  {journal} {\bibinfo
   {journal} {Astrophys. J.}\ }\textbf {\bibinfo {volume} {787}},\ \bibinfo
  {pages} {136} (\bibinfo {year} {2014})}\BibitemShut {NoStop}%
\bibitem [{\citenamefont {Damour}\ \emph {et~al.}(1992)\citenamefont {Damour},
  \citenamefont {Soffel},\ and\ \citenamefont {Xu}}]{Damour:1991yw}%
  \BibitemOpen
  \bibfield  {author} {\bibinfo {author} {\bibfnamefont {T.}~\bibnamefont
  {Damour}}, \bibinfo {author} {\bibfnamefont {M.}~\bibnamefont {Soffel}}, \
  and\ \bibinfo {author} {\bibfnamefont {C.-m.}\ \bibnamefont {Xu}},\ }\href
  {\doibase 10.1103/PhysRevD.45.1017} {\bibfield  {journal} {\bibinfo
  {journal} {Phys. Rev.}\ }\textbf {\bibinfo {volume} {D45}},\ \bibinfo {pages}
  {1017} (\bibinfo {year} {1992})}\BibitemShut {NoStop}%
\bibitem [{\citenamefont {Flanagan}\ and\ \citenamefont
  {Hinderer}(2008)}]{Flanagan:2007ix}%
  \BibitemOpen
  \bibfield  {author} {\bibinfo {author} {\bibfnamefont {E.~E.}\ \bibnamefont
  {Flanagan}}\ and\ \bibinfo {author} {\bibfnamefont {T.}~\bibnamefont
  {Hinderer}},\ }\href {\doibase 10.1103/PhysRevD.77.021502} {\bibfield
  {journal} {\bibinfo  {journal} {Phys. Rev.}\ }\textbf {\bibinfo {volume}
  {D77}},\ \bibinfo {pages} {021502} (\bibinfo {year} {2008})}\BibitemShut
  {NoStop}%
\bibitem [{\citenamefont {Binnington}\ and\ \citenamefont
  {Poisson}(2009)}]{Binnington:2009bb}%
  \BibitemOpen
  \bibfield  {author} {\bibinfo {author} {\bibfnamefont {T.}~\bibnamefont
  {Binnington}}\ and\ \bibinfo {author} {\bibfnamefont {E.}~\bibnamefont
  {Poisson}},\ }\href {\doibase 10.1103/PhysRevD.80.084018} {\bibfield
  {journal} {\bibinfo  {journal} {Phys. Rev.}\ }\textbf {\bibinfo {volume}
  {D80}},\ \bibinfo {pages} {084018} (\bibinfo {year} {2009})}\BibitemShut
  {NoStop}%
\bibitem [{\citenamefont {Damour}\ \emph {et~al.}(2012)\citenamefont {Damour},
  \citenamefont {Nagar},\ and\ \citenamefont {Villain}}]{Damour:2012yf}%
  \BibitemOpen
  \bibfield  {author} {\bibinfo {author} {\bibfnamefont {T.}~\bibnamefont
  {Damour}}, \bibinfo {author} {\bibfnamefont {A.}~\bibnamefont {Nagar}}, \
  and\ \bibinfo {author} {\bibfnamefont {L.}~\bibnamefont {Villain}},\ }\href
  {\doibase 10.1103/PhysRevD.85.123007} {\bibfield  {journal} {\bibinfo
  {journal} {Phys. Rev.}\ }\textbf {\bibinfo {volume} {D85}},\ \bibinfo {pages}
  {123007} (\bibinfo {year} {2012})}\BibitemShut {NoStop}%
\bibitem [{\citenamefont {Hinderer}(2008)}]{Hinderer:2007mb}%
  \BibitemOpen
  \bibfield  {author} {\bibinfo {author} {\bibfnamefont {T.}~\bibnamefont
  {Hinderer}},\ }\href {\doibase 10.1086/533487} {\bibfield  {journal}
  {\bibinfo  {journal} {Astrophys. J.}\ }\textbf {\bibinfo {volume} {677}},\
  \bibinfo {pages} {1216} (\bibinfo {year} {2008})}\BibitemShut {NoStop}%
\bibitem [{\citenamefont {Hinderer}\ \emph {et~al.}(2010)\citenamefont
  {Hinderer}, \citenamefont {Lackey}, \citenamefont {Lang},\ and\ \citenamefont
  {Read}}]{Hinderer:2009ca}%
  \BibitemOpen
  \bibfield  {author} {\bibinfo {author} {\bibfnamefont {T.}~\bibnamefont
  {Hinderer}}, \bibinfo {author} {\bibfnamefont {B.~D.}\ \bibnamefont
  {Lackey}}, \bibinfo {author} {\bibfnamefont {R.~N.}\ \bibnamefont {Lang}}, \
  and\ \bibinfo {author} {\bibfnamefont {J.~S.}\ \bibnamefont {Read}},\ }\href
  {\doibase 10.1103/PhysRevD.81.123016} {\bibfield  {journal} {\bibinfo
  {journal} {Phys. Rev.}\ }\textbf {\bibinfo {volume} {D81}},\ \bibinfo {pages}
  {123016} (\bibinfo {year} {2010})}\BibitemShut {NoStop}%
\bibitem [{\citenamefont {Damour}\ and\ \citenamefont
  {Nagar}(2009)}]{Damour:2009vw}%
  \BibitemOpen
  \bibfield  {author} {\bibinfo {author} {\bibfnamefont {T.}~\bibnamefont
  {Damour}}\ and\ \bibinfo {author} {\bibfnamefont {A.}~\bibnamefont {Nagar}},\
  }\href {\doibase 10.1103/PhysRevD.80.084035} {\bibfield  {journal} {\bibinfo
  {journal} {Phys. Rev.}\ }\textbf {\bibinfo {volume} {D80}},\ \bibinfo {pages}
  {084035} (\bibinfo {year} {2009})}\BibitemShut {NoStop}%
\bibitem [{\citenamefont {Postnikov}\ \emph {et~al.}(2010)\citenamefont
  {Postnikov}, \citenamefont {Prakash},\ and\ \citenamefont
  {Lattimer}}]{Postnikov:2010yn}%
  \BibitemOpen
  \bibfield  {author} {\bibinfo {author} {\bibfnamefont {S.}~\bibnamefont
  {Postnikov}}, \bibinfo {author} {\bibfnamefont {M.}~\bibnamefont {Prakash}},
  \ and\ \bibinfo {author} {\bibfnamefont {J.~M.}\ \bibnamefont {Lattimer}},\
  }\href {\doibase 10.1103/PhysRevD.82.024016} {\bibfield  {journal} {\bibinfo
  {journal} {Phys. Rev.}\ }\textbf {\bibinfo {volume} {D82}},\ \bibinfo {pages}
  {024016} (\bibinfo {year} {2010})}\BibitemShut {NoStop}%
\bibitem [{\citenamefont {Fattoyev}\ \emph {et~al.}(2013)\citenamefont
  {Fattoyev}, \citenamefont {Carvajal}, \citenamefont {Newton},\ and\
  \citenamefont {Li}}]{Fattoyev:2012uu}%
  \BibitemOpen
  \bibfield  {author} {\bibinfo {author} {\bibfnamefont {F.~J.}\ \bibnamefont
  {Fattoyev}}, \bibinfo {author} {\bibfnamefont {J.}~\bibnamefont {Carvajal}},
  \bibinfo {author} {\bibfnamefont {W.~G.}\ \bibnamefont {Newton}}, \ and\
  \bibinfo {author} {\bibfnamefont {B.-A.}\ \bibnamefont {Li}},\ }\href
  {\doibase 10.1103/PhysRevC.87.015806} {\bibfield  {journal} {\bibinfo
  {journal} {Phys. Rev.}\ }\textbf {\bibinfo {volume} {C87}},\ \bibinfo {pages}
  {015806} (\bibinfo {year} {2013})}\BibitemShut {NoStop}%
\bibitem [{\citenamefont {Steiner}\ \emph {et~al.}(2015)\citenamefont
  {Steiner}, \citenamefont {Gandolfi}, \citenamefont {Fattoyev},\ and\
  \citenamefont {Newton}}]{Steiner:2014pda}%
  \BibitemOpen
  \bibfield  {author} {\bibinfo {author} {\bibfnamefont {A.~W.}\ \bibnamefont
  {Steiner}}, \bibinfo {author} {\bibfnamefont {S.}~\bibnamefont {Gandolfi}},
  \bibinfo {author} {\bibfnamefont {F.~J.}\ \bibnamefont {Fattoyev}}, \ and\
  \bibinfo {author} {\bibfnamefont {W.~G.}\ \bibnamefont {Newton}},\ }\href
  {\doibase 10.1103/PhysRevC.91.015804} {\bibfield  {journal} {\bibinfo
  {journal} {Phys. Rev.}\ }\textbf {\bibinfo {volume} {C91}},\ \bibinfo {pages}
  {015804} (\bibinfo {year} {2015})}\BibitemShut {NoStop}%
\bibitem [{\citenamefont {Piekarewicz}\ and\ \citenamefont
  {Fattoyev}(2019{\natexlab{b}})}]{Piekarewicz:2018sgy}%
  \BibitemOpen
  \bibfield  {author} {\bibinfo {author} {\bibfnamefont {J.}~\bibnamefont
  {Piekarewicz}}\ and\ \bibinfo {author} {\bibfnamefont {F.~J.}\ \bibnamefont
  {Fattoyev}},\ }\href {\doibase 10.1103/PhysRevC.99.045802} {\bibfield
  {journal} {\bibinfo  {journal} {Phys. Rev. C}\ }\textbf {\bibinfo {volume}
  {99}},\ \bibinfo {pages} {045802} (\bibinfo {year}
  {2019}{\natexlab{b}})}\BibitemShut {NoStop}%
\bibitem [{\citenamefont {Essick}\ \emph {et~al.}(2020)\citenamefont {Essick},
  \citenamefont {Tews}, \citenamefont {Landry}, \citenamefont {Reddy},\ and\
  \citenamefont {Holz}}]{Essick:2020flb}%
  \BibitemOpen
  \bibfield  {author} {\bibinfo {author} {\bibfnamefont {R.}~\bibnamefont
  {Essick}}, \bibinfo {author} {\bibfnamefont {I.}~\bibnamefont {Tews}},
  \bibinfo {author} {\bibfnamefont {P.}~\bibnamefont {Landry}}, \bibinfo
  {author} {\bibfnamefont {S.}~\bibnamefont {Reddy}}, \ and\ \bibinfo {author}
  {\bibfnamefont {D.~E.}\ \bibnamefont {Holz}},\ }\href {\doibase
  10.1103/PhysRevC.102.055803} {\bibfield  {journal} {\bibinfo  {journal}
  {Phys. Rev. C}\ }\textbf {\bibinfo {volume} {102}},\ \bibinfo {pages}
  {055803} (\bibinfo {year} {2020})}\BibitemShut {NoStop}%
\bibitem [{\citenamefont {Abbott}\ \emph
  {et~al.}(2017{\natexlab{b}})\citenamefont {Abbott} \emph
  {et~al.}}]{Abbott:2016wyt}%
  \BibitemOpen
  \bibfield  {author} {\bibinfo {author} {\bibfnamefont {B.~P.}\ \bibnamefont
  {Abbott}} \emph {et~al.} (\bibinfo {collaboration} {LIGO Scientific,
  Virgo}),\ }\href {\doibase 10.1002/andp.201600209} {\bibfield  {journal}
  {\bibinfo  {journal} {Annalen Phys.}\ }\textbf {\bibinfo {volume} {529}},\
  \bibinfo {pages} {1600209} (\bibinfo {year}
  {2017}{\natexlab{b}})}\BibitemShut {NoStop}%
\bibitem [{\citenamefont {Cutler}\ and\ \citenamefont
  {Flanagan}(1994)}]{Cutler:1994ys}%
  \BibitemOpen
  \bibfield  {author} {\bibinfo {author} {\bibfnamefont {C.}~\bibnamefont
  {Cutler}}\ and\ \bibinfo {author} {\bibfnamefont {E.~E.}\ \bibnamefont
  {Flanagan}},\ }\href {\doibase 10.1103/PhysRevD.49.2658} {\bibfield
  {journal} {\bibinfo  {journal} {Phys. Rev.}\ }\textbf {\bibinfo {volume}
  {D49}},\ \bibinfo {pages} {2658} (\bibinfo {year} {1994})}\BibitemShut
  {NoStop}%
\bibitem [{\citenamefont {Abbott}\ \emph {et~al.}(2020)\citenamefont {Abbott}
  \emph {et~al.}}]{Abbott:2020khf}%
  \BibitemOpen
  \bibfield  {author} {\bibinfo {author} {\bibfnamefont {R.}~\bibnamefont
  {Abbott}} \emph {et~al.} (\bibinfo {collaboration} {LIGO Scientific,
  Virgo}),\ }\href {\doibase 10.3847/2041-8213/ab960f} {\bibfield  {journal}
  {\bibinfo  {journal} {Astrophys. J.}\ }\textbf {\bibinfo {volume} {896}},\
  \bibinfo {pages} {L44} (\bibinfo {year} {2020})}\BibitemShut {NoStop}%
\bibitem [{\citenamefont {Fattoyev}\ \emph {et~al.}(2020)\citenamefont
  {Fattoyev}, \citenamefont {Horowitz}, \citenamefont {Piekarewicz},\ and\
  \citenamefont {Reed}}]{Fattoyev:2020cws}%
  \BibitemOpen
  \bibfield  {author} {\bibinfo {author} {\bibfnamefont {F.~J.}\ \bibnamefont
  {Fattoyev}}, \bibinfo {author} {\bibfnamefont {C.~J.}\ \bibnamefont
  {Horowitz}}, \bibinfo {author} {\bibfnamefont {J.}~\bibnamefont
  {Piekarewicz}}, \ and\ \bibinfo {author} {\bibfnamefont {B.}~\bibnamefont
  {Reed}},\ }\href {\doibase 10.1103/PhysRevC.102.065805} {\bibfield  {journal}
  {\bibinfo  {journal} {Phys. Rev. C}\ }\textbf {\bibinfo {volume} {102}},\
  \bibinfo {pages} {065805} (\bibinfo {year} {2020})}\BibitemShut {NoStop}%
\bibitem [{\citenamefont {Margalit}\ and\ \citenamefont
  {Metzger}(2017)}]{Margalit:2017dij}%
  \BibitemOpen
  \bibfield  {author} {\bibinfo {author} {\bibfnamefont {B.}~\bibnamefont
  {Margalit}}\ and\ \bibinfo {author} {\bibfnamefont {B.~D.}\ \bibnamefont
  {Metzger}},\ }\href {\doibase 10.3847/2041-8213/aa991c} {\bibfield  {journal}
  {\bibinfo  {journal} {Astrophys. J.}\ }\textbf {\bibinfo {volume} {850}},\
  \bibinfo {pages} {L19} (\bibinfo {year} {2017})}\BibitemShut {NoStop}%
\bibitem [{\citenamefont {Lu}\ \emph {et~al.}(2020)\citenamefont {Lu},
  \citenamefont {Beniamini},\ and\ \citenamefont {Bonnerot}}]{Lu:2020gfh}%
  \BibitemOpen
  \bibfield  {author} {\bibinfo {author} {\bibfnamefont {W.}~\bibnamefont
  {Lu}}, \bibinfo {author} {\bibfnamefont {P.}~\bibnamefont {Beniamini}}, \
  and\ \bibinfo {author} {\bibfnamefont {C.}~\bibnamefont {Bonnerot}},\ }\href
  {\doibase 10.1093/mnras/staa3372} {\bibfield  {journal} {\bibinfo  {journal}
  {Mon. Not. Roy. Astron. Soc.}\ }\textbf {\bibinfo {volume} {500}},\ \bibinfo
  {pages} {1817} (\bibinfo {year} {2020})}\BibitemShut {NoStop}%
\bibitem [{\citenamefont {Shapiro}(1964)}]{Shapiro:1964}%
  \BibitemOpen
  \bibfield  {author} {\bibinfo {author} {\bibfnamefont {I.~I.}\ \bibnamefont
  {Shapiro}},\ }\href@noop {} {\bibfield  {journal} {\bibinfo  {journal} {Phys.
  Rev. Lett.}\ }\textbf {\bibinfo {volume} {13}},\ \bibinfo {pages} {789}
  (\bibinfo {year} {1964})}\BibitemShut {NoStop}%
\bibitem [{\citenamefont {Baym}()}]{Baym:2019}%
  \BibitemOpen
  \bibfield  {author} {\bibinfo {author} {\bibfnamefont {G.}~\bibnamefont
  {Baym}},\ }\bibfield  {booktitle} {\emph {\bibinfo {booktitle} {Proceedings
  of the 8th International Conference on Quarks and Nuclear Physics
  (QNP2018)}},\ }\href {\doibase 10.7566/JPSCP.26.011001} {\
  10.7566/JPSCP.26.011001}\BibitemShut {NoStop}%
\bibitem [{\citenamefont {Baym}\ \emph {et~al.}(2018)\citenamefont {Baym},
  \citenamefont {Hatsuda}, \citenamefont {Kojo}, \citenamefont {Powell},
  \citenamefont {Song},\ and\ \citenamefont {Takatsuka}}]{Baym:2017whm}%
  \BibitemOpen
  \bibfield  {author} {\bibinfo {author} {\bibfnamefont {G.}~\bibnamefont
  {Baym}}, \bibinfo {author} {\bibfnamefont {T.}~\bibnamefont {Hatsuda}},
  \bibinfo {author} {\bibfnamefont {T.}~\bibnamefont {Kojo}}, \bibinfo {author}
  {\bibfnamefont {P.~D.}\ \bibnamefont {Powell}}, \bibinfo {author}
  {\bibfnamefont {Y.}~\bibnamefont {Song}}, \ and\ \bibinfo {author}
  {\bibfnamefont {T.}~\bibnamefont {Takatsuka}},\ }\href {\doibase
  10.1088/1361-6633/aaae14} {\bibfield  {journal} {\bibinfo  {journal} {Rept.
  Prog. Phys.}\ }\textbf {\bibinfo {volume} {81}},\ \bibinfo {pages} {056902}
  (\bibinfo {year} {2018})}\BibitemShut {NoStop}%
\end{thebibliography}%
\end{document}